\documentclass[
 reprint, amsmath,amssymb, aps,
]{revtex4-1}

\usepackage{graphicx}
\usepackage{dcolumn}
\usepackage{bm}

\newcommand{\be}{\begin{equation}} \newcommand{\ee}{\end{equation}}
\newcommand{\bea}{\begin{eqnarray}} \newcommand{\eea}{\end{eqnarray}}

\begin{document}

\title{Phase Transitions in Cooperative Coinfections: Simulation Results for Networks and Lattices}

\author{Peter Grassberger}\affiliation{Max Planck Institute for the Physics of Complex Systems, Dresden, Germany}
                          \affiliation{JSC, FZ J\"ulich, D-52425 J\"ulich, Germany}
\email{p.grassberger@fz-juelich.de}
\author{Li Chen}\affiliation{Robert Koch-Institute, 13353 Berlin, Germany}
                \affiliation{Max Planck Institute for the Physics of Complex Systems, Dresden, Germany}
\author{Fakhteh Ghanbarnejad} \affiliation{Max Planck Institute for the Physics of Complex Systems, Dresden, Germany}
                              \affiliation{Robert Koch-Institute, 13353 Berlin, Germany}
\author{Weiran Cai}\affiliation{Medical Faculty}
                   \affiliation{Dept. of Electrical and Information Engineering, Technische Universit\"at Dresden, 01307 Dresden, Germany}

\date{\today}

\begin{abstract}

We study the spreading of two mutually cooperative diseases
on different network topologies, and with two microscopic realizations,
both of which are stochastic versions of an SIR type model studied
by us recently in mean field approximation. There it had been found that
cooperativity can lead to first-order spreading/extinction
transitions. However, due to the rapid mixing implied by the mean field
assumption, first order transitions required
non-zero initial densities of sick individuals.
For the stochastic model studied here the results
depend strongly on the underlying network. First order transitions
are found when there are few short but many long loops: (i) No first order
transitions exist on trees and on 2-d lattices with local contacts
(ii) They do exist
on Erd\H{o}s-R\'enyi (ER) networks, on d-dimensional lattices with $d\geq 4$,
and on 2-d lattices with sufficiently long-ranged contacts;
(iii) On 3-d lattices with local contacts the results depend on the microscopic
details of the implementation; (iv) While single infected seeds can always lead
to infinite epidemics on regular lattices, on ER networks one sometimes
needs finite initial densities of infected nodes; (v) In all cases
the first order transitions are actually ``hybrid", i.e. they display also
power law scaling usually associated with second order transitions.
On regular lattices, our model can also be interpreted
as the growth of an interface due to cooperative attachment of two species
of particles. Critically pinned interfaces in this model seem to be in
different universality classes than standard critically pinned interfaces
in models with forbidden overhangs.
Finally, the detailed results mentioned above hold only when both
diseases propagate along the same network of links. If they use different
links, results can be rather different in detail, but are similar overall.

\end{abstract}

\maketitle

\section{Introduction}

In human history the most fatal threat are infectious diseases \cite{Hays:2005}. 
Accordingly, scientists from various disciplines have studied their spreading,
including epidemiologists, applied mathematicians, statisticians, and 
physicists \cite{Anderson:1992}. 
From the perspective of statistical physics, the most fundamental problem
is to understand epidemics when conditions are just barely favorable for 
their outbreak, since the transition from zero to non-zero chance for a large 
epidemic (in an infinite population pool) is akin to a phase transition. The 
most basic epidemic models, including the SIS epidemic (
susceptible-infected-susceptible) and the SIR (susceptible-infected-removed) 
epidemic model \cite{Kermack:1927,Mollison:1977} show continuous (or ``second order") 
transitions in the sense that an epidemic starting from an infinitesimal ``seed" density
just above threshold never reaches more than an infinitesimal fraction of the 
population. But there exist also models with discontinuous (``first order") transitions
\cite{Dodds:2004,Janssen:2004,Bizhani:2012,Martcheva:2006,Reluga:2008,Buldyrev:2010,Parshani:2010,Son:2012,Chen:2013,hebert2015,miller2015complex} 
where this fraction is finite. In the language of critical phenomena this fraction is called an {\it 
order parameter}.

This distinction between continuous and discontinuous transitions (or, as the 
latter are also called in the mathematical literature, ``backward bifurcations") is 
fundamental. In a continuous transition one has universal scaling laws which follow from 
renormalization group ideas \cite{Amit:1984,Janssen:2004}. In particular, the behavior 
in large but finite systems is governed by {\it finite size scaling} (FSS), and one 
has power laws with computable exponents even in the subcritical regime. In the 
case of SIR epidemics, the universality class is that of ordinary percolation (OP) 
\cite{Grassberger:1983}, while for SIS epidemics it is the directed percolation 
universality class \cite{Hinrichsen:2000}. 

Thus when conditions for the spreading of the epidemic improve, one obtains warning signals 
which can be used to initiate counter measures before the actual outbreak. No such warning 
exists in ``pure" first order transitions, but most first order transitions in epidemic 
and percolation models are ``hybrid" 
\cite{Dorogovtsev:2008,Goltsev:2006,Baxter:2011,Bizhani:2012,azimi-2014,miller2015complex,Grass-SOS,Azimi-2015}.
This means that they show a discontinuous jump of the order parameter, but show also some 
universal scaling laws. As we shall see, the same is also true for most of the transitions 
discussed in the present paper.

One ingredient that can lead to discontinuous transitions is {\it cooperativity}. Cooperativity
can exist in two basic forms: Either different nodes in a network can cooperate to infect a 
common neighbor \cite{Dodds:2004,Janssen:2004,Bizhani:2012}, or two (or more) different diseases can 
cooperate \cite{Martcheva:2006,Chen:2013}. In the latter case, such 
infections are called {\it coinfections}, and the joint epidemics are called {\it syndemics}
\cite{singer:2009}. Well known examples are the Spanish Flu and TB or pneumonia \cite{Brundage:2008,Oei:2012},
and HIV and a plethora of other diseases like hepatitis B \& C \cite{Sulkowski:2008}, TB 
\cite{Sharma:2005}, and malaria \cite{AbuRaddad:2006}. 

Real epidemic outbreaks are of course very complex phenomena involving a huge amount of detail
such as latencies, mobility of agents, age structures, varying degrees of (partial) immunity,
seasonal oscillations, spatial randomness, counter measures
like medication and quarantine, and stochastic fluctuations. One of the most fruitful ideas
in statistical physics, most clearly illustrated by the famous Ising model \cite{Amit:1984} 
was to dismiss most of these complications and to study the simplest model showing the basic
features. This is justified theoretically by the concept of universality and its foundation
in the renormalization group. 

In this spirit, a minimal model for cooperative syndemics of two diseases ($A$ and $B$) of 
SIR type was introduced in \cite{Chen:2013}. In order to reduce it to a set of coupled 
ordinary differential equations which can then be treated analytically or by numerical 
integration, even stochastic fluctuations were neglected in \cite{Chen:2013} and the model
was treated by mean field theory. This basically assumes that agents are well mixed (analogous 
to Boltzmann's molecular chaos assumption). It has the obvious drawback that cooperativity
cannot be effective, if the initial fraction of infected agents is infinitesimal. In a 
more realistic modeling, the initially infected agents could form a local cluster or ``droplet", 
within which cooperativity can act and together with which it can spread. Such nucleation 
phenomena are basic for most real first order phase transitions and explain phenomena 
like supercooling of vapor. But due to the perfect mixing they are not possible in the 
model of \cite{Chen:2013}. In spite of this, first order phase transitions were found there,
but only when the initially infected fraction is finite. 

The aim of the present paper is to treat the model of \cite{Chen:2013} as an interacting 
particle system \cite{griffeath:1979,durrett:1994,marro:2005}. Agents are represented by 
nodes on a graph (or, as special type of graphs, a regular lattice), and  each agent can 
be in one of a finite number of discrete states. Infections occur stochastically between 
neighbors on the graph. Time is assumed to be discrete, and agents who got infected by 
disease $A$, say, stay infective during 
exactly one time step, after which they recover and become immune against $A$. But they 
can still catch disease $B$, and indeed they do this with greater probability than ``virgin" 
agents that had not been infected yet at all.

To our surprise, we were not only able to verify the existence of first order transitions
(starting in some cases even from a single doubly infected agent), but we found a rich 
zoo of scenarios depending on the topology of the network. In particular, we found no
first order transitions on trees, on 2-d lattices with local contacts, and on Albert-Barab\'asi 
networks, but we found 
them on 2-d lattices with long range contacts, on Erd\H{o}s-R\'enyi (ER) networks, and 
on 4-d lattices. All discontinuous transitions found in this paper are indeed hybrid. The 
transitions on ER networks seem to represent
the most striking hybrid phase transitions so far studied in the literature.
But the most strange result was found for 3-d lattices with local contacts. There, the 
existence of first order transitions depends on the microscopic realization of the model.
At first sight this might seem to violate universality. But, actually, universality only 
makes statements about models which both have second order transitions. It makes no 
claim that two models with the same symmetry, dimension, etc., must have transitions of the 
same order.

The paper is organized as follows: In Sec. II, we briefly review
the mean field treatment
\cite{Chen:2013}, which will  be helpful to understand the simulation part. In Sec. III,
the two stochastic model versions are precisely defined. There, also the 
difference between two epidemics spreading along the same set of links and two epidemics 
which use different links (sometimes called multiplex networks \cite{gomez:2013} is discussed.  
Specific network types are discussed in Secs. IV to VI: Trees and ER networks (Sec. IV), regular 
lattices with nearest neighbor infections (Sec. V), 2-dimensional lattices with long range 
infections (Sec. VI), and small-world and Albert-Barab\'asi networks (Sec. VII). 
Multiplex networks are shortly discussed in Sec. VIII. Finally, Sec. IX contains conclusions 
and discusses some open problems. In particular, 
we discuss there ``SIC" (susceptible-infected-coinfective) models and their possible relation
to interdependent networks \cite{Buldyrev:2010,Parshani:2010,Son:2012}. 

Some of the results of the present paper were already presented in a short letter 
\cite{cai2015avalanche}.

\section{Mean Field Predictions} \label{Sec:MF}

As in \cite{Chen:2013} we shall only consider the case of two diseases $A$ and $B$.
We will always denote by capital letters ($A,B$) agents who actually have the respective
disease, and by lower-case letters ($a,b$) those who had it in the past. Thus each 
agent can be in one of nine states: 0 (all susceptible), $A$ (infected with disease $A$ 
but not yet infected with $B$, $AB$ (infected with both), up to $ab$ (immune to both).
We assume that the dynamics is described by a set of nine rate equations 
\be
   \frac{dx_i}{dt} = \sum_j \mu_{ij}(x_j - x_i)  + \sum_{jk} \nu_{ijk} x_k (x_j - x_i),
\ee
where $x_i$ is the fraction of the population in state $i$, and where 
$\mu_{ij}$ and $\nu_{ijk}$ are recovery and infection rates. We assume neither invasion
nor birth or death, thus the total population size is fixed.

\begin{figure} [htp]
\centering
\includegraphics[width=5.3cm]{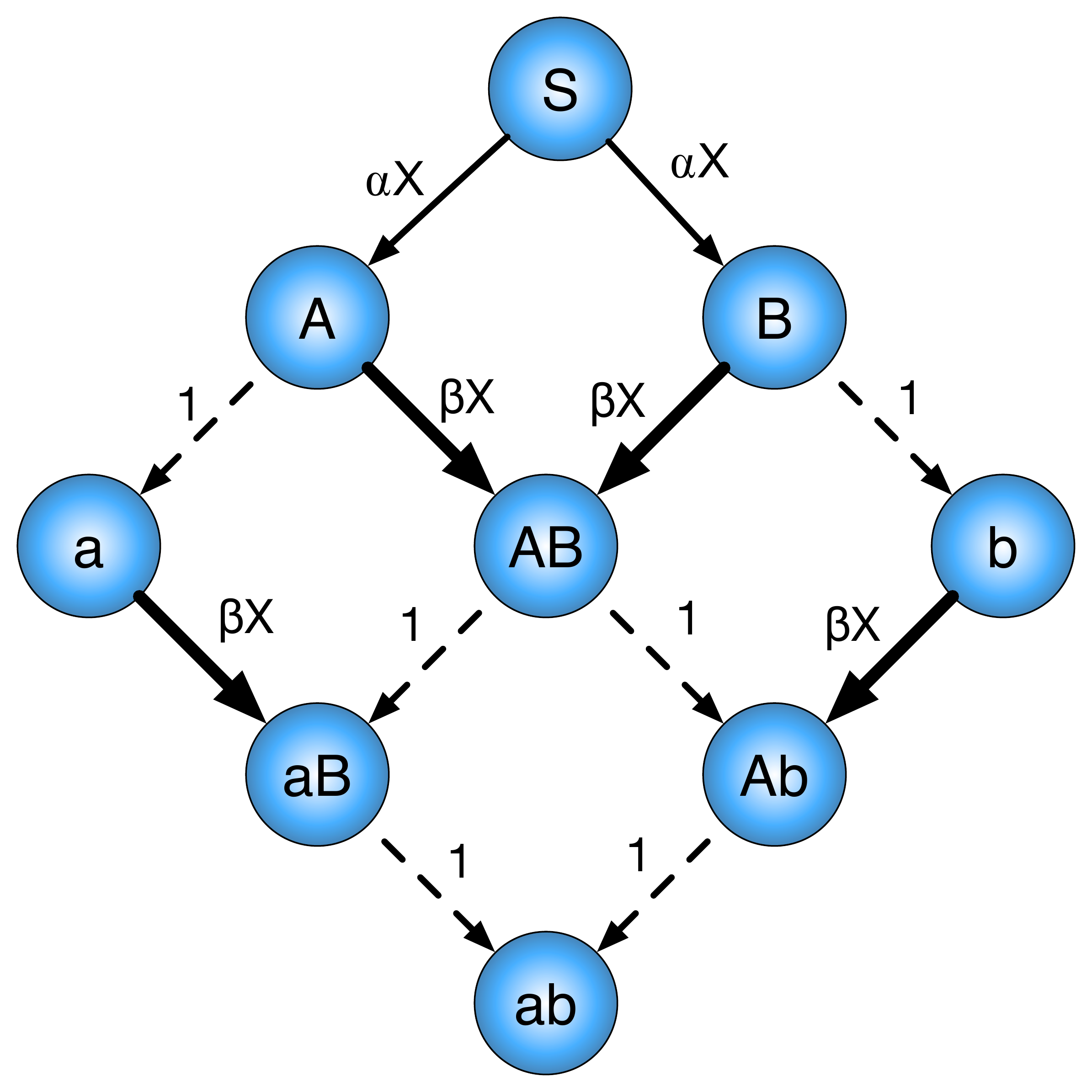}
\caption{Flow chart in two disease coinfection with $A,B$ symmetry and restrictions on the infection 
   rates as discussed in the text. Capital letters $A$ and $B$ represent infective states,
   lower case letters $a$ and $b$ stand for `recovered' ones. Infecting neighbors are not
   indicated explicitly, but it is assumed that all individuals infected with disease $A$, say, have the
   same chances to pass $A$ on to another individual. Thus every infection process occurs with a rate
   proportional to the fraction $X$ of the population having the corresponding disease.
}
\label{Fig:MF:flow}
\end{figure}

In the following we shall also consider only the restricted case of two symmetric 
diseases, where furthermore each agent has the same recovery rate and the same
infectivity. In this case the model can be represented by the flow diagram shown in
Fig.~\ref{Fig:MF:flow}. In particular, the rate with which a susceptible agent 
acquires a disease is then proportional to the fraction of the population that 
carries this disease. In the following we shall denote this fraction by $X =
x_A+x_{Ab}+x_{AB} = x_B+x_{bA}+x_{AB}$, where we have used the $A\leftrightarrow B$
symmetry. In addition we denote by $S = x_0$ the fraction of susceptibles and by $P =
x_A+x_a = x_B+x_b$ those who have or had been infected by one disease but not by 
the other. In terms of these three fractions, the original system of nine ODE's
can be reduced to three ODE's,
\begin{eqnarray} \label{eq:three}
    \dot{S} & = & -2\alpha SX \nonumber \\
    \dot{P} & = &(\alpha S -\beta P)X \nonumber \\
    \dot{X} & = &(\alpha S +\beta P)X -X.
\end{eqnarray}
Here, $\alpha$ is the rate for a primary infection (i.e., for the infection of 
an agent in state 0), while $\beta$ is the rate for secondary infections, and 
the recovery rate was set to unity. 
Cooperativity implies that $\beta > \alpha$, i.e. $C\equiv \beta/\alpha >1$.

\begin{figure} [htp]
\centering
\includegraphics[width=4.2cm]{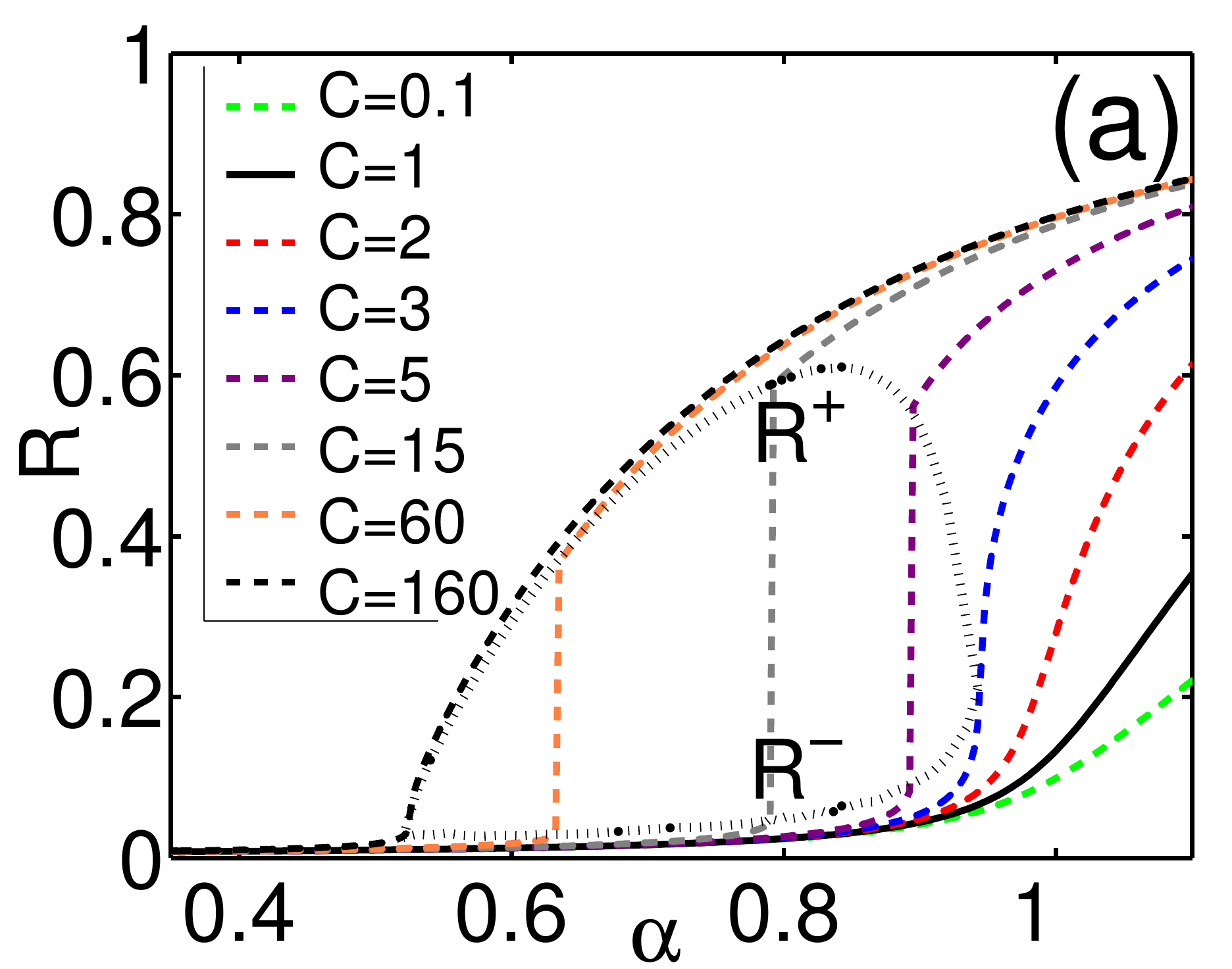}
\includegraphics[width=4.2cm]{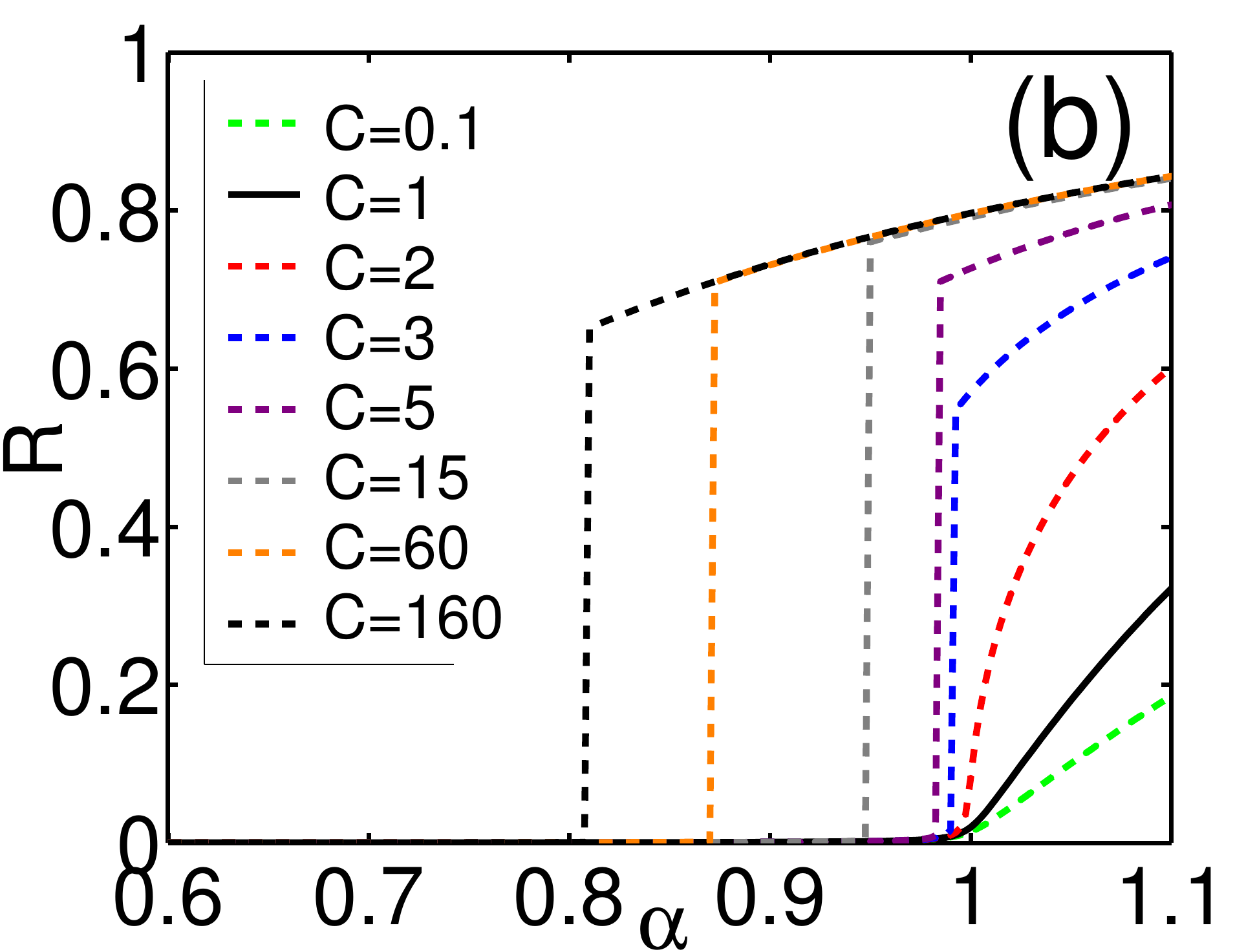}
\caption{(color online) Order parameter $R = 1-S_\infty$ plotted against $\alpha$ for (a) $\epsilon=0.005$
   (where $\epsilon$ is the initial fraction of sick agents) and (b) $\epsilon=10^{-4}$. 
   Each curve corresponds to a different level of cooperativity $C$. The
   dotted lines in panel (a) indicate the upper and lower limits $R^+$ and $R^-$ of the jumps at the 
   first order transitions.
}\label{Fig:MF:phases}
\end{figure}

These equations can then be either integrated numerically or discussed analytically.
The former gives rise to plots like the one in Fig.~\ref{Fig:MF:phases}, 
which suggests that there are discontinuous jumps when $C_{\rm min}<C<C_{\rm max}$,
where both $C_{\rm max}$ and $C_{\rm min}$ depend on the fraction of initially infected agents. The analytic
treatment given in \cite{Chen:2013} leads to exact inequalities which show 
that this interpretation is indeed correct, and that the jumps in 
Fig.~\ref{Fig:MF:phases} are not numerical artifacts.  

An important observation is, however, that first order transitions are seen 
in the direct integration only
when $\epsilon$, the initial fraction of infected agents, is not zero. In the 
limit $\epsilon \to 0$ the fraction of affected agents stays always infinitesimal
when $\alpha$ is below or infinitesimally close to 1 (the threshold for a single 
disease outbreak), and cooperativity cannot become effective as long as $C$ is finite.

\section{Agent based stochastic models}

In our simulations we assume that agents occupy the nodes of a network. They don't move, i.e. we 
can attribute the nine states $0,A,B,a,b,AB,aB,Ab$, and $ab$ directly to the nodes. Time is 
discrete, with every sick node staying sick and infective for exactly one time step. Initially,
all sites are susceptible (state 0), except for the set of {\it seeds}, which are nodes in one 
of the states $A,B,$ or $AB$. We never allow any immune node ($a,b,ab,Ab$ or $Ba$) in the initial 
state. Unless specified differently, we assume that all seed nodes are doubly infected (i.e.,
$AB$).

In principle we could allow both diseases to use different sets of links for their propagation.
This would correspond to multiplex networks \cite{gomez:2013}. We shall discuss this possibility 
later, but in most of the simulations (unless specified differently) both diseases use the same 
set of links. 

The simulations use two data structures: First of all, we store in a character array of size $N$
($N$ is the number of nodes) the state of every node. This array will be updated in each time step.
Secondly, we keep lists of ``active" sites, i.e. of sites in one of the infective states $A,B,AB,aB$ 
and $Ab$. From these lists we can see which nodes can be infected in the next time step and by which
disease(s). Actually, we keep two such lists: One for the sites which are presently active, and one 
for those which will become active in the next time step. At the end of time step, the first is 
replaced by the second.

When implementing this, we have several detailed options, of which we considered two:
\begin{itemize}
\item In the first we assume a latency of exactly one time step. Thus every newly infected node
will not be active (i.e., {\it infective}) until the next time step. Notice that this concerns only
newly acquired diseases. If a node newly infected with disease $A$, say, had already disease $B$,
the latter is not affected. We call this also ``parallel update with delay" or ``SU" (for 
``synchronous updating").\\
\item Alternatively, we can assume that a newly infected site becomes immediately infective. Thus, 
while we work our way through the lists of active sites, we immediately update their disease status.
If done without precautions, this could introduce a dependency on the (arbitrarily but not randomly)
chosen way how we go through the lists, and it could break the $A\leftrightarrow B$ symmetry, if
one type of infection is always done before the other. To avoid such artifacts, we shuffle the 
lists randomly in each time step, and we chose for each doubly infective active site at random 
whether it first infects with $A$ or with $B$. We call this also ``random sequential update 
without delay" or ``AU" (for ``asynchronous updating").
\end{itemize}
Notice that any finite latency period is not supposed to change the universality class of any 
epidemic model (it would only change if latencies can become large, so that a new large time 
scale is introduced). But we should expect that it affects non-universal properties such as
the precise locations of phase transition points. It seems that the difference between the two models 
is for some network topologies sufficient to shift a first order transition outside the range 
allowed by the physical values of the control parameters. 

The two control parameters in our models are:
\begin{itemize}
\item 
$p$, the probability with which an active site
infects a fully immune neighbor (i.e. a neighbor in state 0), and 
\item
$q$ which is the infection
probability for a neighbor who has already (either in the present time step or in the past)
acquired the other disease. 
\end{itemize}
We could of course also differentiate between the latter two possibilities, but we did not
in order to keep the model(s) simple. As we shall see, even with only two different infection 
probabilities there is a rich zoo of behaviors. Cooperativity corresponds obviously to the case 
$q>p$. The opposite case $q<p$ will not be considered in the present paper.

For moderately large networks we let the epidemic proceed until all activity has died out, and measure 
then the properties of the clusters of immunes $a,b,$ or $ab$. In addition we also performed 
simulations on very large networks where this would not be feasible. This concerns mostly regular 
lattices, where we used up to $> 10^9$ nodes. In these cases we stopped the epidemic before it 
could reach the boundary (or, if periodic or helical boundary conditions were used, before it 
could wrap around the torus). In this way we can effectively simulate the finite time behavior 
on infinite lattices, and could compare the growth of the set of active sites with the growth
known for ordinary percolation \cite{marro:2005,Hinrichsen:2000,Grassberger:1983}.

In the following sections we shall only discuss cases with perfect symmetry between the two 
diseases, where also both diseases use the same set of links. This is of course not very 
realistic, as many diseases have their own way of spreading. Alternatively we could consider
multiplex networks (see Sec. VII), where each disease has its own set of links which is independent 
of the links used by the other disease. The fact that this can lead to completely different 
behavior is best illustrated by Erd\H{o}s-R\'enyi (ER) networks. As far as single 
disease are concerned, the spreading on ER networks is of mean field 
type \cite{Bollobas:2001,Newman:2001,Newman:2002}. The same is true for multiple diseases, if 
their link sets are independent. Consider a doubly infected node on an ER network with finite 
mean degree $\langle k\rangle$. If this node is infecting neighbors with probability $p$, then 
the chance for one of these neighbors to become doubly infected will be finite if the same links 
are used by both networks, while it will be $\propto 1/N$ for multiplex networks. Thus a double 
epidemic can spread even from a single infected seed if the same links are used, in contrast to 
the mean field behavior discussed in Sec. II. Ultimately, this is the basis for the intricacies 
found in the next section.  

\section{Trees and Erd\H{o}s-R\'enyi Networks}

Erd\H{o}s-R\'enyi networks are random networks of $N$ nodes where any pair of nodes is linked
with probability $\langle k\rangle/2N$, with $\langle k\rangle$ being the average degree of the nodes. 
In the interesting case 
of finite $\langle k\rangle$ and large $N$ the networks are sparse, and thus there are no small loops. More 
precisely, the chance that a randomly picked node is on a loop of finite length $\ell$ tends to 
zero $\sim 1/N$, when $N\to\infty$ \cite{Bollobas:2001}. Thus ER networks are locally tree-like.
As a consequence, critical phenomena in spin models on trees and on ER networks are usually in the same (mean-field) 
universality class. For percolation, the situation is a bit more complicated, since on trees there exists an
entire critical phase with two critical end points \cite{Hasega:2014}. Yet the situation is similar for trees
and for ER models, since the classical Flory theory
based on Cayley trees \cite{Stauffer:1994} yields for the lower end point $p_{c1}$ (where infinite clusters first 
arise) the same mean field critical exponents as theories 
based on ER networks \cite{Bollobas:2001,Newman:2001,Newman:2002}. As we shall see, this is dramatically
different for cooperative coinfections.

\subsection{Trees, single node seeds}
 
The situation is most simple for epidemics starting from one doubly infected node on a tree, with all other 
nodes being in state 0. Due to the absence of loops and because the epidemic can only spread away from the 
seed (all nodes on the backward path are immune), the spreading of coinfections is qualitatively always the 
same as for the spreading of a single disease. More precisely, in the case of a common network for $A$ and $B$
the threshold for either disease to spread is precisely at $p= p_{c1} \equiv \langle k \rangle/ \langle k(k-1) \rangle$ 
\cite{Newman:2002} \footnote{In order to avoid confusion, we should point out that even above
this value the average fraction of immune sites (the average order parameter) is still zero, and becomes 
non-zero only for $p = p_{c2} = 1$ \cite{Hasega:2014}. In spite of this, $p_{c1}$ is often called the threshold 
\cite{Stauffer:1994}.} for both
models (with and without latency). For the model with latency the threshold for both diseases to spread together (i.e. 
for having a large cluster of doubly infected nodes) is
$p= \sqrt{p_{c1}}$ (independent of $q$), while it is $pq = p_{c1}$ for the model without latency. For two independent 
multiplex networks, the latter two thresholds are replaced by ${\cal O}(p_{c1} N^{1/2})$, i.e. if there are 
no strong hubs the coinfection can survive only when $\langle k \rangle \sim N^{1/2}$. In all these cases the 
transition is continuous, i.e. second order.

Thus trees are trivial from this point of view, but there is still one interesting aspect. The spreading 
of epidemics on trees is mathematically described by a branching process \cite{athreya:1972}. In a standard 
critical branching process, the survival probability decays as $P(t) \sim 1/t$, while the average number $N(t)$
of off-springs at time $t$ is constant. Finally, if the control parameter is a distance 
$\epsilon$ above the critical point, $P(\infty) \sim \epsilon$. All these apply
to the model with latency (for any $q$), and to the model without latency if $q<1$. But the case $q=1$
in the model without latency is different, as it corresponds to a doubly critical process, if $p=p_c$.

To see this, it is useful to reduce the possible node states to three: uninfected (index 0), singly infected
(``$s$") and doubly infected (``$d$"). The model without latency is then described by the following transitions:
\be
     0 \to 0 : \quad 1
\ee
\bea  
     s & \to 0 : & \quad 1-p \nonumber \\ 
     s & \to s : & \quad p
\eea
\bea
     d & \to 0 : & \quad (1-p)^2 \nonumber \\
     d & \to s : & \quad (2-p-q)p \nonumber \\
     d & \to d : & \quad pq
\eea
These are the possible transitions and their rates along any link from a mother to a daughter node. Such
branching processes with two types of particles, where one type can only reproduce itself while the other 
can reproduce and produce particles of the first type, have been studied previously as models for 
cancer growth, where $s$ is a malign cell and $d$ is benign \cite{antal:2011}. 

Since there are in average $\langle k -1\rangle$ links from mother to daughter, an $s$ node has in average 
$x=p\langle k -1\rangle$
descendents of its own type, while a $d$ node reproduces itself $y=pq\langle k -1\rangle$ times. Thus, 
processes starting with one singly infected node are critical when
$x=1$, while processes stating with one doubly 
infected node are critical when $y=1$. Since $q\leq 1$, in the latter case also the spreading of $s$ nodes 
is critical. The average numbers $N_s(t)$ and $N_d(t)$ are easily seen to satisfy 
\be
    N_d(t+1) = y N_d(t),\quad N_s(t+1) = z N_d(t) + x N_s(t)
\ee
with $x=(2-p-q)p\langle k -1\rangle$.  If we start with a $d$ node and take $p=p_c$ and $q=1$ (i.e., when 
$x=y=1$) we have then $N_d(t)=1$ and $N_s(t) =t$. Thus, although the process is critical, $N_s$ is not 
constant but increases linearly with time. This also modifies the extinction probability. Using the standard
generating function trick \cite{athreya:1972}, one finds that the probability for $s$ to die out in a process
that starts with a $d$ is 
\be
   P_{s|d}(t) \sim t^{-1/2}
\ee
at the critical point, while it is 
\be
   P_{s|d}(\infty) \sim (p-p_c)^{1/2}
\ee
when $q=1$ and $p>p_c$ \cite{antal:2011}. As we shall see, this difference between $q=1$ and $q<1$ has also 
consequences for ER networks.

\subsection{ER networks, single node seeds} 

This is not at all the case for ER networks. In that case studying the time course of the disease is 
not very illuminating, as it follows the one for trees up to times when loops become important, and after 
that the behavior is rather complicated and does not scale. More interesting is to study the order 
parameter, i.e. the fraction of doubly immune nodes after the epidemic has died. 

\begin{figure}[!t]
\centering
\includegraphics[width=8cm]{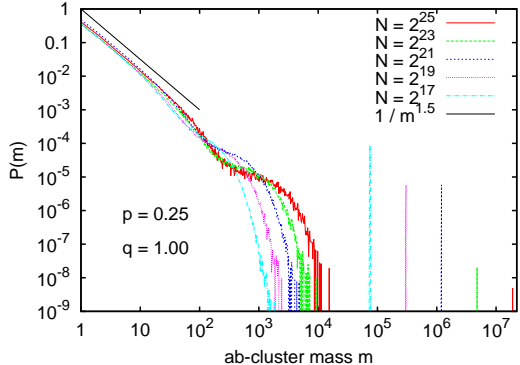}
\caption{(color online) Mass distribution of ER network at $p= 0.25$ and $q=1.0$. Each curve corresponds 
    to a fixed network size. The vertical lines on the right hand side are very narrow peaks, 
    whose widths are smaller than the line width. The straight full line indicates the power law $P(m) \propto 
    m^{-3/2}$. }
\label{ER_massdistrib}
\end{figure}

Throughout this subsection we shall use $\langle k \rangle = 4$. In Fig. \ref{ER_massdistrib}
we show distributions of the masses of the final $ab$-cluster, averaged over a large number of runs on 
the giant connected component. Here the model without latency
was used, but the results for the parallel update with delay are qualitatively the same. For each $N$ we 
first generated a random network by randomly placing $N\langle k \rangle/2$ links and identified the giant 
component. Since this would give 
double links, we then rewired \cite{Maslov:2002} the links sufficiently often to eliminate all such double 
links. Then we run ${\cal O}(N/1000)$ epidemics from randomly chosen seeds, rewiring ${\cal O}(1000)$ times 
after every run. Since this alone would give a frozen degree distribution (rewiring does not change the 
degree distribution), we repeated this entire procedure until we had $\sim 10^9$ starts for each $N$.

\begin{figure}
\centering
\includegraphics[width=8cm]{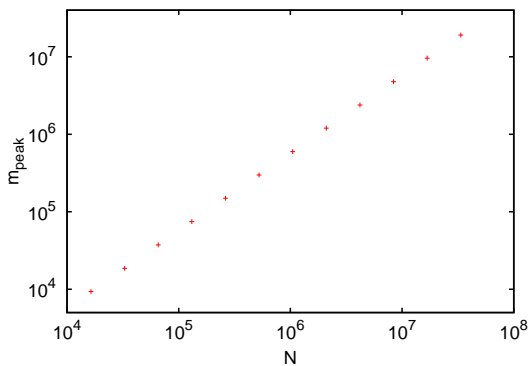}
\caption{(color online) Positions of the right hand peaks in Fig. \ref{ER_massdistrib}, i.e. average masses of the 
``giant" $ab$-clusters, for $p= 0.25$ and $q=1.0$, plotted against $N$. Error bars are smaller than 
the symbol sizes. The data are fitted perfectly by a straight line with slope 1, i.e. giant 
$ab$-clusters contain a fixed fraction of the nodes.}
\label{ER_massdistrib_peaks}
\end{figure}

\begin{figure}[]
\centering
\includegraphics[width=8cm]{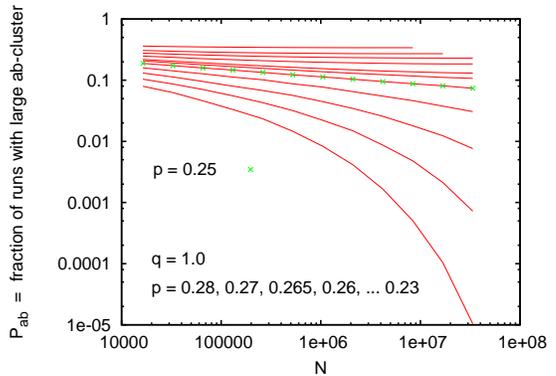}
\caption{(color online) Fraction of events that lead to giant $ab$-clusters which contribute to the peaks in 
   Fig.~\ref{ER_massdistrib}. Each curve corresponds to one value of $p$, while $q=1$ for all curves.
   The curve for $p= p_c$ is indicated by crosses.}
\label{ER_survive}
\end{figure}

\begin{figure}[]
\centering
\includegraphics[width=8cm]{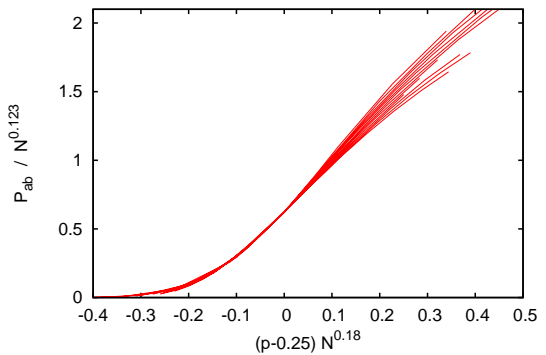}
\caption{(color online) Same data as in Fig. \ref{ER_survive}, but plotted in a way that suggests a data 
   collapse according to the FSS ansatz Eq.~(\ref{FSS}).}
\label{ER_survive-collaps}
\end{figure}

The data shown in Fig. \ref{ER_massdistrib} are for $p=p_c=0.25$ and $q=1$. For these values the single 
disease dynamics would be critical, and indeed the mass distributions for the clusters of singly immune sites
show the power laws $P(m) \sim m^{1-\tau}$ with $\tau = 2.5$ known from ordinary percolation
\cite{Stauffer:1994}, with additional peaks at the high mass end due to events where there were also 
giant coinfection clusters (data not shown). 

For small $m$ the distributions shown in Fig. \ref{ER_massdistrib} 
have the same power law, but this power law breaks down for $m\approx 100$. After a region without clear 
scaling properties there is a wide gap where $P(m)=0$, and finally there is a huge narrow peak for very large 
$m$. A more careful inspection (see Fig. \ref{ER_massdistrib_peaks}) shows that these peaks occur at 
$m\propto N$, i. e. they are due to events where the $ab$-cluster contained a finite fraction of nodes.
This is in striking contrast to critical OP, where the percolating cluster contains a vanishing fraction 
of nodes at the critical point.

\begin{figure}[]
\centering
\includegraphics[width=8cm]{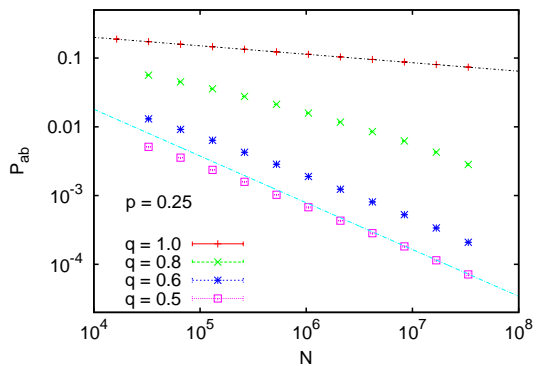}
\vglue -4mm
\caption{(color online) Log-log plot of the probabilities $P_{\rm /ab}$ to generate large $ab$-clusters versus $N$. Each curve 
is for a different value of $q$, while $p$ is fixed at the critical point $p=1/\langle k\rangle$. The two 
straight lines indicate power laws $N^{-0.123}$ (which gives a perfect fit for $q=1$) and $N^{-0.67}$,
which is the asymptotic value for $q<1$.}
\label{P-versus-N}
\end{figure}

One might suspect that this doubly-peaked shape of the mass distribution results simply from the fact 
that the data shown in Fig.~\ref{ER_massdistrib} correspond  already to the supercritical regime. That this 
is not so (at least not in a naive sense!) is seen from Fig. \ref{ER_survive}. There we plot the fraction 
$P_{\rm ab}$ of runs that lead 
to a giant $ab$-cluster, where ``giant" $ab$-clusters are very clearly defined by the very broad valley
separating the peak from the left hand part of the mass distribution \footnote{In a straightforward implementation
most CPU time would be spent in following the evolution of giant $ab$-clusters to the very end. This is not needed,
if one is interested only in the values of $P_{\rm ab}$. Thus in obtaining the data for 
Figs.~\ref{ER_massdistrib}, \ref{ER_survive-collaps}, and \ref{P-versus-N} we interrupted the evolution when the 
size of the $ab$-cluster was above a threshold which had been determined before in runs with lower statistics. 
This introduces a negligible bias, but without this trick it would have been impossible to obtain this high
statistics.}. This fraction is plotted against $N$
for several values of $p$ near $p_c$, and with $q=1$ in all cases. We see that $P_{\rm coinfect}$ decays 
fast with $N$ for all $p<p_c$, while it approaches finite values for all $p>p_c$. At $p=p_c$ it seems to 
obey a power law 
\be
   P_{\rm ab} \sim N^{-\gamma}                                         \label{Pab}
\ee
with $\gamma = 0.123\pm 0.001$.

The data shown in Fig. \ref{ER_survive} can be made to collapse reasonably well by plotting $N^\gamma P_{\rm ab}$
against $(p-p_c) N^{0.18}$ (see Fig. \ref{ER_survive-collaps}, which suggests a finite size scaling (FSS) ansatz
\be
   P_{\rm ab}(N,p) \sim (p-p_c)^{\beta'} \Phi((p-p_c)^\nu N)    \label{FSS}
\ee
with $\nu' = 5.5\pm .3$ and $\beta' = \gamma \nu = 0.67\pm .04$.

For $p=p_c$ we also estimated the probabilities that there exists a giant single disease cluster, without a
large cluster of the other disease. For OP, the chance to hit a giant cluster on an ER network of size $N$ 
decreases as $P_a \sim N^{-1/3}$ \cite{Bollobas:2001}. Our data (for $q=1$) are not very precise, since 
giant single disease clusters are not cleanly separated from small clusters (in contrast to Fig.~\ref{ER_massdistrib}),
but our best estimate is $P_a \sim N^{-0.13(2)}$, i.e. roughly the same decay as for giant $ab$-clusters. Thus 
disease $b$, even if it finally dies out, has a positive effect on the survival of disease $a$, since it 
makes the $a$-cluster grow faster during early times.

All this indicates that $p=p_c$ is also the critical point for coinfections, and that $P_{\rm ab}$
shows qualitatively the same scaling (although with different exponents) as the probability for a random 
seed in OP to grow into the infinite incipient cluster. Indeed, for $N\to\infty$ and $p>p_c$, the latter is 
given for OP by $P(p=p_c+\epsilon) \sim \epsilon^\beta$, which is in that case also the scaling of the 
probability with which an infinite incipient cluster infects a random node. Since OP is a purely geometric 
problem, both are simply related to the density of the 
infinite cluster. This is no longer true in the present case. As we have seen, the density of the 
infinite $ab$-cluster is independent of $N$, so that the order parameter exponent measured via its density is 
$\beta = 0$. Models were the density of an infinite cluster and the chance to generate this cluster 
scale with different powers $\beta$ and $\beta'$ are well known \cite{mendes:1994,Janssen:2004,grassberger:2006},
but usually $\beta$ and $\beta'$ are both non-zero. The novel feature here is that 
one of them vanishes. Thus, while the transition looks like second order from the point of view of 
cluster growth {\it dynamics}, it looks like first order from the point of view of cluster {\it geometry}. This is a 
striking case of a hybrid transition \cite{Dorogovtsev:2008,Goltsev:2006}.

\begin{figure}[]
\centering
\includegraphics[width=8cm]{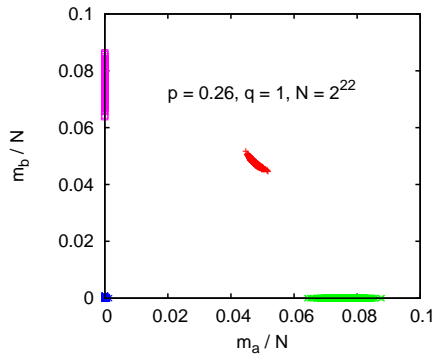}
\caption{(color online) Scatter plot of the joint distribution of single disease clusters, for $p=0.26$ and $q=1.0$. Each
of the four clusters of points is well separated from the others, and corresponds to the cases of no giant
outbreak, giant outbreaks of only one disease, and outbreaks of both diseases. In the latter case (and only
then) also the cluster of nodes which have both diseases is large.}
\label{ER-joint_masses}
\end{figure}

Finally, we should point out that mass distributions for clusters of {\it singly} infected nodes have 
in general two peaks at high masses (i.e. three peaks altogether). One of these peaks is due to events where 
only one of the diseases survives, while the other is from events where both diseases survive. This is even 
more clearly seen when looking at joint distributions of $m_a$ and $m_b$, as shown for one particular set 
of control parameters in Fig.~\ref{ER-joint_masses}. There we see clearly four components: (1) without any
giant outbreak, (2) with only an $a$-outbreak, (3) with a $b$-outbreak, and (4) with both outbreaks. The 
fourth component is also characterized by giant $ab$-outbreaks, while $m_{ab}$ is small in the first three
components. When $p$ is close to $p_c$, single disease clusters are larger in the components (2,3) without 
giant $ab$-outbreaks than in component (4). This is reversed for large $p$, where most nodes get both 
diseases, if there is a giant $ab$-outbreak.

For $q<1$ results are similar, as long as $q>q^*$, where the value of $q^*$ depends on the detailed model.
For random sequential update without delay, $q^*\approx 0.35$ (more precise estimates will be given later). 
In this regime there exist still large $ab$-clusters containing non-vanishing fractions of all nodes. The 
behavior of $P_{\rm ab}$ is still similar to Fig.~\ref{ER_survive}, but attempted data collapses as in 
Fig. \ref{ER_survive-collaps} are even less perfect. Indeed, as shown in Fig.~\ref{P-versus-N}, there is
a cross-over from $N^{-\gamma}$ for $q=1$ to $N^{-2/3}$ for $q\to q^*$. The latter is the power law 
expected for two independent critical diseases spreading on ER networks \cite{Bollobas:2001}. Also the 
dependence of $P_{\rm ab}$ on $p$ for $N\to\infty$ and $q<1$ is as expected for two independent diseases 
\cite{Bollobas:2001}, $P_{\rm ab}\sim (p-p_c)^{2/3}$. The fact that there are different power laws for $q=1$
and $q<1$ are directly related to the discussion for trees in the previous subsection.

Thus, as far as the probabilities to lead to giant clusters is concerned, two cooperating diseases with $q<1$ 
are essentially independent. This is not true for the sizes of the giant clusters, where we find results 
similar to those shown in Fig.~\ref{ER-joint_masses}. In particular, whenever both single diseases have 
giant outbreaks, there is also a giant cluster of $ab$-nodes. The probability that there are two coexisting 
giant single disease clusters without large overlap is zero. This difference is easily explained by the 
fact that the decisions whether there are giant outbreaks or not are made at early times, when the 
networks still look tree like. The structures of the giant clusters are, however, decided at late times
when (large) loops are abundant.

\begin{figure}[]
\centering
\includegraphics[width=8cm]{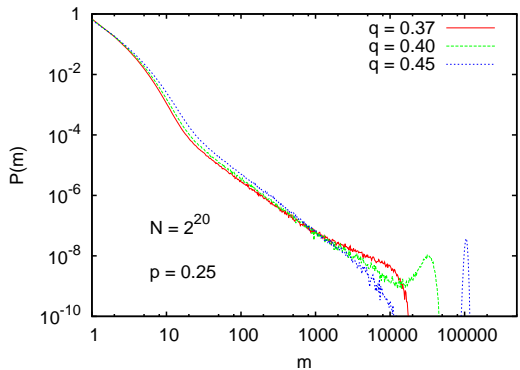}
\vglue -4mm
\caption{(color online) Mass distributions analogous to Fig.~\ref{ER_massdistrib}, but for three values of $q$ slightly 
larger than $q^*$. One sees clearly that the two peaks coalesce as $q^*$ is approached. Data for 
$q=q^*$ are not shown due to the very larger finite size effects which make their interpretation 
difficult.} 
\label{P_m-N}
\end{figure}

As $q^*$ is approached from above (keeping $p=1/\langle k\rangle$), the double-peak structure of the mass 
distribution $P(m)$ becomes more shallow (see Fig.~\ref{P_m-N}), until it disappears when $q=q^*$, and $P(m)$
has a single maximum for $q<q^*$. Thus $q=q^*$ is a (multi-)critical point.

\subsection{ER networks, multiple node seeds}

The data shown in the previous subsection suggest that there exists already the possibility for having 
a giant $ab$-cluster even for $p<p_c$, but that this cluster just cannot be infected by a single node 
seed. This would also be suggested by the analogy with the mean field model, where multiple seeds are needed
to see the full phase structure.

\begin{figure}[]
\centering
\includegraphics[width=8cm]{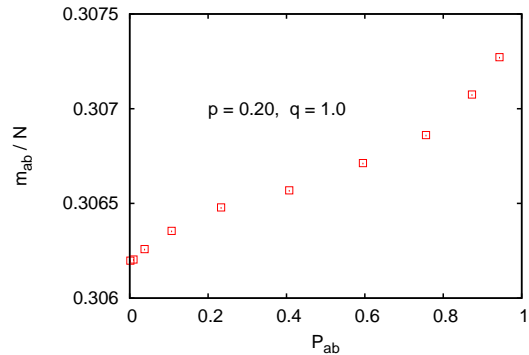}
\vglue -4mm
\caption{(color online) Average masses of ``giant" $ab$-clusters for $N=2^{20}, p = 0.20$, and $q=1$, plotted against the 
probability to obtain such giant $ab$-clusters. Each point corresponds to a different value of $n$, where $n$
is the size of the ``seed" (the number of initially infected nodes). The values of $n$ are $200, 240, 280, 
\ldots, 560$, with $n$ increasing from left to right.}
\label{ER-mass_versus_m0}
\end{figure}

Indeed, if we use a value of $p$ slightly below $p_c$ and start with a ``seed" of $n$ randomly located doubly infected 
nodes, we find qualitatively similar behavior to that seen in Fig. \ref{ER_massdistrib}, but with a much higher 
peak on the right hand side. Unless one goes to $p\ll p_c$ and to very large $n$, this peak is still well
separated from the rest of the distribution, and its position is essentially independent of $n$ (see 
Fig.~\ref{ER-mass_versus_m0} -- for larger $N$ even less dependence on $n$ is found). 
Notice that we did not check that $ab$-clusters whose masses are shown in Fig.~\ref{ER-mass_versus_m0} are 
connected. But the very weak dependence on $m_0$ proves that they are, up to very small disconnected components. 

\begin{figure}[]
\centering
\includegraphics[width=8.6cm]{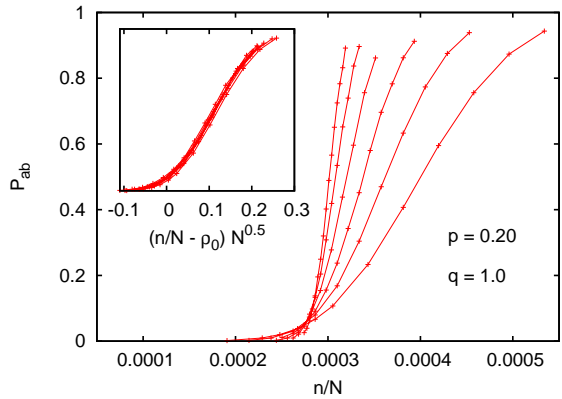}
\vglue -4mm
\caption{(color online) Probabilities $P_{\rm ab}$ versus seed size $n/N$, for $p=0.20$ and $q=1$. Each curve corresponds to one value
of $N$, with $N=2^{20}, 2^{21},\ldots, 2^{25}$ (steeper curves correspond to larger $N$). The insert shows a data
collapse, obtained by plotting these data against $(n/N-\rho_0(p,q))N^x$ with $\rho_0 = 0.000282$ and $x=0.5$.}
\label{ER-P_versus_m0}
\end{figure}

In mean field theory \cite{Chen:2013}, it was found that the seed had to contain a finite fraction of all 
nodes, in order to obtain a first order transition. To check whether this is also true on ER graphs, we plotted 
$P_{\rm ab}$ in  Fig.~\ref{ER-P_versus_m0} versus seed size $n/N$, for $q=1$ and for one particular value of $p < p_c$.
We see that indeed the curves become steeper with increasing $N$, and that they all cross an one particular 
seed density $n/N = \rho_0(p,q) \equiv 0.000282(5)$. Plotting these data against $(n/N-\rho_0(p,q))N^x$ gives 
a very good data collapse if $x=0.50(3)$, as shown in the insert in Fig.~\ref{ER-P_versus_m0}.

\begin{figure}[]
\centering
\includegraphics[width=8.6cm]{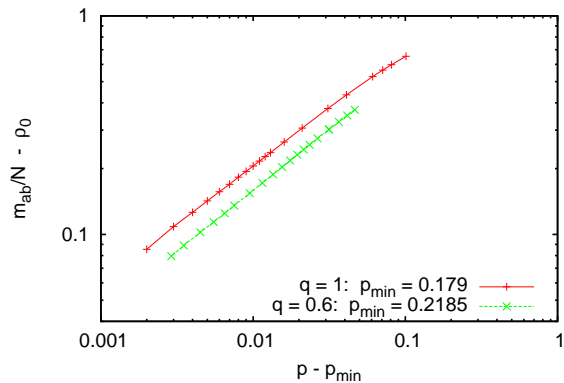}
\vglue -4mm
\caption{(color online) Relative sizes of giant $ab$-clusters versus $p-p_{\rm min}$ for two values of $q$. Both curves 
are compatible with power laws with exponents $0.57(3)$, provided one subtracts the density of seed nodes.
}
\label{ER-ab-sizes}
\end{figure}

The (average) masses of giant $ab$-clusters does depend strongly on $p$. 
Results for $q=1$ and for $q=0.6$ are shown in Fig. \ref{ER-ab-sizes}, where $n$ was chosen for 
each $p$ such that the right hand peak contained roughly between 1 and 10 per cent of all events. Within
this range and for the values of $N$ ($>10^6 - 10^7$) used in these simulations, the 
peak positions were stable within statistical fluctuations, at least for the values of $p$ shown. 
Figure \ref{ER-ab-sizes} suggests that the fraction of the nodes in the next $ab$-cluster becomes equal 
to the fraction $\rho_0$ of initially infected nodes 
at some value $p_{\rm min}$, which is about 0.179 for $q=1$ and 0.2185 for $q=0.6$. The value of $p_{\rm min}$
increases with decreasing $q$, until $p_{\rm min}=p_c$ is reached for $q=q^*$. For $p$ close to $p_{\rm min}$,
the relative giant $ab$-masses satisfy (after subtraction of the seed density) a power law. The observed 
power depends slightly on $q$, but this could be a systematic error, in which case the common power for 
both values of $q$ is $0.57(3)$.

For $p \searrow p_{\rm min}$, the peak becomes wide and the valley separating it from the rest of the distribution
becomes narrow and shallow, until finally for $p=p_{\rm min}$ the entire mass distribution is single humped.
Thus it seems that $p_{\rm min}$ is a further critical point, where the coinfection cluster looses its 
identity. Qualitatively, it resembles the point $\alpha = 0.5$ in Fig. \ref{Fig:MF:phases}.

\subsection{Trees, multiple node seeds}

With the hindsight obtained from studying ER networks, we can go back to trees and look whether we find 
there first order transitions, if we use multiple node seeds. For ER networks, first order transitions were 
always related to the existence of large loops. Since these loops are absent on trees, we expect that 
we will not find any sign of first order transitions, even if we consider multiple node seeds. As we 
shall see, this is verified by our simulations.

We studied Cayley trees with degree $k$=3
for all central (non-leaf) nodes. On such trees the critical value for single diseases, above which an 
infinite cluster can exits, is $p_c=1/2$. Notice, however, that even for $p>p_c$ all epidemics hit only 
a vanishing fraction of nodes in the limit $N\to\infty$, as long as $p<1$. 

As in the case of ER networks, we started with $n$ doubly infected nodes randomly located 
on the network (results are qualitatively the same, if we favor or disfavor leaves at 
this stage). For any $p\leq p_c$, the distribution of $ab$-masses was always found to be unimodal. 
Thus, if we look for non-trivial results, we have to consider $p>p_c$.

\begin{figure}[!t]
\centering
\includegraphics[width=8.7cm]{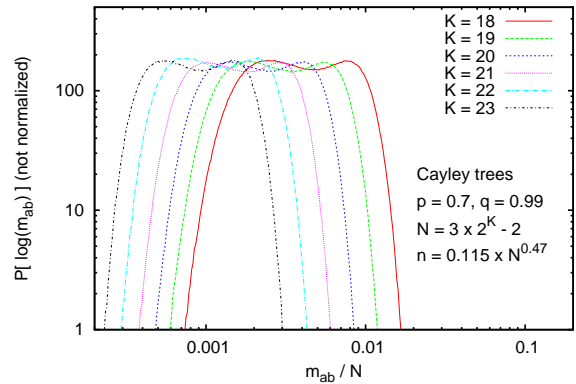}
\vglue -4mm
\caption{(color online) Distributions of the number $m_{ab}$ of $ab$ nodes in the final state for Cayley trees with $k=3$.
More precisely, $P(\log m_{ab}) = m_{ab} P(m_{ab})$ is plotted against $\log [m_{ab}/N]$ for Cayley trees
with $K$ generations, $K = 18,\ldots, 23$. All curves correspond to $p=0.7$ and $q=0.99$. The number $n$
of seed nodes is chosen such that both maxima have the same height. Overall normalization is arbitrarily 
chosen such that also all curves have the same maximal height.}
\label{Cayley}
\end{figure}

Indeed, we find bimodal distributions of $ab$-masses for $p>p_c$, if we choose $n$ properly. Results 
for $p=0.7$ and $q=1$ and for the algorithm without latency are shown in Fig.~\ref{Cayley} (for the 
algorithm with latency, results are quantitatively different, but qualitatively the same). In this 
figure, each curve corresponds to one value of $N$. The number of seed nodes was chosen for each $N$ 
that both maxima have the same height, which gives $n=0.115 N^{0.47}$ within the statistical errors. 
If the observed double peak distribution were to 
indicate a first order transition, their positions should tend to constant densities when $N\to\infty$.
Instead we see, however, that both peaks shift to the left as $N$ increases. Indeed, there is a decent 
data collapse if we plot $m_{ab}P(m_{ab})$ against $m_{ab}/N^D$ with $D = 0.55(2)$, just as we would 
expect for a critical phenomenon. Also, the scaling $n\sim N^\alpha$ of seed sizes used in Fig.~\ref{Cayley} 
indicates a standard critical phenomenon. Finally, if we use smaller seed sizes, not only the height but also
the position of the right hand peak decreases significantly, suggesting that the ``giant" $ab$-cluster
is not really well defined and connected, as it was for ER graphs.

We also looked at different observables, and they all confirmed our conclusion that there are no first 
order (or hybrid) transitions on trees.

\section{Regular lattices}

On regular lattices, one can either study the properties of the giant $ab$-cluster after the epidemics 
have dies out, or one can follow the spreading as it evolves in time. Both strategies have advantages
and disadvantages. In the former case it becomes infeasible to use too large lattices, whence one has
to be careful about finite size corrections. In the latter one can stop the evolution before the finiteness
of the lattice is seen, in which case there are no finite size corrections at all. But there are then
finite {\it time} corrections. Fortunately, only a small fraction of the entire lattice is touched. This 
allows -- eventually together with hashing \cite{Grassberger:2003} which we did not use, however,
in the present work -- the use of extremely large lattices, for which the finite time corrections 
can be made small as well.

In time dependent simulations, the ``classic" observables \cite{Grassberger:1983,Grassberger:2003} are
the average number of infectious at time $N(t)$, the probability $P(t)$ that there exists still at least one 
infectious site, and $R(t)$, the r.m.s. distance of the infectious sites from the seed. In the following we 
shall not only consider seeds consisting of a single site, but also the 
spreading from an entire hyperplane. This gives much more precise results in cases with first order transitions,
since it avoids the bottleneck in the nucleation phase that occurs in starts from a single seed. It also 
is more natural in such cases, since the growth of the infected cluster is then related to the growth 
of a rough interface that gets pinned at the critical point. Finally, we shall also measure various 
quantities related to the fact that we now have two agents $A$ and $B$. This includes $N_{AB}(t)$, the 
number of doubly infected sites, and $\langle \Delta t\rangle$, the average time lag between first and 
secondary infection for sites that finally get both diseases (the precise definition is given later).

A crucial difference between sparse random networks (like, e.g., ER networks) and regular lattices is that 
the latter contain small loops. Due to the absence of small loops on ER networks, 
the critical point for spreading from a single node seed was the same as for single
diseases, $p_c = \langle k \rangle / \langle (k-1)k \rangle$. Thus the cooperativity did not lead to a 
renormalization of the threshold, unless multiple seeds were used. This is not so for any finite dimensional
lattice. Assume that $p$ is slightly below the critical value for single diseases. Then there is a 
finite chance that both diseases survive for a short time $\tau$. During this time, they will help each other
and thus their chance of further survival is enhanced. In other words, for each disease the presence of the 
other disease renormalizes the growth rate. 

\begin{figure}[!t]
\centering
\includegraphics[width=7.5cm]{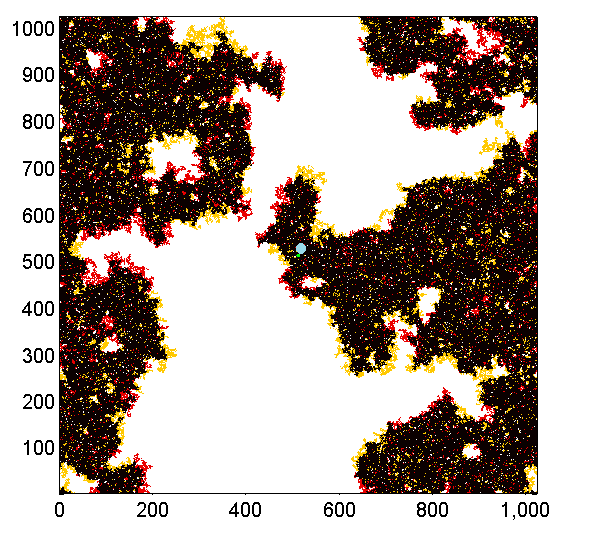}
\caption{(color online) Final state of a coinfection outbreak on a 2D $1024\times 1024$ lattice at 
    $q= 0.99$  and $p = 0.4504 \approx p_c(q)$ with periodic boundary conditions and using the algorithm with delay. 
    The single seed of infection is located in the center of the lattice, as marked by the blue disc.}
\label{LPatt}
\end{figure}

As an illustration, we show in Fig.~\ref{LPatt} the final configuration on a square lattice with nearest
neighbor contacts. On this lattice, the threshold for a single disease is at $p_c = 1/2$ \cite{Stauffer:1994}.
For coinfections with the parameters used in Fig.~\ref{LPatt} it is at $p_c(q\!=\!0.99) \approx 0.4504$. The reason for 
this shift is easily seen from the structure of the cluster. It consists of a ``backbone" of doubly infected 
sites, surrounded by two ``halos" with single infections. The latter have a finite thickness, of order 
$(p_c-p_c(q))^\nu$, where $\nu$ is the correlation length exponent. Therefore, there is always a site with 
disease $b$ close to any site with disease $a$ and vice versa. Thus cooperativity is always at work.

Notice that this argument would break down at $p\geq p_c$, where giant clusters of single diseases exist.
Below $p_c$, outbreaks can only be small for both diseases or large for both. In the following, we will 
restrict our attention to $p<p_c$.

\subsection{Two-dimensional lattices, short range infections}

\begin{figure}[]
\centering
\includegraphics[width=7.5cm]{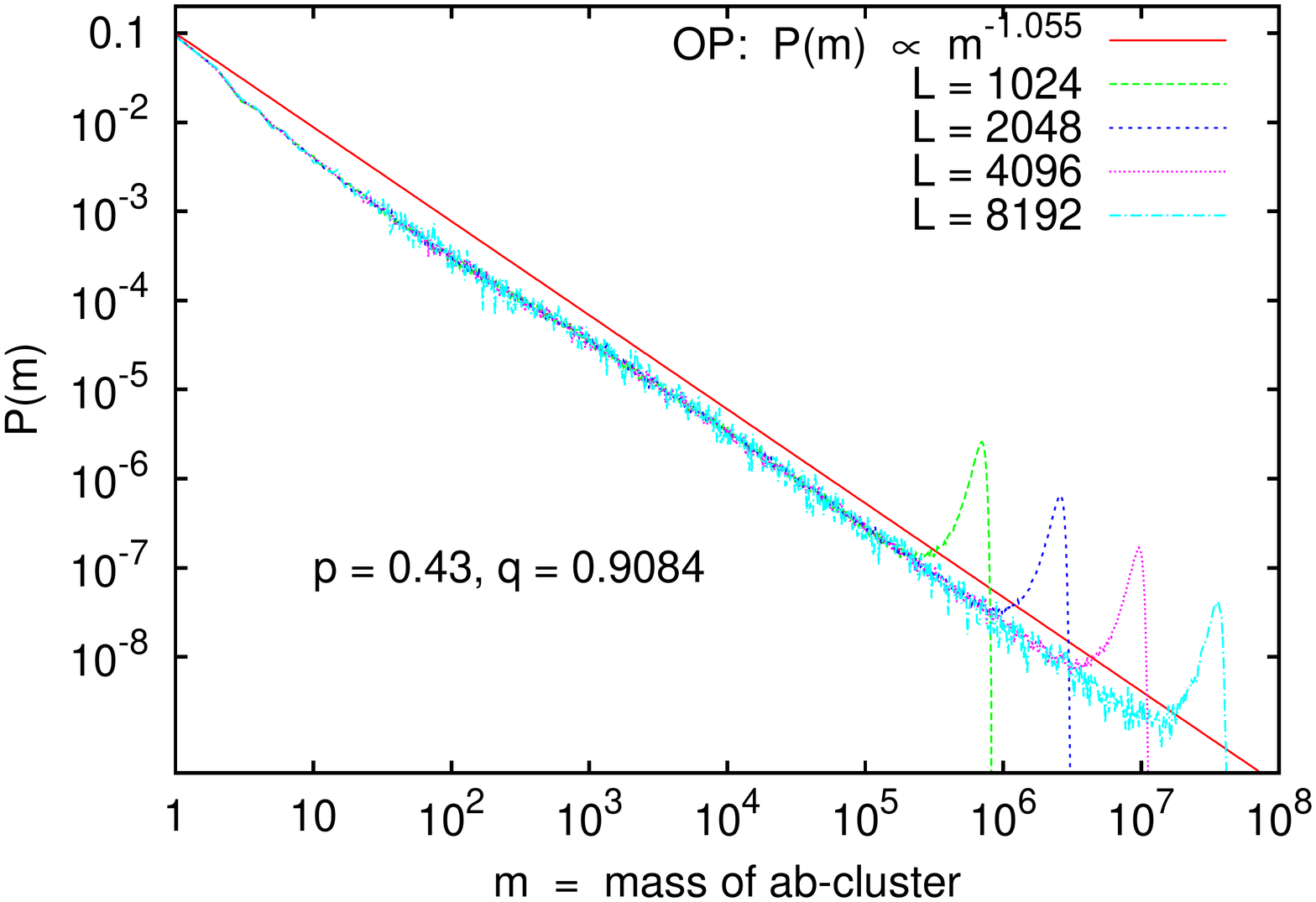}
\vglue -4mm
\includegraphics[width=7.5cm]{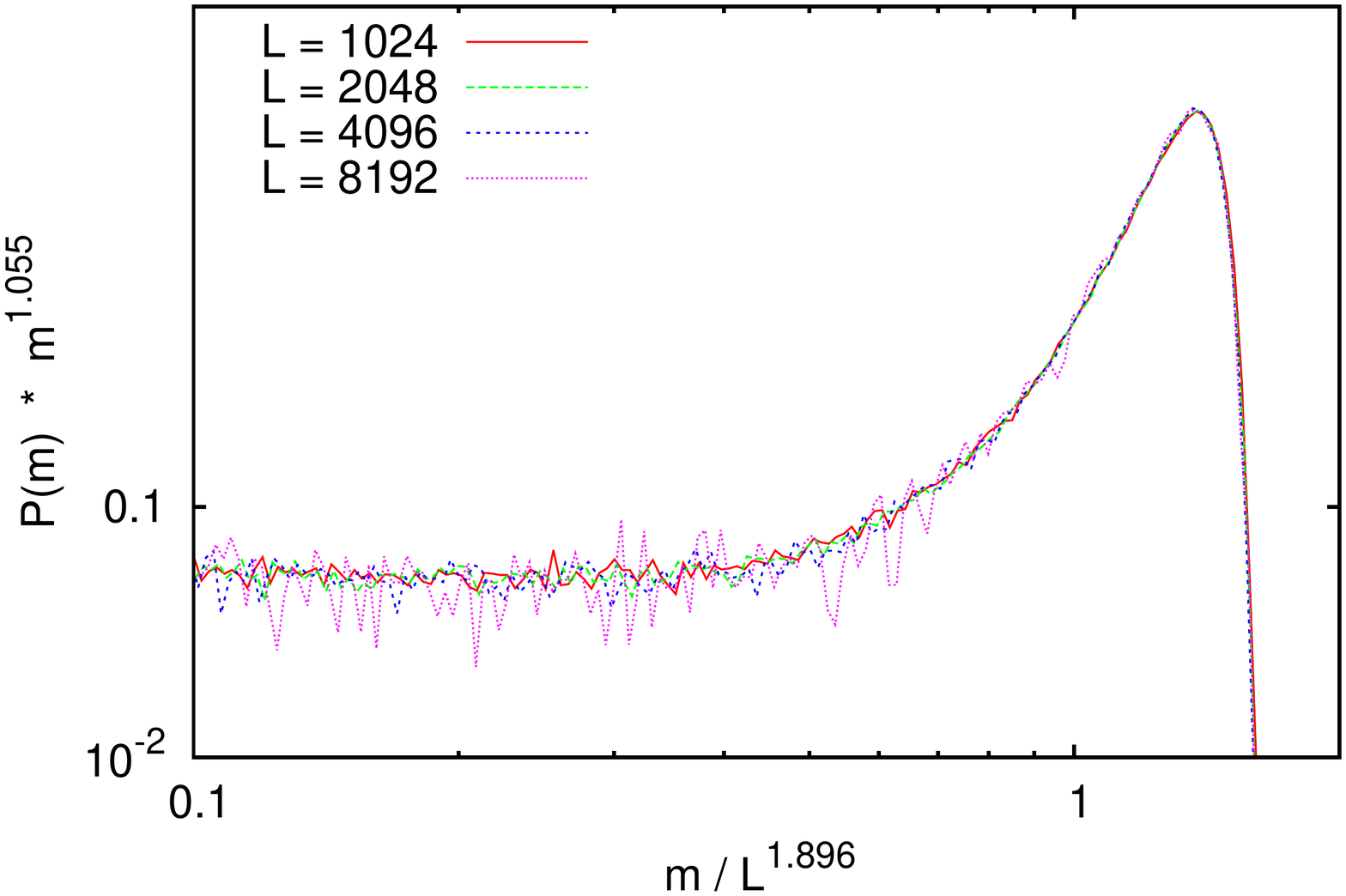}
\vglue -4mm
\caption{(color online) Upper panel: Mass distributions of $ab$-clusters obtained with the algorithm without delay at the critical 
    point for $q=0.9084$. Lattice sizes are $L\times L$ with $L=2^{10},\ldots, 2^{13}$, and helical b.c. were used.
    The straight line indicates the scaling $P(m) \sim m^{1-\tau}$ with $\tau = 187/91$ for ordinary percolation.
    Lower panel: Data collapse of the high mass data, obtained by plotting $m^{\tau-1}P(m)$ against $m/L^{D_f}$, 
    where $D_f = 91/48$ is the fractal dimension of the giant cluster in OP \cite{Stauffer:1994}.}
\label{2d-massdistrib}
\end{figure}

Mass distributions of $ab$-clusters obtained at the critical point for a randomly chosen large value of $q$ are 
shown in Fig.~\ref{2d-massdistrib}. The upper panel shows that the bulk of the data show the well known scaling 
$P(m) \sim m^{1-\tau}$ of OP \cite{Stauffer:1994}, while the lower panel shows that the right hand peaks 
correspond to giant clusters whose masses scale exactly as $m \sim L^{D_f}$, where $D_f$ is fractal dimension
for OP. These data not only show that there is no first order transition, but they suggest also strongly 
that the critical point is in the OP universality class. We should add that similar data were obtained for 
other values of $q$ (including $q=1$) and for the algorithm with delay.

\begin{figure}[]
\includegraphics[width=8cm]{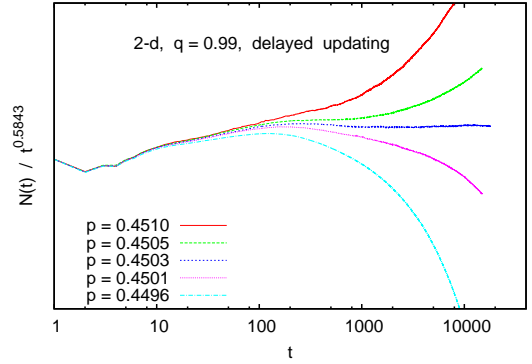}
\vglue -4mm
\caption{(color online) Log-linear plots of $t^{-0.5843} N(t)$ versus $t$ for $q=0.99$ and for five values of $p$ close to the 
   critical point. The exponential prefactor is chosen such that the curve should be asymptotically flat 
   for $p=p_c$.}
\label{2d-N_t}
\end{figure}
 
Values of $N(t)$ for $p$ near the critical point for $q=0.99$ are shown in Fig.~\ref{2d-N_t}. More precisely,
for easier comparison with OP we plotted there $N(t)/t^\eta$ versus $t$, where $\eta = 0.5843(5)$ is the value 
for percolation in 2 dimension \cite{Grassberger:1992}. Lattice sizes were so large that the cluster never 
touched the boundary, and the algorithm with latency was used. We find of course important finite time 
corrections, but the results for $p = p_c(0.99) = 0.45030(3)$ are fully compatible with the expected asymptotic
scaling. We should add that this value of $p_c$ is also compatible with the estimate $p_c(0.99) \approx 0.4504$
obtained from mass distributions. 

This absence of a first order transition and universality with OP was confirmed for the algorithm without latency,
for other values of $q$, and for the square lattice with next-nearest and also with next-next-nearest neighbors
(i.e. with 8 resp. 12 neighbors). In all cases it was verified that $N(t)\sim t^\eta$ for large $t$.

There are, however, two novel scaling laws for $p$ in the vicinity of the single disease critical point $p_c = 1/2$.
Both can be most easily understood by referring to Fig.~\ref{LPatt}. As we said, the thickness of the ``halos" of 
singly infected sites around the doubly infected backbone is equal to the correlation length of OP. When $p\to p_c$,
this correlation length diverges.

\begin{figure}[]
\includegraphics[width=8cm]{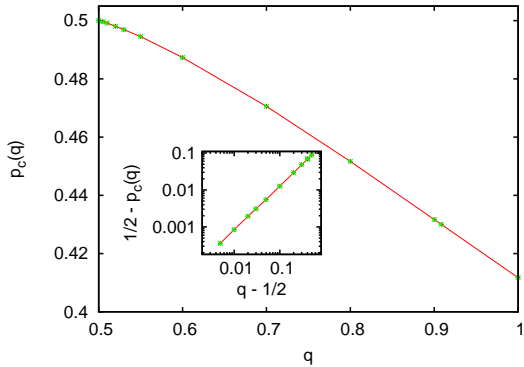}
\vglue -4mm
\caption{(color online) Critical value of $p$ versus $q$, for square lattice and algorithm with delay. The insert shows a 
   log-log plot which suggests Eq.~\ref{2d-pc} for $q\approx 1/2$.}
\label{2d-p_versus_q}
\end{figure}

Consider now the limit $q\to 1/2$ where there is no cooperativity. In this limit also $p_c(q) \to p_c = 1/2$,
and increasingly larger portions of the total cluster are made up by singly infected sites. This means however 
that also cooperativity should become less and less effective in this limit, implying that $dp_c(q) /dq \to 0$. This 
is indeed verified in Fig.~\ref{2d-p_versus_q}. As seen from the insert in this figure, a decent fit is obtained
by the power law with a new independent exponent
\be
   p_c(q) - 1/2 \sim (q-1/2)^\zeta \quad {\rm with} \quad \zeta = 1.19(3).    \label{2d-pc}
\ee

\begin{figure}[]
\includegraphics[width=8cm]{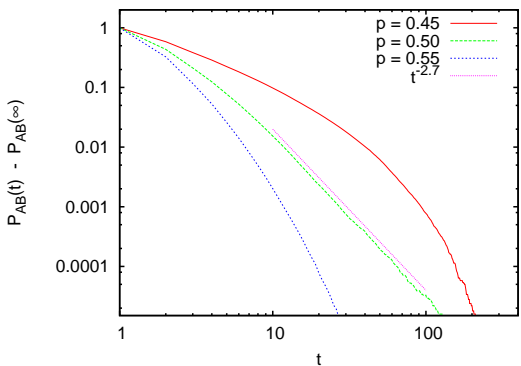}
\vglue -4mm
\caption{(color online) Log-log plots of $P_{AB}(t) - P_{AB}(\infty)$ versus $t$ for 2-d lattices with $q=1$.
   The central curve corresponds to the critical point for single diseases. 
   The straight line indicates its exponent for large $t$.}
\label{2d-P_AB}
\end{figure}
 
For the other scaling law, consider an arbitrary value $q>1/2$. For any $p>p_c(q)$ there will be a non-zero 
chance of survival, i.e. $P(t) \to P_\infty >0$ for $t\to\infty$. Due to universality with OP we expect that 
this asymptotic value is reached faster that with a power law, 
\be
   P(t) - P_\infty < t^{-\Delta}
\ee
for any exponent $\Delta$ (as was also verified numerically for $p\neq 1/2$, see Fig.~\ref{2d-P_AB}). 
Consider now the case $p>1/2$, where there is also a non-zero chance that 
single infected clusters survive forever (if the other disease had died out earlier). Exactly at $p=1/2$ 
the probability that only one of the diseases, say $A$, survives should decay as $P_A(t) \sim t^{-\delta}$ with 
$\delta$ known from OP. Moreover, there will be a small chance that $A$ survives for some time by spreading
into one direction, and $B$ survives by spreading into the opposite direction. Such an epidemic would look 
superficially like an epidemic of double infection, but since there is no cooperativity (since both diseases
survive in different regions), it has a much higher chance to die. Let us define as $P_{AB}(t)$ the 
probability that both diseases have not yet died out at time $t$. In Fig.~\ref{2d-P_AB} we show a log-log 
plot of $P_{AB}(t)-P_{AB}(\infty)$ versus $t$. The data clearly suggest a power law 
\be
   P_{AB}(t) = P_{AB}(\infty) + a/t^\Delta
\ee
with exponent $\Delta = 2.6(2)$. 

\begin{figure}[]
\includegraphics[width=8cm]{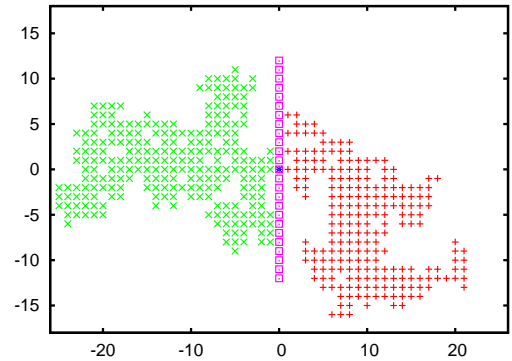}
\vglue -4mm
\caption{(color online)  Two ordinary (bond) percolation clusters, both starting at ${\bf x}=0$, with a killing wall
   ($x = 0$) between them. A wall is killing, if any epidemic that tries to infect it is entirely deleted.}
\label{2d-wall_pict}
\end{figure}

A simple upper bound on this exponent is obtained as follows. We first define a boundary as ``killing",
if any epidemic that tries to infect a site on it is killed. Such a boundary has a much stronger effect 
on clusters than normal boundaries, where only the branch that would pass through the boundary is deleted.
Clusters which start on a killing boundary have therefore a much smaller chance to survive. Numerically,
we found by simulations that $P_{\rm wall} \sim t^{-\delta_{\rm wall}}$ with $\delta_{\rm wall} = 1.76(1)$.
Consider now the situation shown in Fig.~\ref{2d-wall_pict}, where two OP clusters start from the same 
site on a killing wall, and are forced to grow into opposite directions. Any such configuration would 
contribute to $P_{AB}(t)-P_{AB}(\infty)$, which gives immediately
\be
   \Delta \leq 2 \delta_{\rm wall}.
\ee

In this paper we always decide ``on the fly" whether a site or node can be infected. But we could also 
have decided this before the simulation, since any node can be infected at most once by either one of the 
two diseases. The results would be identical. In the latter case we are dealing with frozen randomness.
There is a rather general theorem \cite{Aizenman:1989,Cardy:1999} that forbids first order transitions
in two-dimensional systems with quenched randomness. Although it is not clear whether this theorem applies 
strictly spoken to the present model, the general ideas should. It definitely applies to the cooperative 
percolation model of \cite{Janssen:2004,Bizhani:2012} and to the zero-$T$ random field Ising model, since 
these can be mapped onto the Potts model, and it explains why in this case critical pinning is in the OP
universality class \cite{drossel:1998,Bizhani:2012}). It strongly suggests that there exists only one 
universality class of critically pinned interfaces in isotropic 2-dimensional media, namely that of ordinary 
percolation.

\subsection{Four dimensions and above}

\subsubsection{$d=4$, Point seeds}

The results of the last subsection might suggest that in general, there are no first order transitions 
on regular lattices. To show that this would be wrong, we present simulations for the simple hypercubic 
lattice with $d=4$, as a typical high dimensional lattice. Lattice sizes are in each case sufficiently
large that we have no finite-size corrections at all.

\begin{figure}[]
\includegraphics[width=8cm]{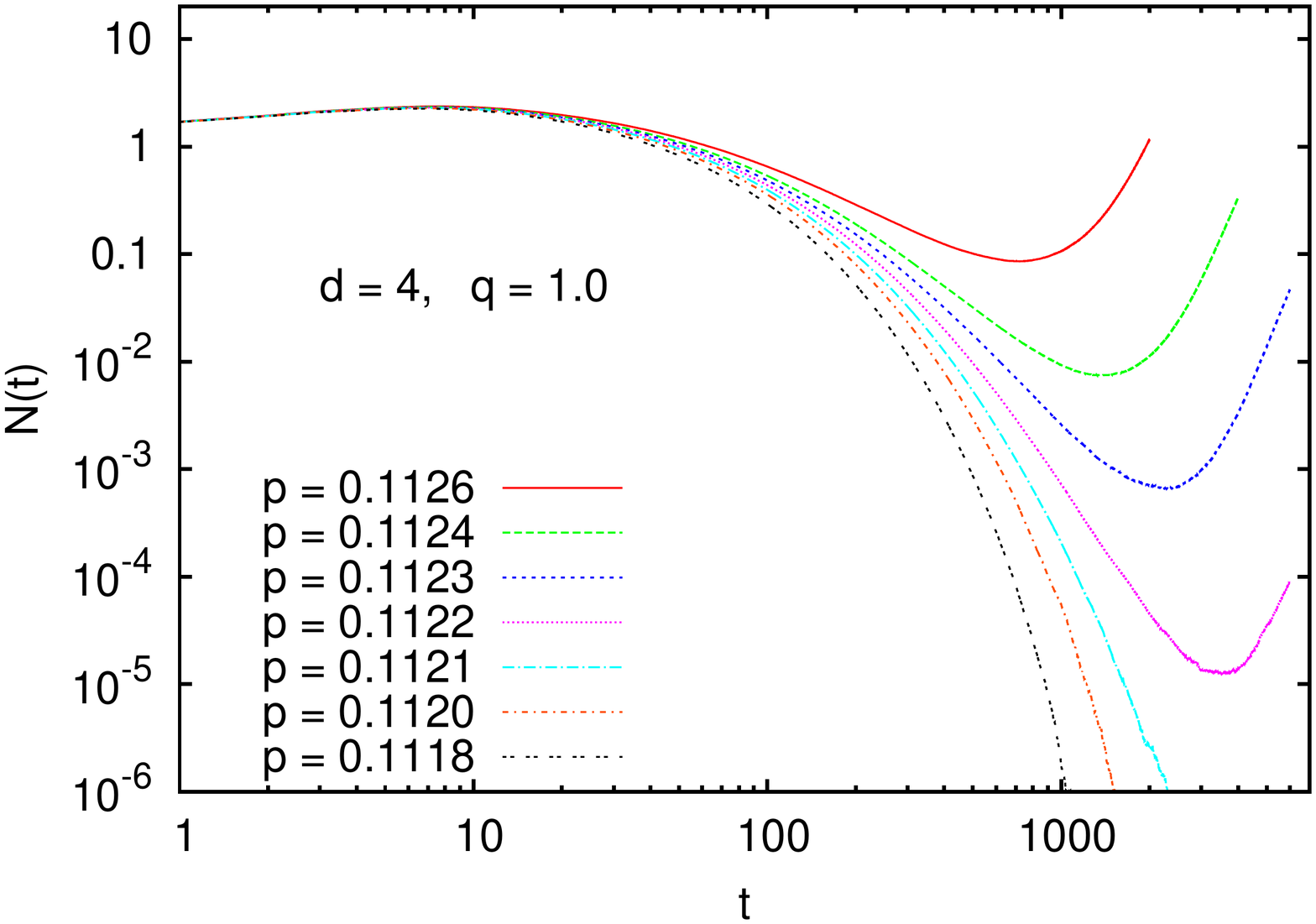}
\vglue -6mm
\includegraphics[width=8cm]{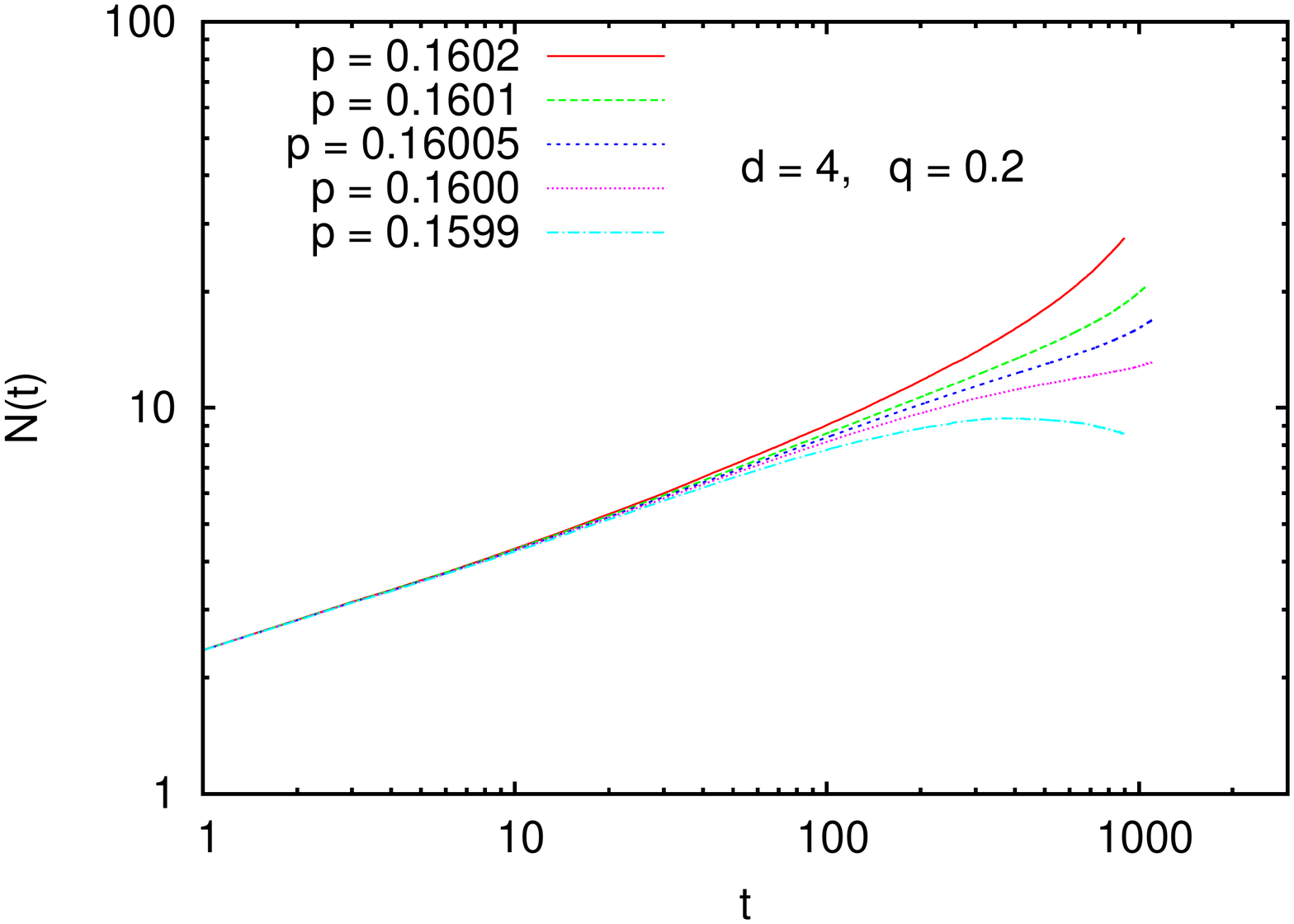}
\vglue -4mm
\caption{(color online) Log-log plots of $N(t)$ versus $t$ for 4-d simple hypercubic lattices. For 
   this lattice, ordinary bond percolation has $p_c = 0.16013(1)$ 
   \cite{Grassberger:2003}, thus all curves except the uppermost one in panel (b) correspond to 
   $p < p_c$. \\
   Panel (a) shows results for $q=1.0$. The best estimate for $p_c(q=1.0)$ from these data is 
   $\approx 0.112$, but a more precise estimate will be given later.\\
   Panel (b) is for $q=0.2$. Here cooperativity is much 
   weaker, and thus our estimate $p_c(0.2) = 0.15997(5)$ is much closer to the value for 
   OP. Superficially, this plot might suggest a power law and thus a second order transition,
   but all structures seen in this plot are real (statistical errors are much smaller than 
   the line thicknesses) and suggest also a (weak) first order transition.}
\label{n_t-4d-q}
\end{figure}

In Fig.~\ref{n_t-4d-q} we show results for $N(t)$, for two rather extreme
values of $q$ and for several $p\approx p_c(q)$. For $q=1.0$ the transition is obviously 
first order. For $p \approx p_c(q)$ the epidemic seems first to die out (faster than with
a power law!), but finally -- if $p > p_c(q)$ it turns around and increases with a power 
much large than that for critical OP. This is very reminiscent of nucleation where clusters 
have to become large before they can grow further with high probability. As long as the 
cluster size is small, it is much more likely that the cluster dies than that it grows.
For $q=0.2$ (which is only very little above the OP value $p_c = 0.16013(1)$ 
\cite{Grassberger:2003}) the situation {\it seems} to be different. At a rough glance,
the data suggest a power law for $p\approx 0.160$, which then would suggest a second order
transition. But actually none of the curves in Fig.~\ref{n_t-4d-q}b is asymptotically 
straight (all structures seen in this plot are real, since stochastic errors are much smaller 
than the line thicknesses), and a closer look shows that also now curves bend down and 
pass through a (much less pronounced) nucleation phase.

\begin{figure}[]
\includegraphics[width=8cm]{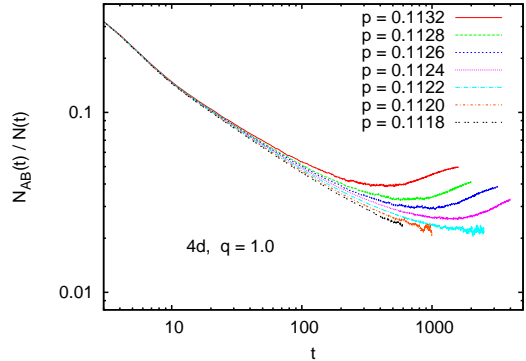}
\vglue -4mm
\caption{(color online) Log-log plots of $N_{AB}(t)/N(t)$ versus $t$ for $q = 1.0$. For short times,
   all four curves decrease -- suggesting that $A$ and $B$ spread independently without
   much overlap. For $p>p_c(q)$ this decrease finally stops and is reversed, because then
   cooperativity creates a compact cluster.}
\label{nAB_t-4d-q100}
\end{figure}
 
In the next subsection we will present more clear numerical evidence for the absence 
of a tricritical point and for the transition being discontinuous for all $q>p$. 
In the following we will give more heuristic arguments, supported by indirect numerical 
evidence.

To understand the mechanism behind this scenario, it is helpful to look at $N_{AB}(t)$, the 
number of doubly infected sites. This is shown in Fig.~\ref{nAB_t-4d-q100} for $q=1.0$
and for four values of $p$. More precisely, we show there $N_{AB}(t)/N(t)$, i.e. the 
probability that an infected site is doubly infected. This at first decreases
steeply for all four values of $p$. It is only in the supercritical case $p\geq 0.115$ that
this decrease stops and finally even turns around. This suggests that at first $A$ and 
$B$ were spreading into different directions, with little overlap between them. This 
little overlap is not enough to generate enough cooperativity which would prevent them
from spreading further apart -- and dying finally because $p$ is subcritical for 
single epidemics and because two infinite clusters cannot coexist anyhow. It is only for 
$p>p_c(q)$ and for large $t$ that occasionally
two clusters with sufficient overlap developed so that they continue to spread coherently.
Notice that this did not happen in $d=2$, since there it is extremely unlikely that two 
clusters can grow without having much overlap.

\begin{figure}[]
\includegraphics[width=8cm]{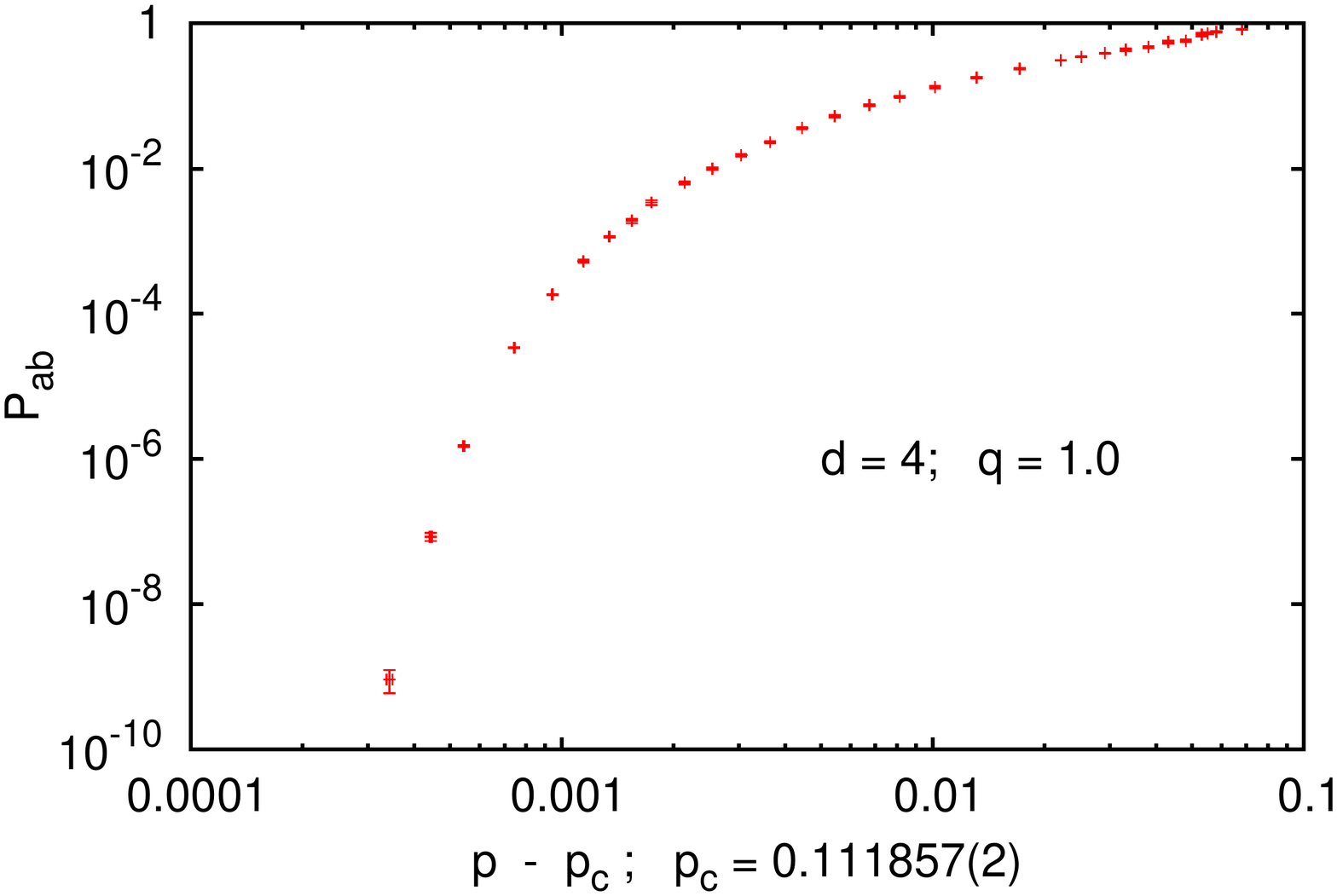}
\caption{(color online) Log-log plot of $P_{\rm ab}$, the probability that a single doubly infected site
evolves into a giant epidemic, plotted against $p-p_c$. As in the previous plots, AU was used with $q=1$.}
\label{4d-P_ab}
\end{figure}

\begin{figure}[]
\includegraphics[width=8cm]{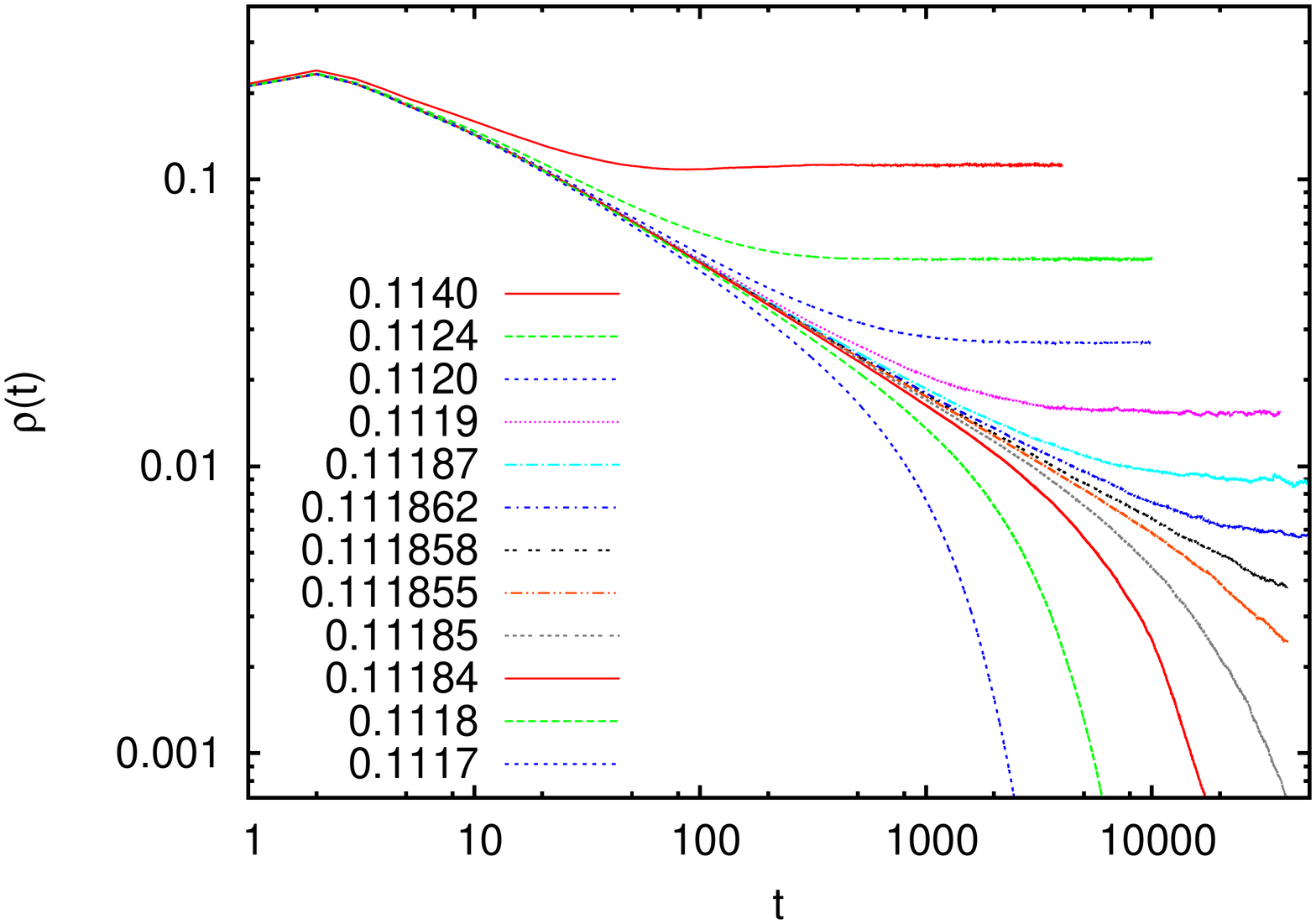}
\vglue -8mm
\includegraphics[width=8cm]{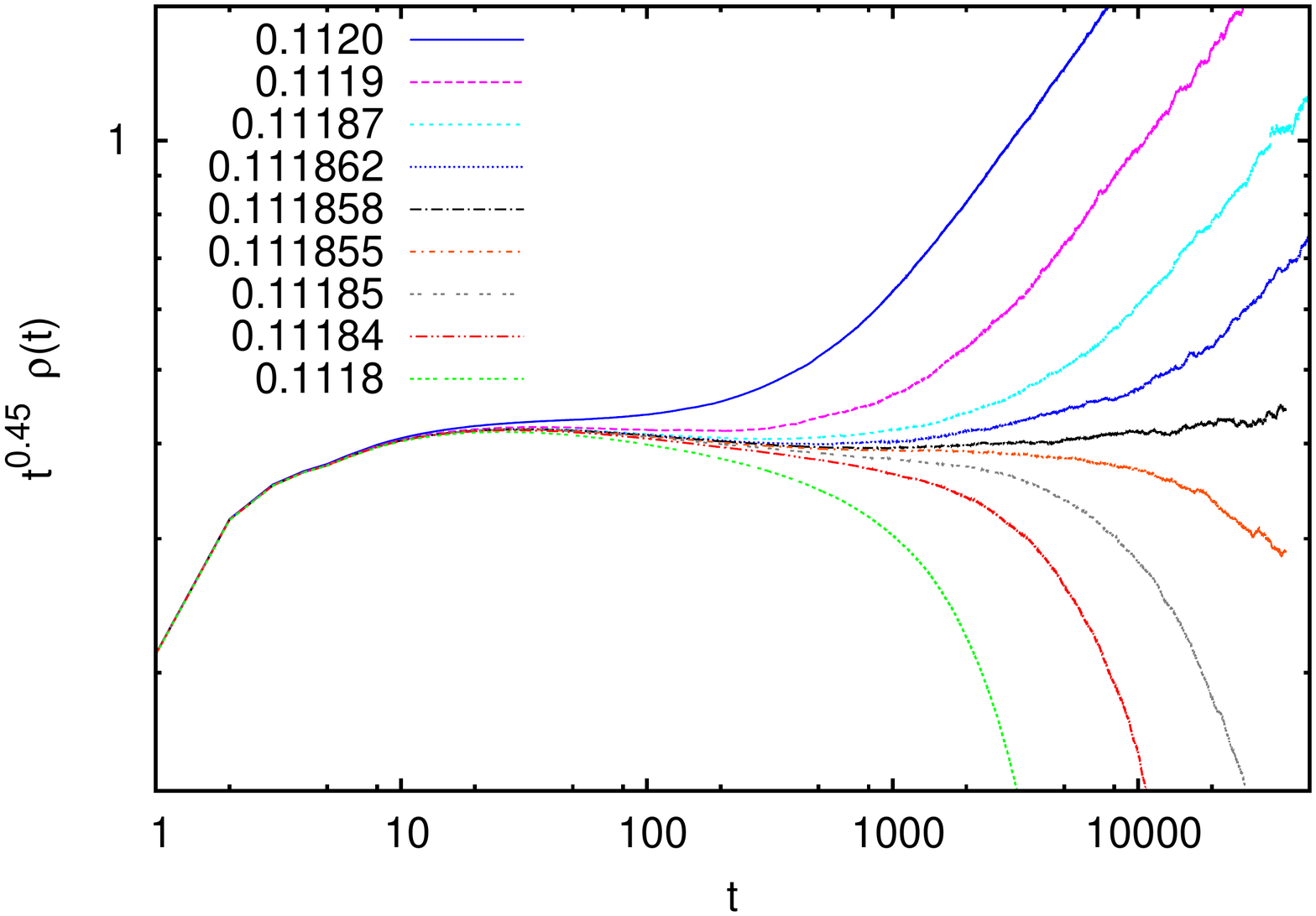}
\vglue -4mm
\caption{(color online) (a) Log-log plots of $\rho(t) = N(t)/L^3$ versus $t$ for $d=4$ and $q = 1.0$, 
   when using all sites of 
   an entire hyperplane $z=0$ as seeds. Lateral boundary conditions are helical, and the diseases 
   are allowed to spread into the positive $z$ region only.\\
   (b) Part of the same data, bu shown as $t^{\delta}\rho(t)$ against $t$, with $\delta=0.45$. Since the 
   data for small $t$ have obviously large non-scaling corrections, only data for $t> 10$ are plotted.}
\label{4d-hyper-N_t}
\end{figure}

In the next subsection we will see that $p_c$ can be estimated much more precisely by using infected
hyperplanes as seeds rather than single points. In this way we will find $p_c = 0.111857(2)$ for $q=1$.
Using this value, we plot in Fig.~\ref{4d-P_ab} how $P_{\rm ab}(p)$, the probability that a single infected
site creates an infinite epidemic, depends on $p-p_c$. We see that the curve becomes steeper and steeper
on a log-log plot as $p-p_c\to 0$, indicating that $P_{\rm ab}(p)$ has at threshold an
essential singularity. This reminiscent of nucleation phenomena where the chance for small droplets to 
become macroscopic in metastable phases behave similarly \cite{Debenedetti}. As we shall see in the 
next section, the behavior in three dimensions is very different.

\subsubsection{Hyperplane seeds}
    \label{hyperplane}

In order to avoid the nucleation phase (which, among others, prevents a precise estimate of 
$p_c(q)$), we also made simulations where we started with an entire infected hyperplane as 
seed. In this case the boundary between healthy and sick regions is formed by a propagating 
interface which starts off flat and becomes increasingly rough. For $p<p_c(q)$ the growth 
finally stops and the interface gets pinned, while it continues to move forever for $p>p_c(q)$.
Exactly at $p=p_c(q)$, one might expect it to be in the universality class of critically pinned interfaces
in isotropic random media \cite{family:1991,barabasi:1995,Bizhani:2012,Bizhani:2014}.
In Fig.~\ref{4d-hyper-N_t}a we show $\rho(t) = N(t)/L^3$ versus $t$ for $q=1.0$ and several values of $p$ 
close to $p_c(q)$. All data in this plot were obtained from lattices of size $L^3\times L_z$ with laterally 
periodic (more precisely, helical) boundary conditions. The diseases started at the base surface $z=0$ and 
$L_z$ was so large that the upper boundary at $z=L_z$ was never reached. The base surface had size $L^3$ 
with $L$ between 256 and 512. This is big enough so that finite size corrections are small (for a more 
detailed discussion see below). We see that there is a clear power law 
\be
    \rho(t) \sim t^{-\delta}   \quad {\rm with} \quad \delta = 0.45(2)
\ee
when $p=p_c=0.1118565(15)$, which is therefore our best estimate of $p_c(q\!=\!1.0)$. 

To demonstrate the 
quality of the data on the one hand and the fact that this power law has important corrections on the other 
hand, we show in Fig.~\ref{4d-hyper-N_t}b the same data plotted as $t^{\delta}\rho(t)$ against $p$.
Notice the much enlarged resolution on the $y$-axis. We see that there is actually no single curve
which is clearly a horizontal straight line. The error bars on $\delta$ and $p_c$ are a naive attempt to 
take into account this uncertainty.

\begin{figure}[]
\includegraphics[width=8cm]{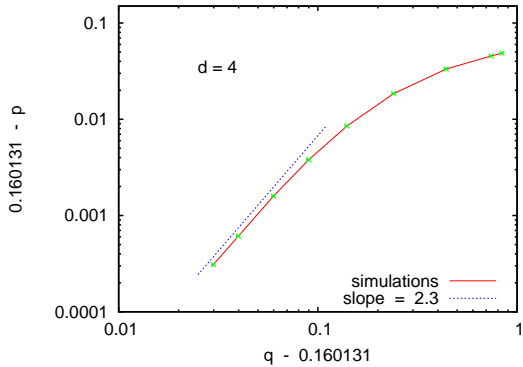}
\vglue -4mm
\caption{(color online) Log-log plot of $p_c - p_c(q)$ against $q-p_c$, where $p_c = 0.160131$ \cite{Paul:2001}
   is the (bond) percolation threshold on the simple 4-d hypercubic lattice.}
\label{4d-hyper-pc}
\end{figure}

The most dramatic result from this plot is the vastly improved estimate of $p_c(q)$. It agrees with
the previous estimate from point seeds, but it is about four orders of magnitude more precise. Similar 
plots were also produced for other values of $q$ in the range $0.19 \leq q \leq 1$. They are all 
qualitatively similar, and they suggest that indeed the percolation transition is discontinuous in this 
entire range (estimates in the range $0.160131 < q < 0.19$ were inconclusive). The results for 
$p_c(q)$ are shown in Fig.~\ref{4d-hyper-pc}. They suggest a power law $p_c - p_c(q) \sim (q-p_c)^{2.3}$.
Notice that this is in contrast to the case with cooperativity between different neighbors 
\cite{Dodds:2004,Janssen:2004,Bizhani:2012}. In that case, there exists a tricritical point $q^* \in ]0,1[$ 
such that the transition id continuous for $q<q^*$ and discontinuous for $q>q^*$. 
The absence of such a tricritical 
point in the present model might be related to the appearance of a new divergent length
scale when $p\to p_c(q)$, as discussed in connection with Fig.~\ref{LPatt}.

\begin{figure}[]
\includegraphics[width=8cm]{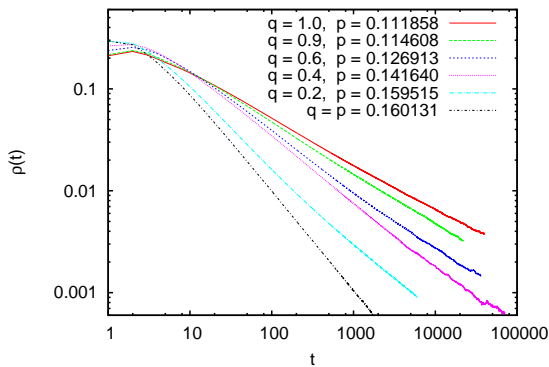}
\vglue -4mm
\caption{(color online) Log-log plot of $\rho(t)$ versus $t$ for five values of $q$. For each $q$, the value 
   of $p$ was chosen such that the curves are most straight for large $t$.}
\label{N_t_allq}
\end{figure}

Approximately, the data shown in Fig.~\ref{4d-hyper-N_t} obey a finite time scaling (FTS) ansatz 
\be
    \rho(t,p) = t^{-\delta} F[(p-p_c(q))t^{1/\nu_t}]        \label{n-FTS}
\ee
with $\nu_t = 1.04$. Similar ansatzes describe also reasonably well all data for $0.2 \leq q < 1$. We 
believe that this actually would describe the asymptotic behavior, and that the 
estimate of $\nu_t$ is correct up to about 5 percent. But the fit is far from perfect and -- what is much
worse -- if no corrections to scaling were applied, similar FTS ansatzes for the data obtained at different 
values of $q$ yield substantially different critical exponents. This is e.g. seen by plotting on the 
same log-log plot values of $N(t)$ versus $t$ for different values of $q$, choosing for each $q$ that value of 
$p$ where the curve is straightest. Such a plot is shown in Fig.~\ref{N_t_allq}. We see that actually none
of the curves is straight, their average slope decreases $q$, and they all seem to become parallel to 
the curve for $q=1$ (which is the least curved one) for very large $t$. For small $t$ the slope becomes
steeper with decreasing $q$, in agreement with our claim that we see a very slow cross-over from OP.
The latter would correspond to $q=0.160131$, and the exponent there is $\delta = 0.97(3)$ \cite{grassberger:1986}.

\begin{figure}[]
\includegraphics[width=8cm]{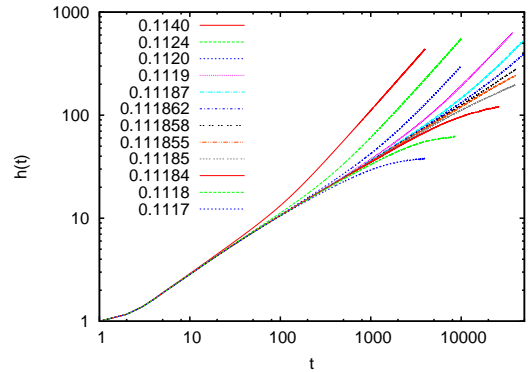}
\vglue -8mm
\includegraphics[width=8cm]{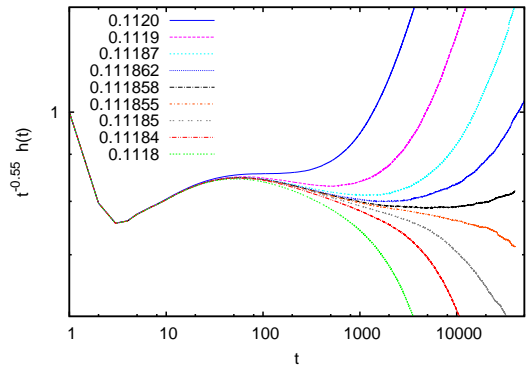}
\vglue -4mm
\caption{(color online) (a) Log-log plots of $h(t,p)$ versus $t$ for $d=4$ and $q = 1.0$, when using all 
   sites of an entire hyperplane $z=0$ as seeds. These data are based on the same runs as those in 
   Fig.~\ref{4d-hyper-N_t}. Part of these data are also shown in panel (b) as
   $t^{-\nu/\nu_t} h(t,p)$ versus $t$.}
\label{4d-hyper-h_t}
\end{figure}

A similar FTS ansatz holds also for the height $h(t,p)$ of the interface. There 
are several ways to define this hight. The data shown in Fig.~\ref{4d-hyper-h_t} use just the 
average value of the $z$ coordinates of the presently infected (``active") sites, $h(t,p) = \langle z\rangle$. 
They can be fitted by the ansatz
\be
    h(t,p) = t^{\nu/\nu_t} G[(p-p_c(q))t^{1/\nu_t}]     \label{h-FTS}
\ee
with $\quad \nu/\nu_t = 0.55(1)$. Again this is far from perfect, as can be seen from a similar blow-up
as for the densities, see Fig.~\ref{4d-hyper-h_t}b.

Scaling laws like Eqs.~(\ref{n-FTS},\ref{h-FTS}) apply also to ordinary percolation \cite{Grassberger:1983}, 
where the exponents are however different. In the present case the cluster behind the growing surface is 
compact, i.e. its height grows proportionally to its mass. The latter is given by $M(t) =\sum_{t'<t} N(t')$, 
from which we obtain 
\be
    1-\delta = \nu/\nu_t,
\ee
which was indeed imposed as a constraint on the values used in Figs.~\ref{4d-hyper-N_t}b and 
\ref{4d-hyper-h_t}b. For OP the cluster is fractal with dimension $D_f$, 
so that at the critical point $M(t) \sim h(t)^{D_f-3}$, leading to $1-\delta = (D_f-d+1)\nu/\nu_t$).
But things are not entirely clean. First of all, for no value of $p$ is $h(t)$ a clean power law. 
Secondly, for $100 < t < 1000$ the curves decrease both in Figs.~\ref{4d-hyper-N_t}b and \ref{4d-hyper-h_t}b.
This could mean that the densities of the grown cluster are not constant in this region, but we 
will show later that this is not the case. 
Most embarrassing is that the estimates for $p_c$ obtained from $\rho(t)$ and from $h(t)$ are 
slightly different. The difference is very small (it is within the error bars quoted above), but 
it is statistically significant. The nominal value of $p_c$ from $\rho(t)$ is slightly higher than
that obtained from $h(t)$. The only explanation for this are finite size corrections. Indeed, 
one expects finite size corrections to be positive for $h(t)$ and negative for $\rho(t)$. This 
was also verified explicitly by making runs at smaller values of $L$. 

All this shows that: (i) Yes, there are clear indications for finite size corrections. But the very 
fact that they are seen and qualitatively as expected makes us sure that they are well under control;
(ii) They cannot be responsible for the deviations from the expected FTS and for the observed
$q$-dependence of the critical exponents, which most likely are a very slow cross-over from OP. As 
a result, all estimates of critical exponents in this subsection have to be taken with some caution.

\begin{figure}[]
\includegraphics[width=8cm]{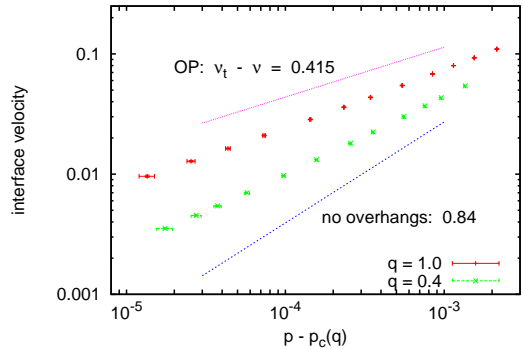}
\vglue -4mm
\caption{(color online) Log-log plots of $v(p)$ versus $p-p_c(q)$ for $d=4$. The two data sets are 
   for $q=0.4$ and $q = 1.0$. In this 
   plot we used $p_c(0.4) = .1416425(20)$ and $p_c(1.0) = 0.1118565(15)$.  }
\label{4d-speed}
\end{figure}
 
In the following figures we show several more observables, all of which show scaling laws and 
demonstrate thereby that the percolation transition is actually hybrid. They also show that
at least the rough features of the scenario depicted so far are consistent.

(1) In Fig.~\ref{4d-speed} 
we show $v(p)$, the velocity by which the height grows for very large times, when $p>p_c(q)$. It is 
simply obtained from the straight lines in the upper right part of Fig.~\ref{4d-hyper-h_t}a.
These data were obtained by using the fact that the (hyper-)surfaces for $p>p_c$ are rather 
smooth in $d=4$. Thus the simulation box can be much wider than high. Moreover, we always
checked that the height difference between the highest and lowest active site is $<L_z$. As 
long as this is guaranteed we can ``recycle" the part of the simulation box below the lowest
active site. That means we erase in this part the old configuration and overwrite it with 
the new growing part on top of the surface. Effectively, this means that we replace the 
simulation box by a torus, and let the surface circle around it. Two data sets, for $q=0.4$
and for $q=1.0$ are shown in Fig.~\ref{4d-speed}. 
Also indicated in this figure are the results for OP and the predictions of the ``standard"
model for pinned rough interfaces, where overhangs are neglected \cite{le_Doussal:2002,tang:2009}.

Both data sets are compatible with power laws with similar exponents. If we accept the FTS 
ansatz, we obtain indeed 
\be
   v \sim (p - p_c(q))^{\nu_t - \nu} \sim (p - p_c(q))^{0.47\pm 0.02}.
\ee
This is indeed compatible with the data, although there are also 
important corrections to scaling. These corrections are larger for $q=0.4$ than for $q=1.0$,
in agreement with our previous discussion. Whatever the true exponent is, it seems very 
unlikely that the model is in the same universality class as the model without overhangs.

\begin{figure}[]
\includegraphics[width=8cm]{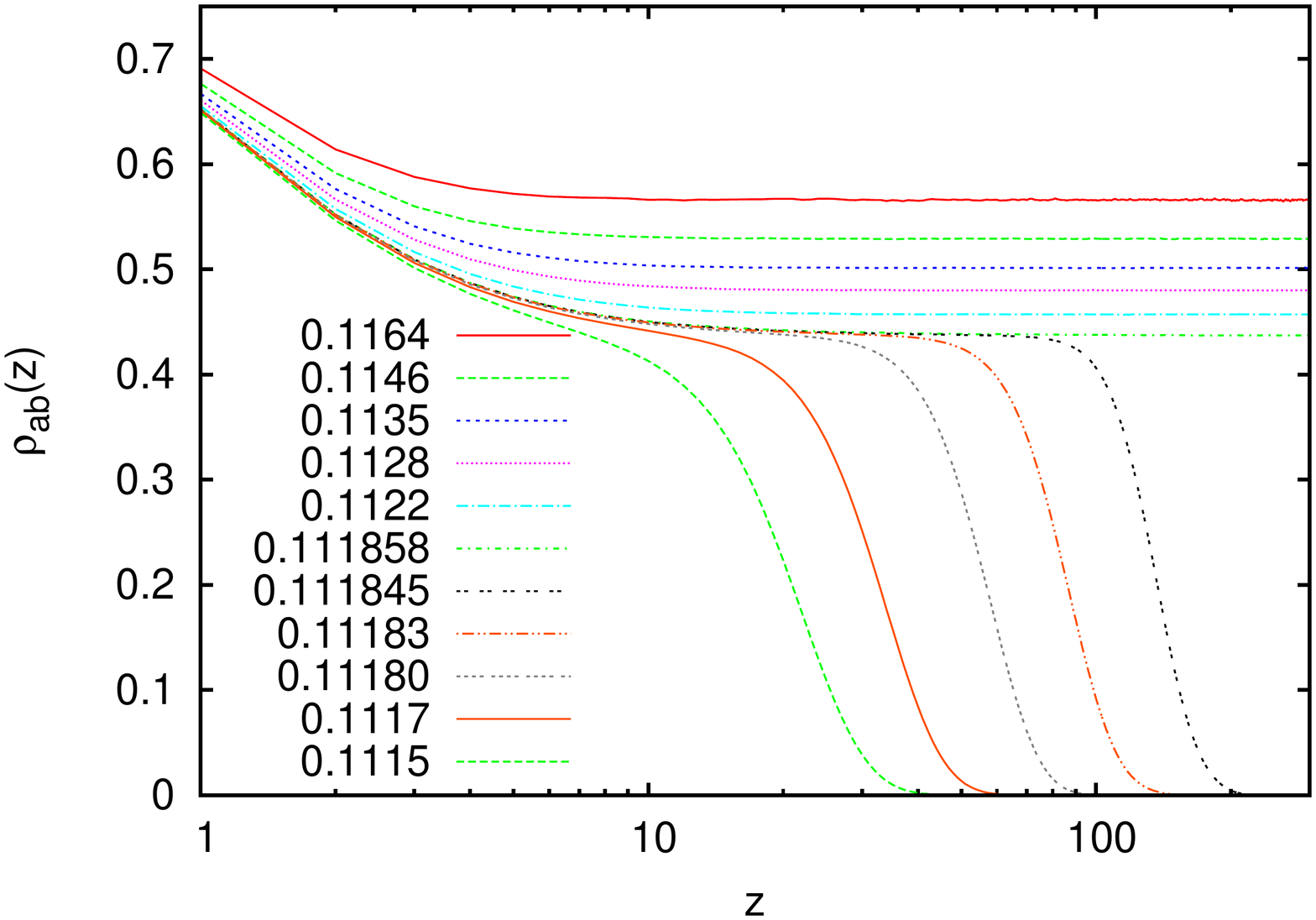}
\vglue -8mm
\includegraphics[width=8cm]{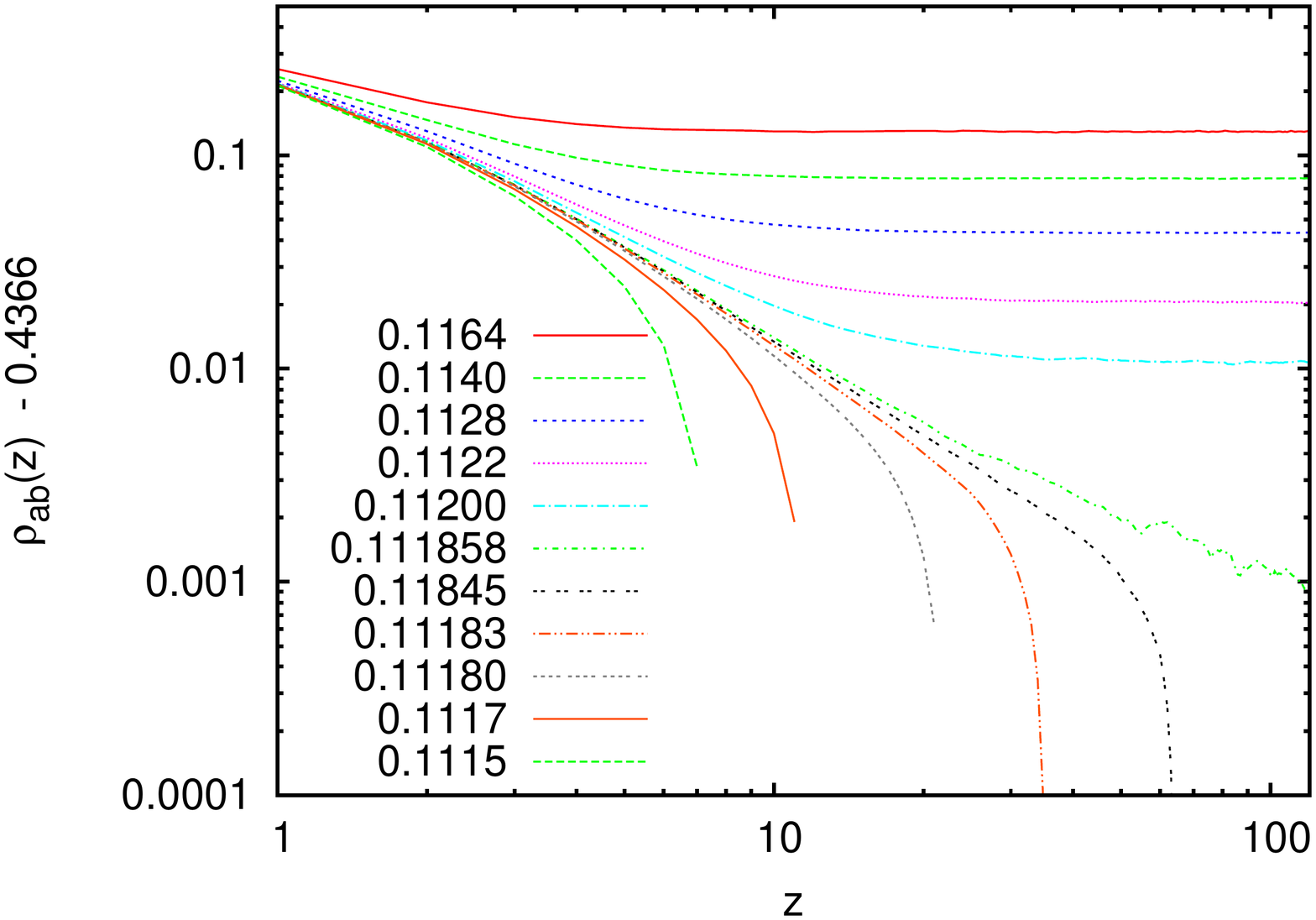}
\vglue -4mm
\caption{(color online) (a) Densities of $ab$-sites at given height $z$, after the cluster 
   had stopped growing at this height, for $q=1$ and various values of $p$. Panel (b) shows
   the same data re-plotted as a log-log plot of $\rho_{ab}(z) - \rho_{ab,c}$ against $\ln z$.}
\label{4d-dens}
\label{4d-dens-log}
\end{figure}

(2) The cluster below the growing surface is compact for $p>p_c(q)$, but it does contain holes.
Thus the densities $\rho_0(z), \rho_a(z), \rho_b(z)$, and $\rho_{ab}(z)$ are all non-zero.
Here, $\rho_\alpha(z)$ is the density of the cluster at height $z$, after the interface 
either has stopped growing (for $p<p_c(q)$) or has passed far beyond $z$ (for $p>p_c(q)$). 
Results for $\rho_{ab}(z)$ are shown in Fig.~\ref{4d-dens} for $q=1.0$. We see a clear 
distinction between sub- and supercritical values of $p$. The density at $p_c(q=1.0)$ 
is $\rho_{ab,c}(q=1.0) = 0.4366(2)$.

(3) Fig.~\ref{4d-dens} shows indeed several scaling laws. The most obvious maybe (but the 
least interesting, because this scaling is already inferred by the results given above)
is how the pinning height scales with the distance from $p_c$. More interesting for us
now is the convergence to $\rho_{ab,c}$ at small times seen in Fig.~\ref{4d-dens}b.
There the data of Fig.~\ref{4d-dens}a are just re-plotted as $\rho_{ab}(z) - \rho_{ab,c}$ 
against $\ln z$, instead of $\rho_{ab}(z)$ against $z$. We see for $p=p_c$ a power 
law with exponent $\approx -1.46(5)$. 

\begin{figure}[]
\includegraphics[width=8cm]{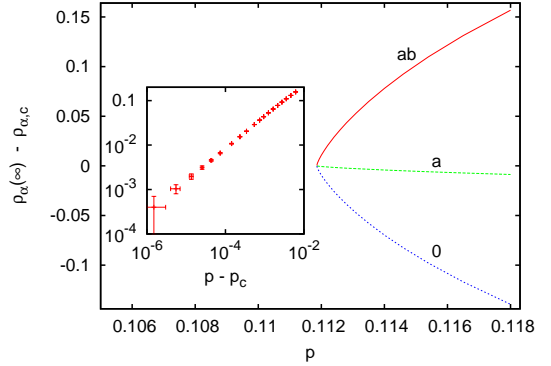}
\vglue -4mm
\caption{(color online) Asymptotic (for $z\to\infty$) densities of singly, doubly, and 
   not at all infected sites plotted against $p$, for $p > p_c(1.0)$. To show all three 
   curves on the same plot, we actually show $\rho_\alpha(\infty)-\rho_{\alpha,c}$, i.e. we 
   subtracted the critical densities. For the latter we used $\rho_{ab,c}=0.4366, \rho_{a,c}=0.03865$,
   and $\rho_{0,c} = 0.4861$. The inset shows a log-log plot of 
   $\rho_{ab}(\infty)-\rho_{ab,c}$ against $p-p_c$.}
\label{4d-dens-infty}
\end{figure}

(4) In Fig.~\ref{4d-dens-infty} we show the limiting densities $\rho_0(\infty), 
\rho_a(\infty)$ and $\rho_{ab}(\infty)$ as functions of $p$ ($\rho_b=\rho_a$ by symmetry).
As seen from the inset, we have for the $ab$-density again a power law,
\be
   \rho_{ab}(\infty)-\rho_{ab,c} \sim (p-p_c)^\beta     \label{order-param}
\ee
with $\beta = 0.73(2)$. The same power law (with the same exponent) is also seen for 
$\rho_0(\infty)$. For the density of $a$-sites, either the amplitude in this power law 
is very small or the exponent $\beta$ is zero.
   
(5) According to our scenario, the percolation transition is first order in $d=4$, while
it is continuous in $d=2$, because there is a bottleneck similar to nucleation in the former
that is absent in the latter. This bottleneck appears because the two diseases grow first 
into different directions in $d=4$, which is much less likely in $d=2$. Thus in $d=2$
the growth of the two diseases is more or less synchronized, while this is much less so
in $d=4$. Therefore we expect also the average time lag between first and second infection
to be large for those sites which finally become doubly infected.

\begin{figure}[]
\includegraphics[width=8cm]{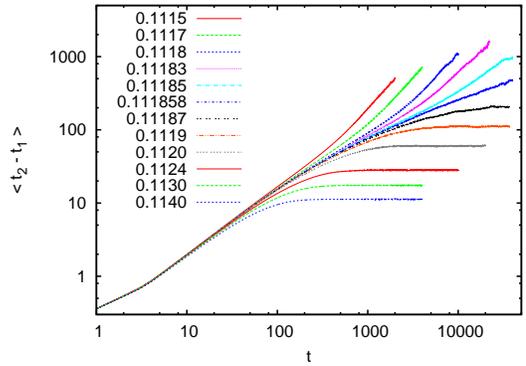}
\vglue -4mm
\caption{(color online) Average time lags between first and second infections of sites whose 
   second infection happened at time $t$, plotted against $t$.}
\label{hyper-tAB-p100}
\end{figure}

For all sites whose secondary infection happens at time $t$, we denote as 
$\langle \Delta (t) \rangle = \langle t_2-t_1 |t_2=t\rangle$ the average time lag 
between primary and secondary infection times $t_1$ and $t_2$. Data for $\langle \Delta (t) \rangle$ 
versus $t$ are shown in Fig.~\ref{hyper-tAB-p100}. We see indeed the expected behavior:
While this quantity is finite in the supercritical phase (where both diseases propagate 
together), it increases very fast in the subcritical phase, while its growth is (very)
roughly described by a power law $\langle \Delta (t) \rangle \sim t^{0.5}$ at the critical
point. For $p>p_c$ its asymptotic value seems to scale roughly as $\langle \Delta (\infty) \rangle 
\sim (p-p_c)^{-0.5}$. But we should also point out that the interpretation of these data is far from 
trivial. First of all, $\langle \Delta (t) \rangle$ increases also in $d=2$ at the critical 
point (although only logarithmically). And secondly, the increase in the supercritical region
is at large $t$ faster than linear with $t$, which cannot hold on forever, as 
$\langle \Delta (t) \rangle$ is strictly less than $t$.

\begin{figure}[]
\includegraphics[width=8cm]{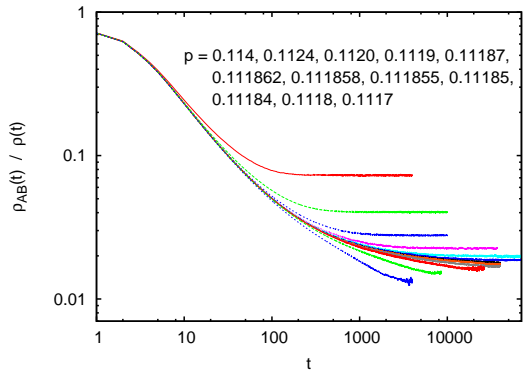}
\vglue -4mm
\caption{(color online) Fractions of sites which get both infections, and where these infections
   arrive at the same time $t$. This is analogous to Fig.~\ref{nAB_t-4d-q100} where data were 
shown for point seeds.}
\label{hyper-rAB-p100}
\end{figure}

(6) Finally we show in Fig.~\ref{hyper-rAB-p100} fractions $\rho_{AB}(t) / \rho(t)$ of 
infected (i.e. active) sites that are doubly infected, similar to the data shown in 
Fig.~\ref{nAB_t-4d-q100} for point seeds. We see a qualitatively similar behavior, except that 
now it is clear that $\rho_{AB}(t) / \rho(t)$ tends at $p=p_c$ to a finite positive value 
(equal to $0.020(1)$) when $t\to\infty$. This seems at first to be at odds with the fact that 
the average time lag between the two infections diverges in this limit, but it has an easy
intuitive explanation: Assume that $A$ and $B$ start at some time from the same site. When
they meet again at some other site, one of them (say $A$) will most likely arrive earlier  
then the other. So $B$ will find an easily infectable $a$-cluster and it will run after 
$A$. But whenever $A$ makes a detour instead of taking the shortest path, $B$ can also 
take the shortcut, and thus it will slowly catch up. Finally, there will be a small chance that 
$B$ arrives at some site simultaneously with $A$, and the whole repeats.

\begin{figure}[]
\includegraphics[width=8cm]{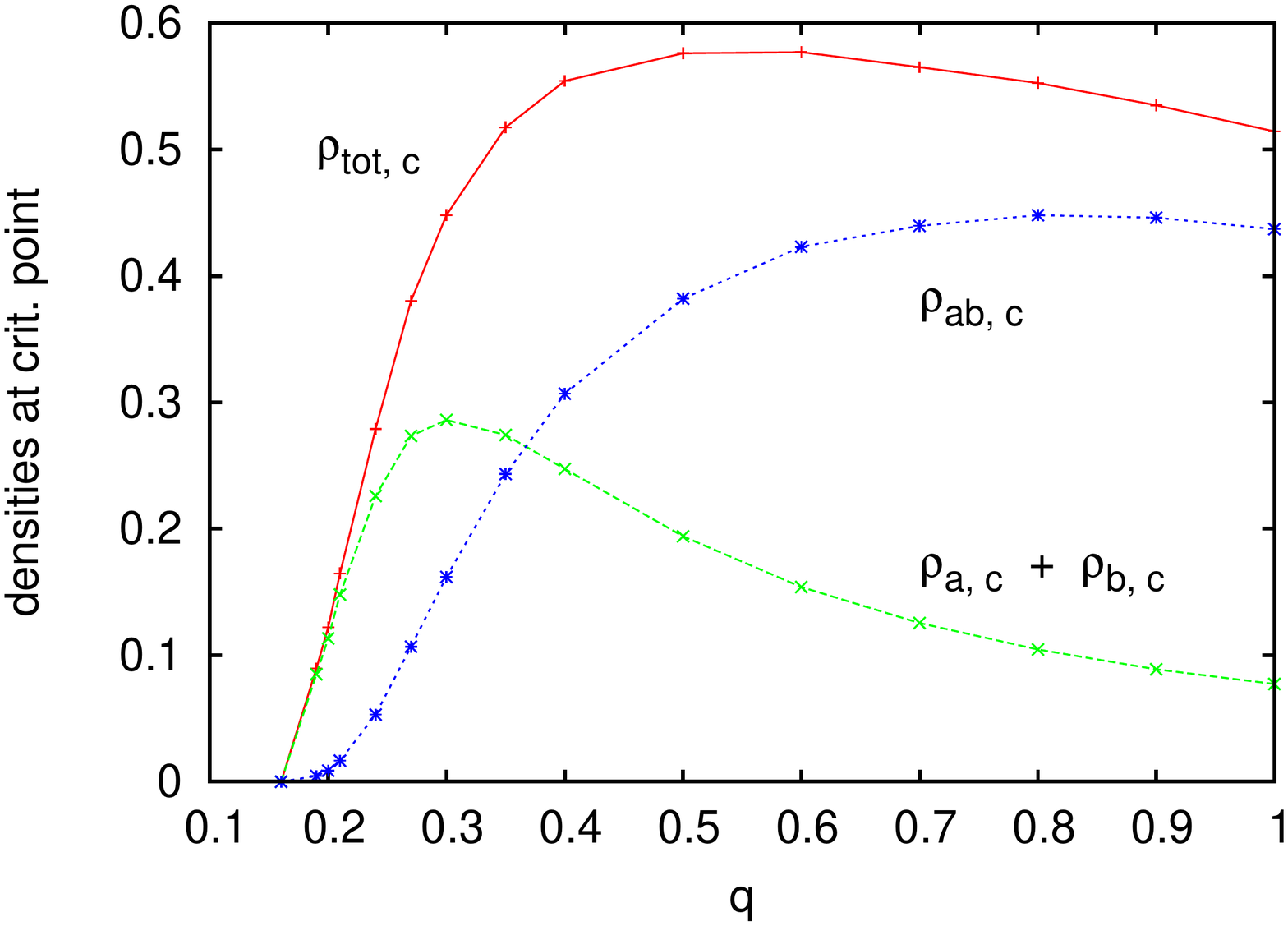}
\vglue -4mm
\caption{(color online) Densities of singly and doubly infected sites at the critical point.
They vanish only at $q=0.16013$ which is the critical point for single epidemics.}
\label{4d-dens_ab}
\end{figure}

The single site seed simulations of the previous subsection have already suggested that
the transition is discontinuous for all $q>p$, and that there is no tricritical point. 
This claim can be made much more strong by using hyperplane seeds and estimating the 
threshold densities as in Figs.~\ref{4d-dens} to \ref{4d-dens-infty}. Results 
obtained in this way are shown in Fig.~\ref{4d-dens_ab}. We see that all densities are 
indeed non-zero at threshold for all $q>p$. They are very small for small $q$ (in 
particular, $\rho_{ab}$ is tiny), but the data show clearly that there is no tricritical
point.

\subsubsection{$d>4$}

We have not made simulations in $d>4$. For $d=5$ we expect qualitatively the same result.
There, the chance that $A$ and $B$ can spread for some finite time without interfering 
is even larger, but finally they should overlap, and then a compact cluster is formed.
This is no longer true for $d>6$, which is the upper critical dimension for OP. For 
$d>6$ multiple infinite clusters can coexist, and thus $A$ and $B$ can spread forever 
without cooperating. This might lead to a scenario with a tricritical point $q_{\rm tri}$.
For $q>q_{\rm tri}$ cooperativity would dominate, leading to $p_c(q) < p_c$. This would 
then prevent single diseases from spreading at $p_c(q)$, and one has a first order 
transition. For $q<q_{\rm tri}$, in contrast, it would be entropically favorable for 
the epidemics to spread apart, leading to $p_c(q) = p_c$ and second order transitions.

This seems to be at odds with the fact that there are first order transitions on random
(ER) networks, but this is easily explained by the different limits taken in the two cases.
On finite-dimensional lattices we always consider first the thermodynamic limit of 
infinite system size  before we take the limit $t\to\infty$. On random graphs, in contrast, 
we take first the infinite time limit and let then the system size diverge. If we would 
also take the limit of infinite system size first for sparse random graphs, we would 
end up with trees for which there are indeed no first order transitions.

\subsection{Three dimensions}

We left the case $d=3$ to the end -- not because it is the least interesting, but because it is the 
most puzzling. And we wanted first to be sure that we can numerically distinguish first and second 
order transitions, and that we understand the basic mechanisms behind them.

\subsubsection{Point seeds}

\begin{figure}[]
\includegraphics[width=8cm]{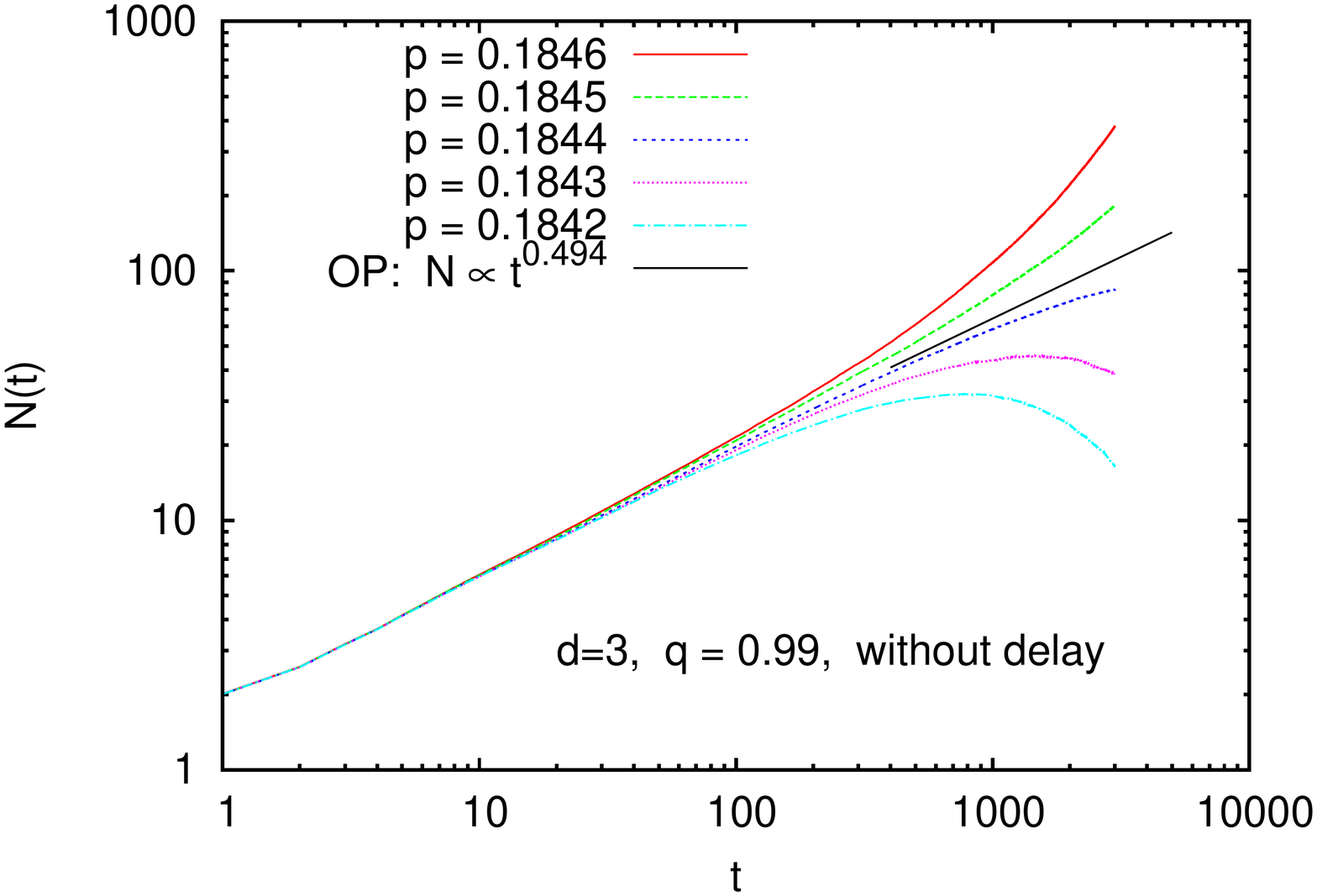}
\vglue -8mm
\includegraphics[width=8cm]{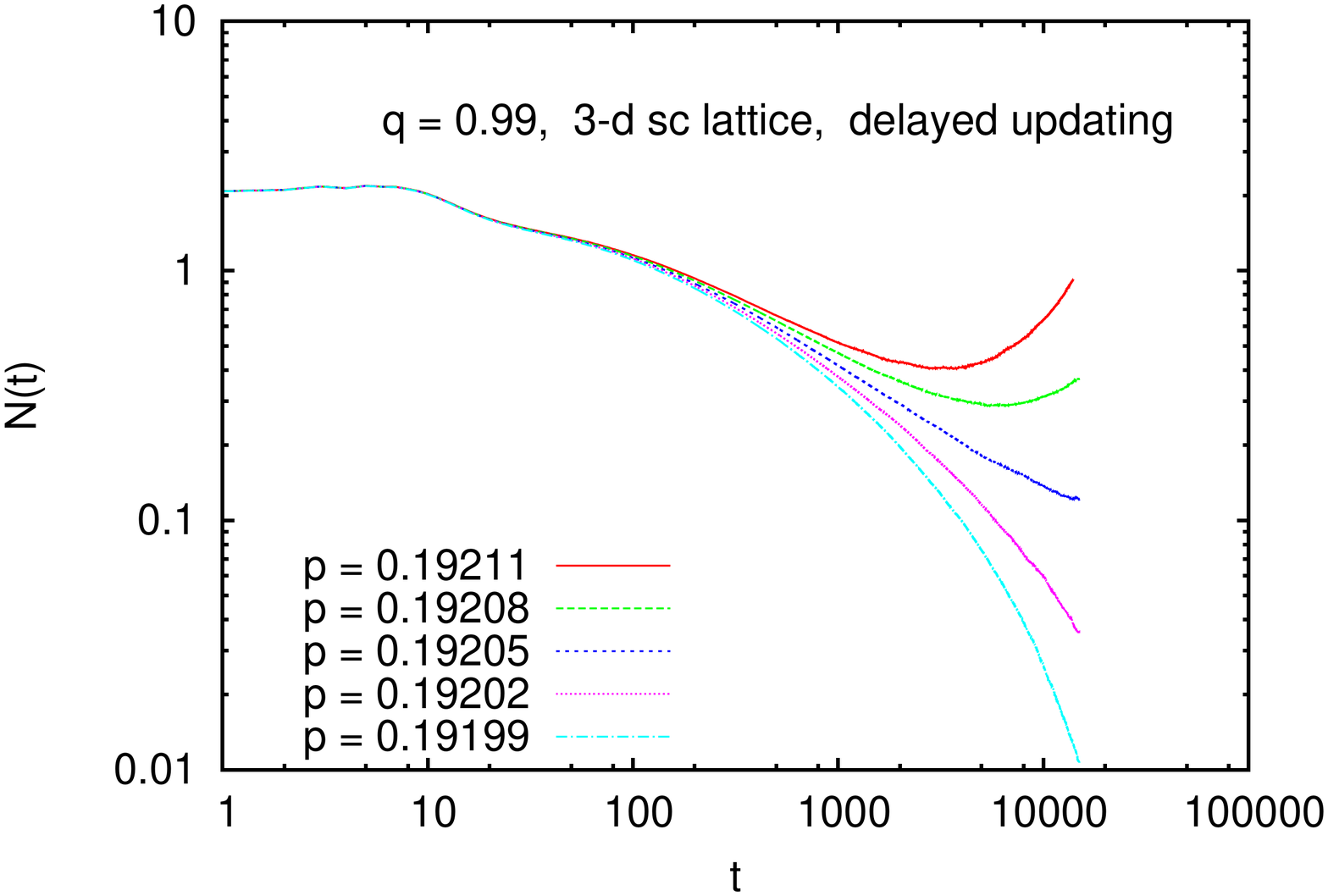}
\vglue -4mm
\caption{(color online) Log-log plot of $N(t)$ against $t$, for cooperative percolation on the simple 
   cubic $3-d$ lattice with $q=0.99$. Each curve is based on many runs with double infected point seeds, 
   and care was taken that the infected cluster never reached the boundary. Error bars are smaller than the
   line thicknesses. \\
   Panel (a): Data obtained by the algorithm without delay. The solid straight line represents the 
   scaling for OP. 
   Panel (b) shows data for the algorithm with delay.}
\label{3d-N_t_99}
\end{figure}

In Fig.~\ref{3d-N_t_99}a we show $N(t)$ versus $t$, for $q=0.99$ and the algorithm without delay. 
Seeds were single points. The solid straight line shows the scaling $\rho(t) \sim t^{0.494}$ for OP 
\cite{grassberger:1992a}. Obviously this presents a perfect fit for $p_c = 0.18443(2)$. Thus we 
conclude that even with very strong cooperativity, the percolation transition is second order and 
in the OP universality class. The same conclusion was drawn from the survival probability $P(t)$ 
and from runs starting with plane seeds (data not shown).

The situation is more complicated for the algorithm with delay. Data analogous to those in 
Fig.~\ref{3d-N_t_99}a are shown in Fig.~\ref{3d-N_t_99}b. Again the simple cubic lattice 
is used with point seeds. These data indicate clearly a first order transition at 
$p_c(q) = 0.19202(5)$. The experience of the 4-d simulations might warn us that this is slightly 
overestimated, but at least a second order transition seem definitely ruled out.

\begin{figure}[]
\includegraphics[width=8cm]{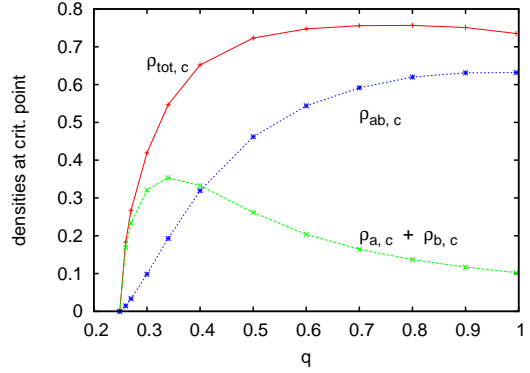}
\vglue -4mm
\caption{(color online) Analogous to Fig.~\ref{4d-dens_ab}, but now for the sc lattice updated with
   algorithm SU (i.e., with delay). In contrast to Fig.~\ref{4d-dens_ab}, these data were obtained 
   by starting from single seeds, but using the more precise critical point estimates obtained with 
   planar seeds.}
\label{3d-dens}
\end{figure}

As for the $d=4$ case, we expect a better estimate for $p_c(q)$ from seeds which form an entire plane
of size $L\times L$. We will show such data later. But before that, we shall use the precise $p_c(q)$ 
values obtained in that way for the entire range $p_c < q \leq 1$ to estimate the critical densities. 
A plot analogous to Fig.~\ref{4d-dens_ab} (which shows the data for $d=4$) is given in 
Fig.~\ref{3d-dens}. Again we see clearly that there is no tricritical point, and the transition 
is discontinuous in the entire range $0.24881< q \leq 1$.

\begin{figure}[]
\includegraphics[width=8cm]{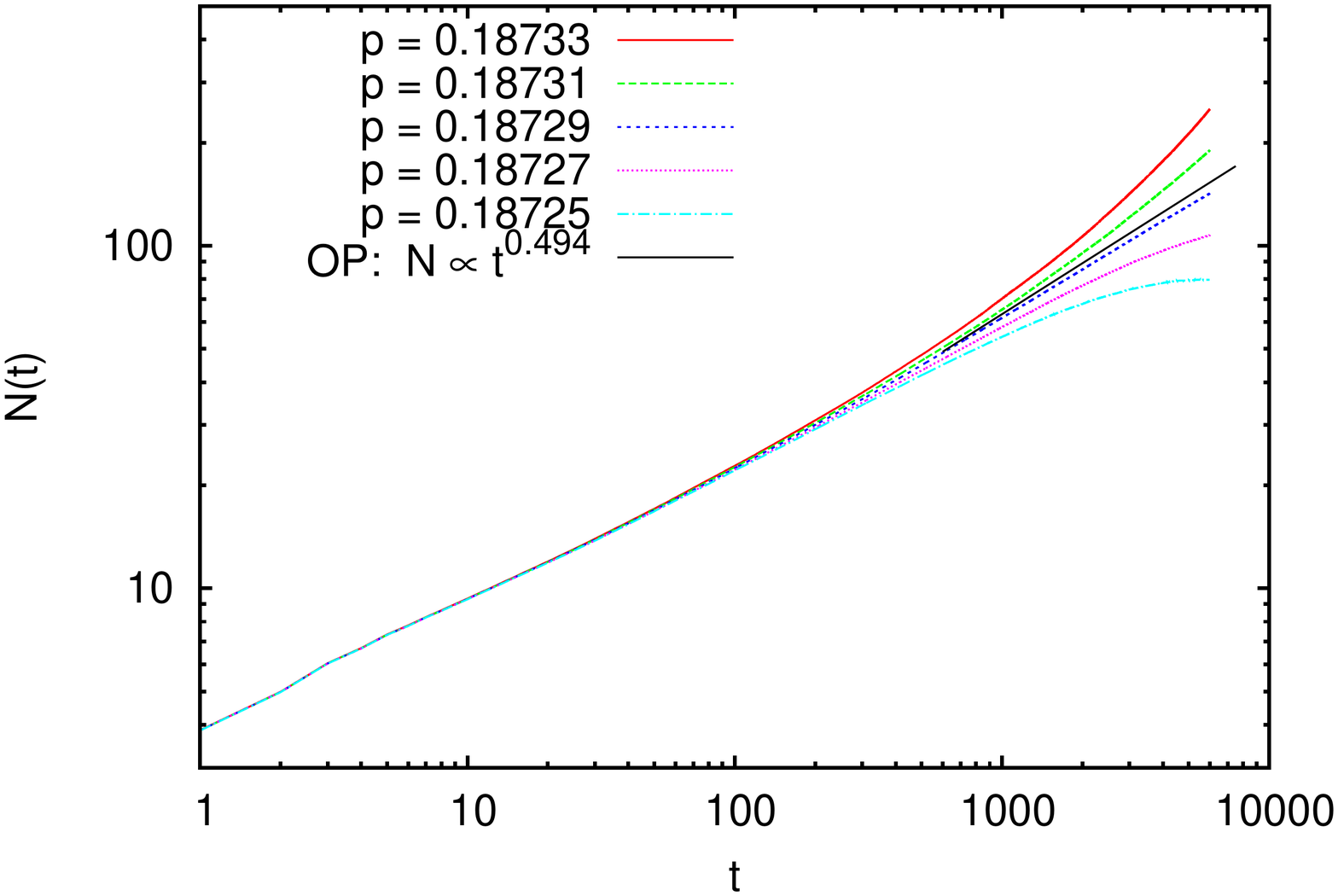}
\vglue -4mm
\caption{(color online) Analogous to Fig.~\ref{3d-N_t_99}b, but now the epidemics start at
   two neighboring sites on the simple cubic lattice. As in Fig.~\ref{3d-N_t_99}a, the solid straight line 
   represents the scaling for OP.}
\label{3d-twoseeds}
\end{figure}

Consider now the case where 
the seed consists not of one doubly infected site, but of two singly infected 
neighboring sites ${\bf x}$ and ${\bf x}+{\bf e}_x$. One of these sites has disease $A$, and the other 
has disease $B$. One should naively expect not much difference, but the data -- shown in 
Fig.~\ref{3d-twoseeds} -- look completely different. There is no longer any indication of a first 
order transition. Rather, the data are again -- as in the model without latency -- perfectly in agreement
with OP, as is also indicated by the same straight line as in Fig.~\ref{3d-N_t_99}a.

\begin{figure}[]
\includegraphics[width=8cm]{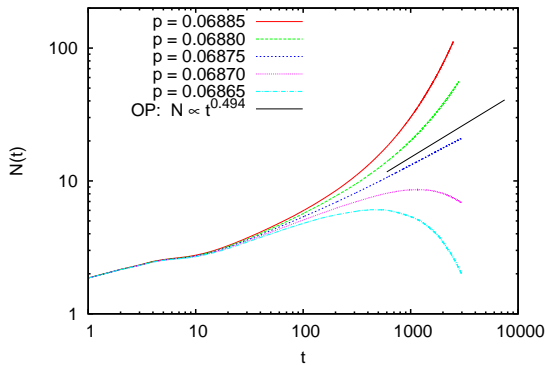}
\vglue -4mm
\caption{(color online) Log-log plot of $N(t)$ for the algorithm with delay, using again a single
   doubly infected seed site (but, for a small change, $q=1$). But now any site could infect 14 other
   sites as described in the text.}
\label{bcc}
\end{figure}

Before we go to explain this puzzling behavior, we show another puzzle in Fig.~\ref{bcc}. There,
again $N(t)$ for epidemics starting from a single point simulated with delay are shown as in 
Fig.~\ref{3d-N_t_99}a, but in contrast now the lattice is not the simple cubic lattice with nearest 
neighbor contacts. Rather, each site ${\bf x}$ can now infect 14 neighbors ${\bf x}+{\bf e}$, where
${\bf e} \in \{(\pm 1,\pm 1,\pm 1), (\pm 2,0,0), (0,\pm 2,0), (0,0,\pm 2)\}$. The last 8 of 
these are neighbors in a body-centered (bcc) lattice, while the first six are next-nearest neighbor bonds
on the bcc lattice. This time there is again 
no indication of a first order transition, and the data are again fully compatible with OP.

After these findings we went back to the algorithm without latency, to check whether similar complications
arise also there. They don't. In that case the transition is robustly of second order, independently 
of the seed and of the lattice type. Also, on the bcc lattice with next-nearest neighbor bonds the 
transition remains second order, if the two diseases start on neighboring points. On the other hand,
on the sc lattice the transition seems always second order when the epidemics start from two sites
with odd Manhattan distance, while it is first order whenever the distance is even.

Although this looks all very strange, an explanation is easy -- although, as we shall see, it can only 
be part of the story. A first hint comes from the fact that 
the sc lattice is, in contrast to the bbc lattice with next-nearest neighbor bonds, bipartite. Therefore
sites on the sc lattice can, like sites on a checker board, be colored black and white or odd \& even.
If the origin is even, then any path from the origin to any even site has an even length, while
all paths to odd sites have odd lengths.

In the algorithm with delay, cooperativity is active only when both diseases try to infect a site 
at {\it different} times. When they arrive at the same time, then there is no cooperativity due to 
the latency. Consider now an even site $i$, when the seed
is the doubly infected origin. Then cooperativity is not effective, if both diseases reach $i$ along
paths of equal lengths. If infections propagate largely along shortest paths, this then reduces
cooperativity substantially. This should not be relevant for very late times, since then most paths
will be longer than minimal ones. But it should be relevant at intermediate times, where it reduces
the effective cooperativity. Thus spreading passes through a difficult intermediate ``bottleneck" phase, 
resulting in a first order transition.

This argument obviously does not apply when the two diseases start at different points which are 
separated by an odd Manhattan distance (i.e., on sites of different parity). In that case they arrive
at any site at different time anyhow, and the distinction between the two algorithms no longer plays 
a big role. 

On non-bipartite lattices, finally, different paths between the same two points can have both
even and odd lengths, and thus the diseases can arrive with any time lag. Now there is still a 
difference between the algorithms with and without delay, but it is much reduced when compared to 
bipartite lattices. Our finding that the transition is second order on the nnn bcc lattice is thus 
non-trivial (a priori, it could have been different), but not very surprising either.

\begin{figure}[]
\includegraphics[width=8cm]{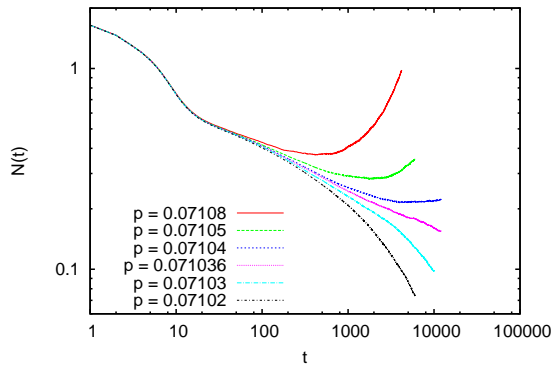}
\vglue -4mm
\caption{(color online) Log-log plot of $N(t)$ for the algorithm with delay, using again a single
   doubly infected seed site with $q=1$. But now any site can infect 12 other 
   sites which are distances 1 and 2 away on the coordinate axes (see text).}
\label{nnn}
\end{figure}

While all this sounds convincing, we should warn the reader that things are actually no so clear.
This is seen by replacing the sc lattice with nearest neighbor links by yet another lattice:
The sc lattice where each site ${\bf x}$ has 12 neighbors ${\bf x}+{\bf e}$ and ${\bf x}+2{\bf e}$, 
where ${\bf e} \in \{(\pm 1,0,0), (0,\pm 1,0), (0,0,\pm 1)\}$.
As the bcc lattice with additional next-nearest neighbors, this is not bipartite and thus 
according to our arguments this should have no ``nucleation" phase and should show thus a second order 
transition in the OP universality class. But the data shown in Fig.~\ref{nnn} definitely do not 
show the latter. Rather they suggest a first order transition with a
very weakly pronounced bottleneck (notice the different y-axis scales in Figs.~\ref{nnn} 
and \ref{3d-N_t_99}b).

\subsubsection{Plane seeds}

For the algorithm without delay we verified that indeed the transition is continuous and in the 
OP universality class, as expected from the point seed simulations. We do not discuss this further,
and consider only the algorithm with delay.
We first discuss simulations on the sc lattice with nearest neighbors only.

If we start with an entire doubly infected plane as in subsection \ref{hyperplane}, every point is connected 
to the seed by paths of even and odd length. By the above argument we expect that there will be a
second order transition in this case. This was indeed verified (data not shown). In order to obtain
a first order transition we  
must make sure that all paths from any seed site to a fixed target site have the same parity. This is 
the case if we color the base plane $z=0$ like a checkerboard and start with all black sites 
doubly infected, while all white sites are susceptible. When simulating this, we of course have to 
make sure that bipartivity is not broken by the lateral boundary conditions. This would be the case
for naive helical b.c. (in which case we indeed observed a cross-over from one asymptotics to the other, 
when $h \approx L$). But bipartivity is conserved by helical b.c. in the horizontal plane, if we 
use an even number of sites (we used planes with $N_0 =2^k$ sites, with $k=19\ldots 23$), but 
use as neighbors of site $i$ the sites $i\pm 1$ and $i\pm L$ (both modulo $N_0$) with odd $L$.
More precisely, we used $L = \sqrt{N_0}-1$ when $k$ is even, and the closest odd number to 
$\sqrt{N_0}$ when $k$ is odd.

\begin{figure}[]
\includegraphics[width=8cm]{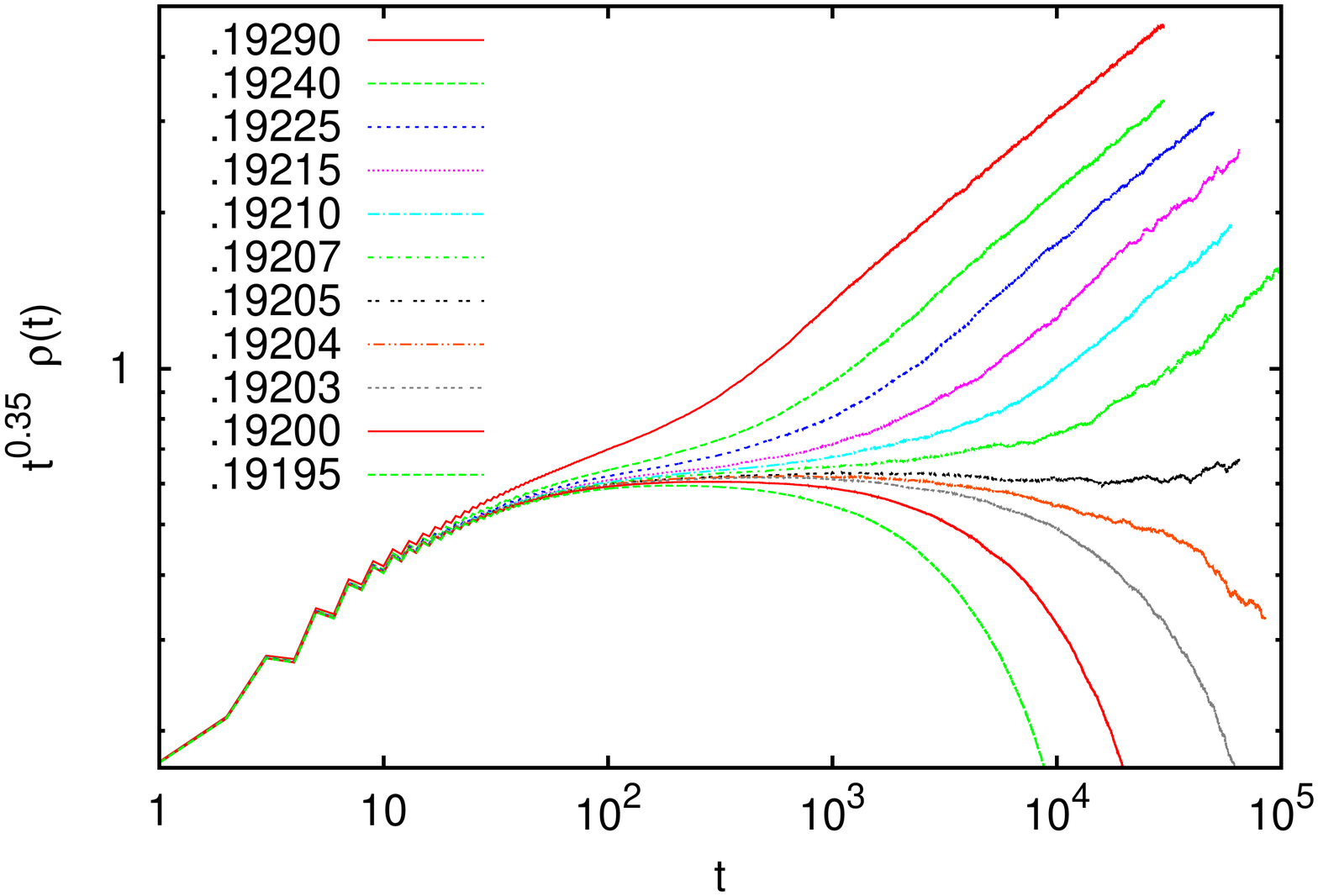}
\vglue -8mm
\includegraphics[width=8cm]{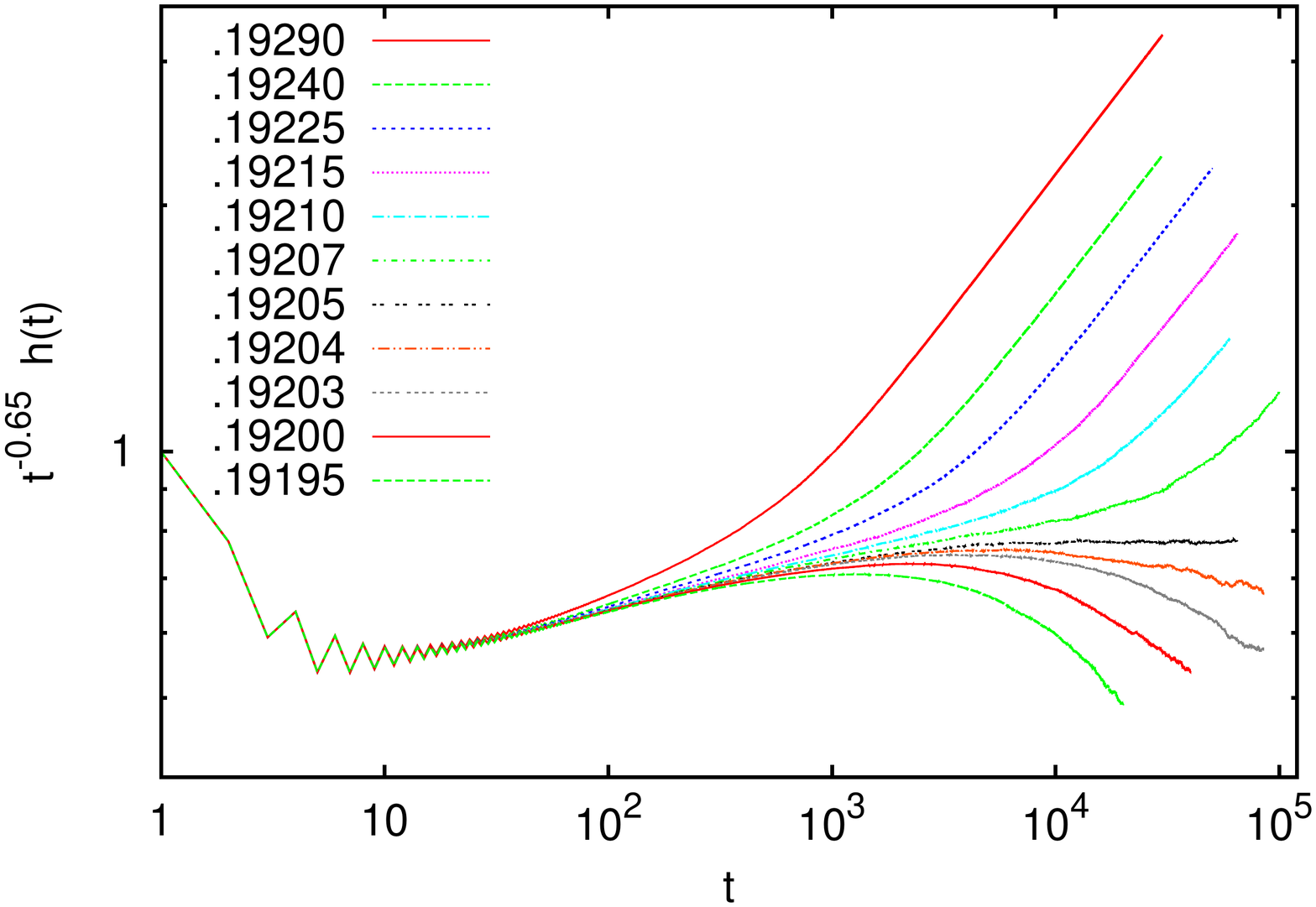}
\vglue -4mm
\caption{(color online) Log-log plot of densities of infected sites and of their average heights versus $t$.
   We used here the algorithm with delay and with $q=0.99$. The sc lattice was used as in 
   Figs.~\ref{3d-N_t_99}b and \ref{3d-twoseeds}, but now the
   seed consisted of the set of all even points on an entire $L\times L$ base surface of an 
   $L\times L\times L_z$ cuboid  with helical lateral boundary conditions as described in the text. Sizes 
   of the base plane ranged from $2^{20}$ to $2^{23}$ sites. As in the 4-dimensional case, we actually 
   plotted the data multiplied by a suitable power of $t$ which makes the critical curve roughly horizontal
   for large $t$. Panel (a) shows $\rho(t) = N(t)/L^2$, while panel (b) shows $\rho(t)$.}
\label{3d-plane-Nh_t}
\end{figure}

Data for $\rho(t)$ and $h(t)$ obtained in this way are shown in Fig.~\ref{3d-plane-Nh_t}
In order to compare with the point seed simulations we used again $q=0.99$. 
As in Figs.~\ref{4d-hyper-N_t}b and \ref{4d-hyper-h_t}b
we multiplied the data by suitable powers $t^\delta$ and $t^{-\nu/\nu_t}$ to make the curves for $p=p_c$
(approximately) horizontal at large $t$. The powers were again constraint to satisfy $\delta+\nu/\nu_t=1$,
as required by the compactness of the infected cluster. The chosen values $\delta=0.35(2), \nu/\nu_t=0.65(2)$
are seen to give a decent fit, although it is -- as in four dimensions -- far from perfect.
As in four dimensions, these simulations gave a much more precise estimate of $p_c$ than the 
point seed simulations. Our best estimate is $p_c(q=0.99) = 0.192047(3)$.

As in $d=4$, these data are compatible with the FTS ansatzes Eqs.~\ref{n-FTS} and \ref{h-FTS}, but we have 
not yet done a very detailed analysis and -- what is even more important -- we have not yet checked 
carefully that all values of $q$ give rise to transitions in the same universality class. Since surfaces
near the pinning point are more rough in $d=3$ than in $d=4$, also finite size corrections are more 
important for those sizes that are presently feasible. We hope to make a more complete analysis of  the 
$3-d$ model in a future publication.

Here we add just a few more remarks:

\begin{figure}[]
\includegraphics[width=8cm]{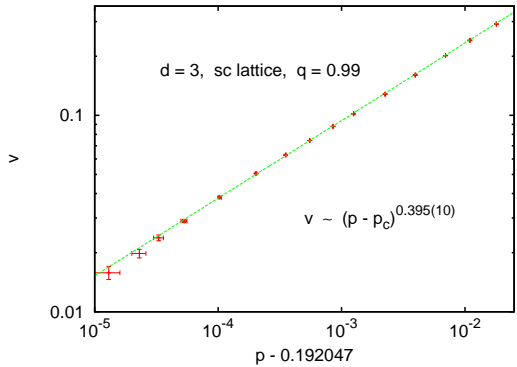}
\vglue -4mm
\caption{(color online) Log-log plot of the speed $v = \lim_{t\to\infty}h(t)/t$ against $p-p_c$. The straight 
   line represents a power law fit for $0.19215\leq p \leq 0.1922$. For $p<0.1921$ the data are 
   unreliable due possible finite size and finite time corrections.}
\label{3d-speed}
\end{figure}

(1) We measured quite carefully the scaling of the interface velocity of the interface in the 
supercritical phase. It is again obtained by using the data in the upper right corners of 
Fig.~\ref{3d-plane-Nh_t}b. Results are shown in Fig.~\ref{3d-speed}, and show a very clean power law,
\be
   v \sim (p - p_c(q))^{\nu_t - \nu}
\ee
with $\nu_t - \nu = 0.395(10)$. As in $d=4$, this seems to rule out the possibility that our model is 
in the universality class of critically pinned rough interfaces with a single field and no overhangs 
\cite{le_Doussal:2002,tang:2009}, where $\nu_t - \nu = 0.64(2)$ \cite{Nattermann:1992,Leschhorn:1997}.

(2) We found again that $\langle \Delta (t) \rangle$ seems to approach finite positive values in 
the supercritical phase and that these values scale with the 
distance from the critical point. In the subcritical phase (including the critical point itself) they 
diverge as $t\to\infty$.

\begin{figure}[]
\includegraphics[width=8cm]{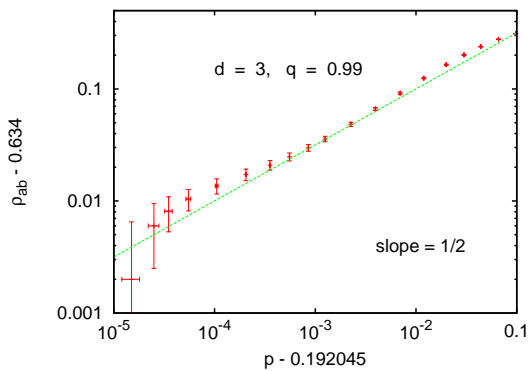}
\vglue -4mm
\caption{(color online) Log-log plot of $\rho_{ab}(\infty)-\rho_{ab,c}$ against $p-p_c$, similar 
   to the inset of Fig.~\ref{4d-dens-infty}. The huge error bars on the points for small $p$ are 
   due to possible finite size \& time corrections.}
\label{3d-dens-ab}
\end{figure}

(3) After the infection has died out, the densities at given height $z$ show a 
behavior qualitatively similar to Figs.~\ref{4d-dens}, although the data 
were much less clean due to the larger finite size corrections. In particular, due to huge
corrections to scaling we were not able
to give a precise estimate of the order parameter exponent $\beta$ defined in Eq.~\ref{order-param}.
We can only say for sure that it is $<0.5$ (see Fig.~\ref{3d-dens-ab}).

Finally, we made also simulations with an infected hyperplane for the last model discussed in the 
previous subsection, where the infection can pass to sites that are distances 1 and 2 away in 
the six coordinate directions. The precise threshold (for $q=1$) turned out to be $p_c = 0.071040(1)$,
and $\delta = 0.38(2)$. The latter is compatible with the value on the sc lattice, which suggests
that both models might be in the same universality class in spite of the big differences between
Figs.~\ref{3d-N_t_99}b and \ref{nnn}.

\subsubsection{Further indications for hybridicity}

Although it should be clear by now that all ``first order" transitions discussed in this paper are 
indeed hybrid, there is one aspect in which the 3-d model with delay is strikingly different 
from the situation in 4-d. For point seeds in $d=4$, the behavior of $N(t)$, $P(t)$, and $P_{\rm ab}$
are all reminiscent of nucleation (for example, see Figs. \ref{n_t-4d-q}a for $N(t)$ and 
\ref{4d-P_ab} for $P_{\rm ab}$; the behavior of $P(t)$ is, near the transition point, very similar 
to that of $N(t)$ and is not shown here). All these observables do not show power laws but
rather exponentials or stretched exponentials (with our precision, we cannot distinguish
between these).

\begin{figure}[]
\includegraphics[width=8.5cm]{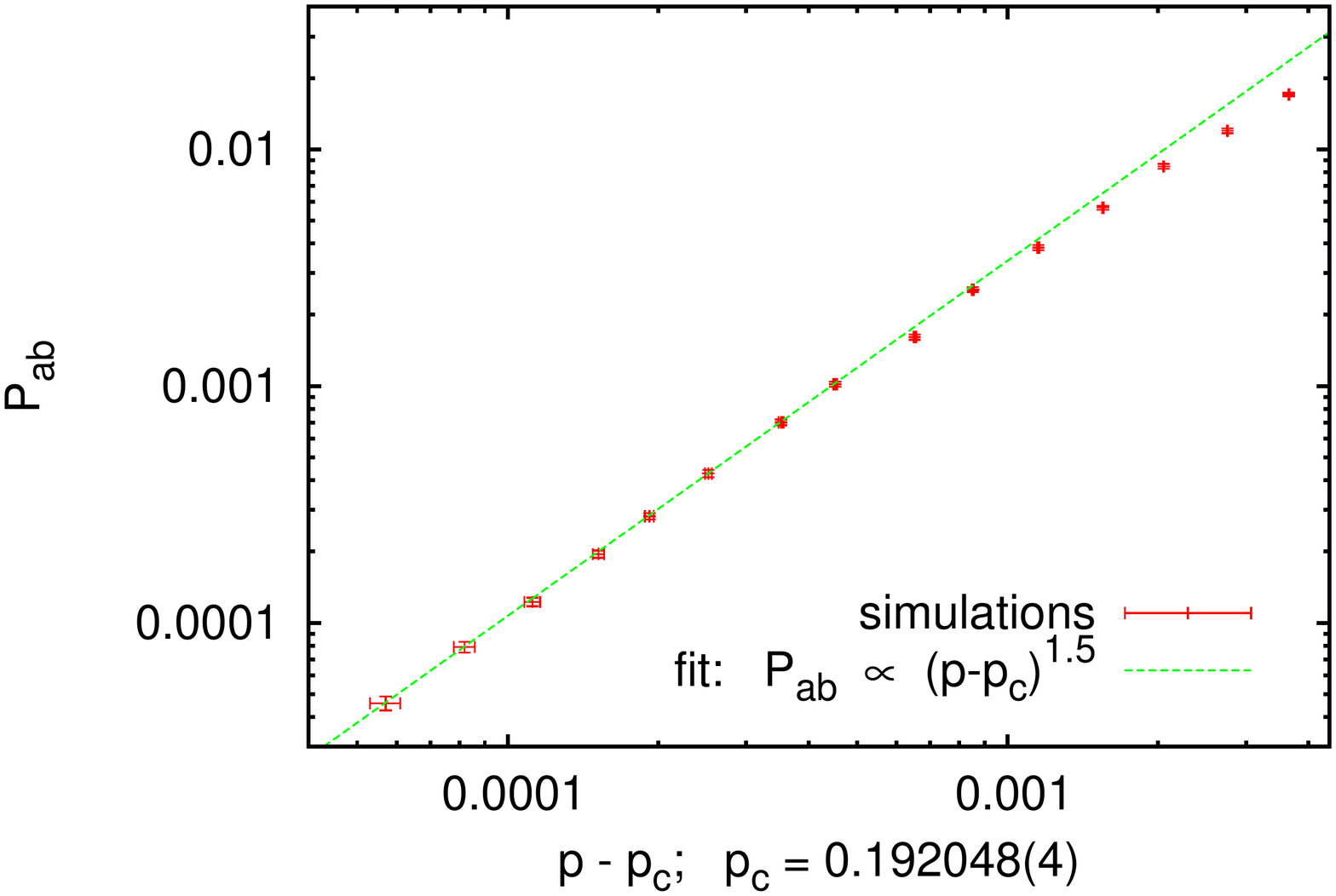}
\vglue -4mm
\caption{(color online) Log-log plot of $P_{\rm ab}$ against $p-p_c$. In contrast to the 4-d case
   (Fig.~\ref{4d-P_ab}) we now see a clear power law, with exponent compatible with $3/2$.
   For this and the following two figures we used $q=0.99$.} 
\label{3d-P_ab}
\end{figure}

\begin{figure}[]
\includegraphics[width=8cm]{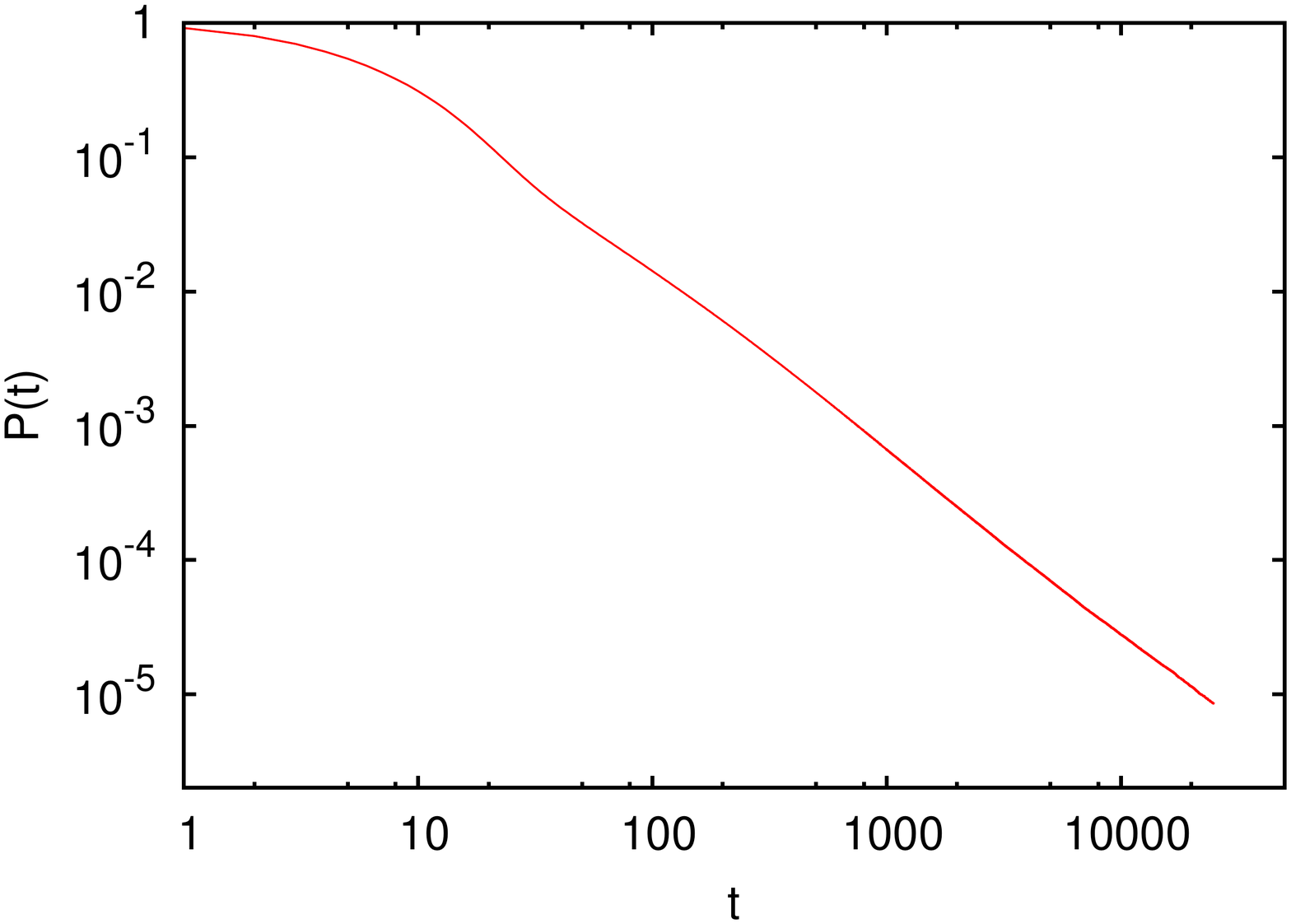}
\vglue -4mm
\caption{(color online) Log-log plot of $P(t)$ against $t$, at $p=p_c$. The data are compatible 
   $t^{-1.37(5)}$, while the analogous plot for $d=4$ would have given a faster decay, similar 
   to the $p=p_c$ curve in Fig.~\ref{n_t-4d-q}a. }
\label{3d-P_t}
\end{figure}

\begin{figure}[]
\includegraphics[width=8cm]{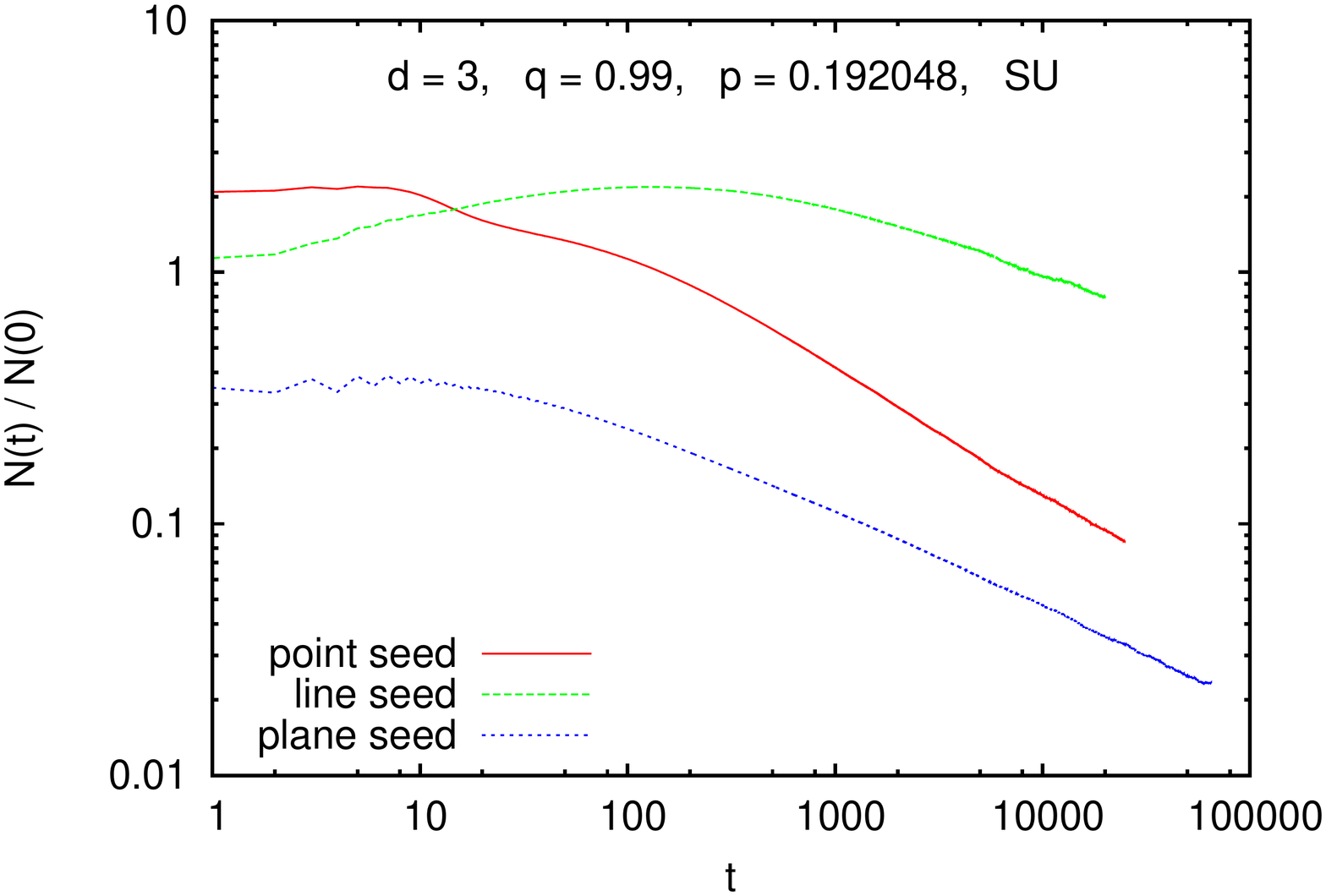}
\vglue -4mm
\caption{(color online) Log-log plot of $N(t)/N(0)$ against $t$, at $p=p_c$, for three types
   of initial conditions: Point seeds, planar seeds, and seeds consisting of every second 
   point on a long straight line. In spite of the large corrections to scaling, the data
   seem compatible with a power law with common exponent $-0.38(3)$.}
\label{strips_etc_N_t}
\end{figure}

For $d=3$, in contrast, all three observables seem to show power laws. For $P_{\rm ab}$ and $P(t)$ 
this is seen in Figs.~\ref{3d-P_ab} and \ref{3d-P_t}. For $N(t)$ we find (approximate)
power laws not only for point seeds, but also for plane and line seeds, see 
Fig.~\ref{strips_etc_N_t}. Although there are large corrections, all these are consistent
with a power law with the {\it same} exponent, $N(t)/N(0) \sim t^{-0.38(3)}$. Thus we 
do have a bottleneck in the spreading of coinfections on the sc lattice with the SU
algorithm, but this bottleneck seems not to be associated with the essential sigularitites
typically found in nucleation \cite{Debenedetti}. Very similar behavior will be 
seen in the next section.

\section{Long range infections}

While high dimensions provide the standard cross over from local to mean field behavior, 
another well known path is to go via long range interactions. In the present case of 
epidemics, this means long range infections. 

Assume that agents are placed on the sites of a $d$-dimensional regular lattice (in the 
present paper we shall only deal with $d=2$), and that the probability for an infected
site ${\bf x}$ to infect another site ${\bf y}$ follows asymptotically a power law,
\be
   p({\bf x}-{\bf y}) \sim |{\bf x}-{\bf y}|^{-\sigma-d},   \label{px}
\ee
so that the probability to infect at least one site at a distance $>r$ decays as $r^{-\sigma}$.
When $\sigma$ is large, we recover the local model, while mean field behavior holds for 
$\sigma=0$. The border between these two regimes has been studied in detail for OP, with the 
most recent and detailed simulations reported in \cite{Grassberger:2013,Grassberger:2013a}.
For critical 2-dimensional OP, mean field behavior (as far as critical exponents are concerned)
holds for $\sigma < 2/3$, while local OP behavior holds for $\sigma >2$. In between there is 
a region where the critical exponents depend on $\sigma$. Indeed, it is still an open question
whether local OP behavior holds only down to $\sigma=2$ or continues to hold down to $\sigma 
\approx 1.79$ \cite{linder:2008,Grassberger:2013a}.

In view of the dramatic differences seen in three dimensions between the models with and without
delay, we first made test runs with both schemes. We found the results again to be rather different
(scaling sets in much earlier for the update without delay), but it seemed that the transitions
were in both cases discontinuous for large $q$. Thus there does not really seem to be as much 
a difference as in $d=3$, and we did not study the model with delay any further.

In the following simulations we used the model without delay and the precise form of $p({\bf x})$ 
used in \cite{linder:2008,Grassberger:2013a}. For each site we have three potential contacts 
distributed according to Eq.~(\ref{px}), and 
the diseases are transmitted through each contact with probability $p$. Initial conditions
were such that one site had disease $A$, while one of its neighbors had disease $B$. We used 
lattices with $N=2^{31}$ sites and helical b.c. (notice that we could have gone to much
larger lattices by using hashing as in \cite{Grassberger:2013a}, but we wanted to keep
the codes simple).

\begin{figure}[]
\includegraphics[width=8cm]{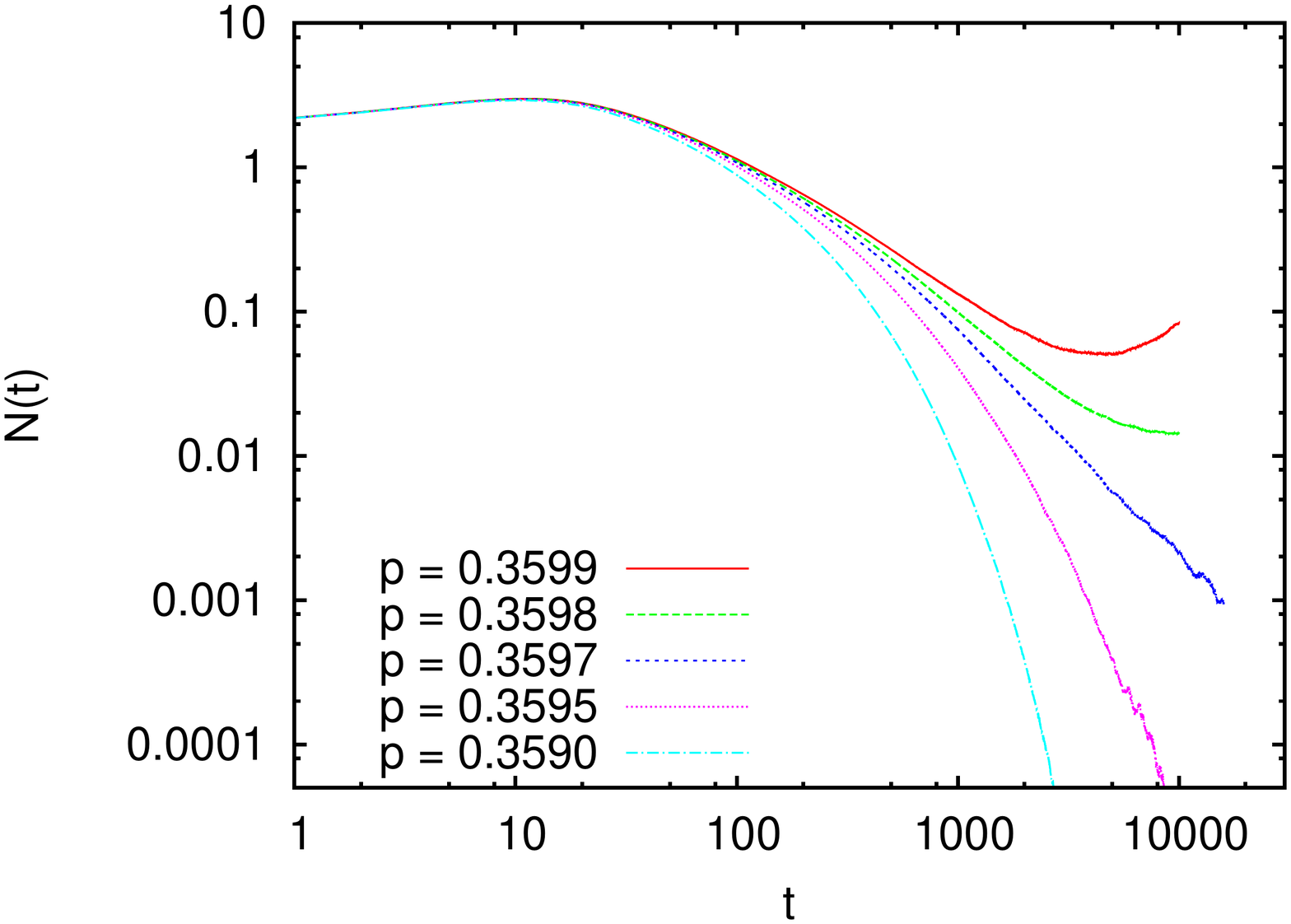}
\vglue -8mm
\includegraphics[width=8cm]{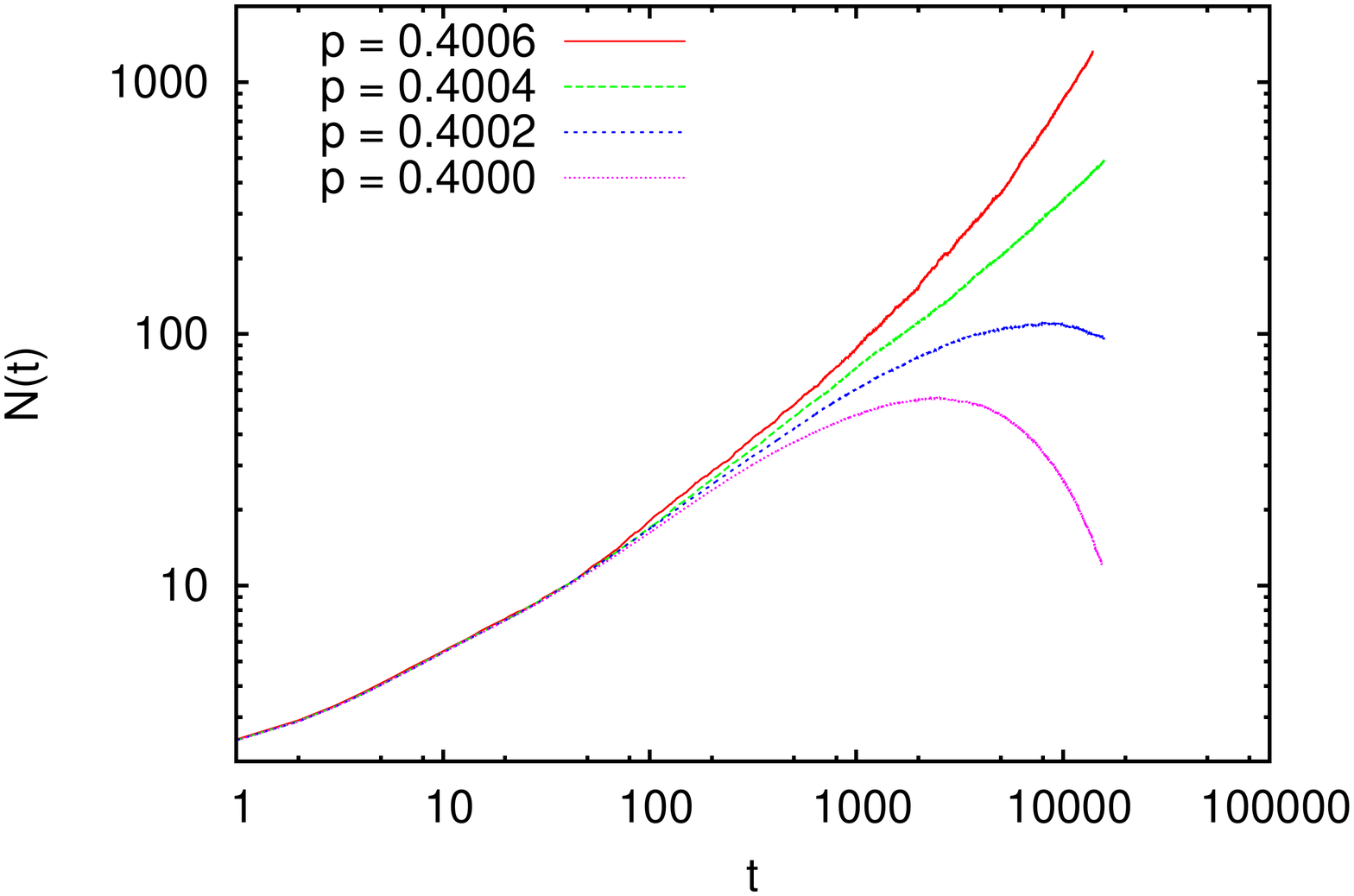}
\vglue -4mm
\caption{(color online) Log-log plots of $N(t)$ versus $t$ for $d=2$, $q = 1$, and several values 
   of $p$ close to the critical point. In each panel we used two initially infected neighbor sites, 
   one infected by $A$ and the other by $B$. In panel (a) the contacts are power law distributed 
   with $\sigma=1.1$, while $\sigma=1.5$ for panel (b).}
\label{levy-sigma}
\end{figure}

\begin{figure}[]
\includegraphics[width=8cm]{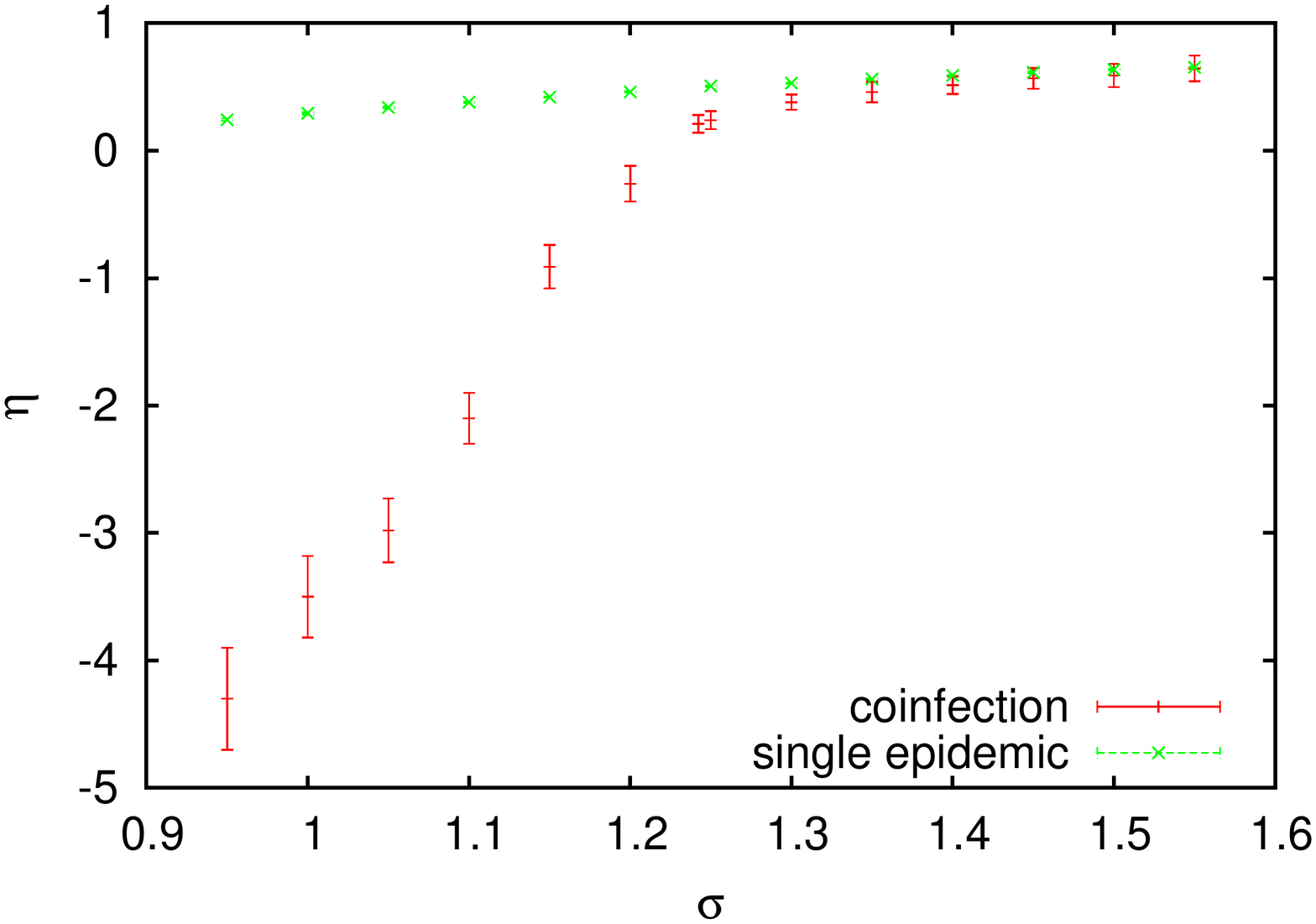}
\caption{(color online) Exponents $\eta$ for the growth of the number of infected sites, plotted
against the Levy exponent $\sigma$ controlling the decrease of the infection probability with
distance. For comparison, we also show the analogous exponents for single epidemics, taken from 
\cite{Grassberger:2013a}. The exponents were estimated from plots like the two previous ones, by 
finding for each $\sigma$ the critical value $p_c$ of $p$ where the curve looks most likely to 
become straight for $t\to\infty$. The error bars reflect essentially the uncertainty in the 
determination of $p_c$}
\label{eta-sigma}
\end{figure}

Plots of $N(t)$ versus $t$ for $q=1$ are shown in 
Figs. \ref{levy-sigma}a (for $\sigma=1.1$) and \ref{levy-sigma}b (for $\sigma=1.5$).
In each plot results are shown for several values of $p$ close to $p_c$. In both panels 
we see large corrections to scaling, but both are compatible with power laws 
\be
   N(t) \sim t^{\eta(\sigma)}               \label{eta-sig}
\ee
at the respective critical points. While $\eta >0$ for $\sigma = 1.5$, as for ordinary percolation, 
this exponent is negative for $\sigma = 1.1$, indicating a first order transition. 

This suggests that there should be a tricritical point for some value of $\sigma$ in between. 
In order to test this, we plot in Fig. \ref{eta-sigma} the estimated exponents $\eta(\sigma)$
against $\sigma$, together with the exponents for the single epidemics. As expected, the two
curves are very close for large $\sigma$ (i.e., for relatively short range contacts), since we 
have already seen that the coinfection transition is in the OP universality class when the contacts
are short range. For $\sigma \to 2/3$ we expect $\eta$ to diverge to $-\infty$, since in that limit
we should obtain the result for random graphs. Our data are compatible with this.

Our data are not precise enough to study the tricritical point in detail. It could coincide
with the point $\sigma \approx 1.25$ where the two curves in Fig. \ref{eta-sigma} seem to separate
\footnote{Naive fits to the raw data would suggest that the two curves already separate already 
for  $\sigma \geq 1.35$. But one should bear in mind that there should be strong cross-over
corrections near a tricritical point. Indeed, we found large finite-$t$ corrections for $1.25 < 
\sigma < 1.35$ which suggest that our estimates of $\eta$ may be underestimated in this region.}.
Alternatively, we could assume that the transition becomes first order when $\eta <0$, which
would give $\sigma \approx 1.24$. If the latter
is to be identified with the tricritical point, then the two curves are presumably different
for all $\sigma$ in the plotted range, but the difference is extremely small for $\sigma > 1.3$.
In any case we should stress that Eq.~(\ref{eta-sig}) describes the data for all intermediate
values of $\sigma$, including the point where $\eta(\sigma)=0$. In contrast to typical 
tricritical phenomena in other systems, the scaling is not qualitatively different 
at the tricritical point. But we should warn the reader that we do see large scaling corrections
(see footnote [61]), and as in the four-dimensional case (see Fig.~\ref{n_t-4d-q}b) and as in
single-disease infection with long range \cite{Grassberger:2013a}, this might
indicate that the true asymptotic behavior is quite different.

We also made simulations for $q<1$. We verified that OP is reached when $q$ becomes close to $p$,
and in each case the data were compatible with Eq.(\ref{eta-sig}). For all $\sigma < 1.2$ the 
values of $\eta$ smoothly passed from positive to negative values when $q$ was increased.

\section{Scale-free and small world networks}

\subsection{Scale-free networks}

Most real world networks have strong hubs and have heavy-tailed degree distribution. They 
are often modeled as scale-free, i.e. the degree distributions satisfy approximate power laws.
The most popular model leading to scale-free networks is due to 
Barab\'asi and Albert \cite{barabasi:1999} (BA). Therefore it is of practical importance to ask whether 
cooperative epidemics can show first order transition on it.

It is well known that the BA model leads to continuous percolation transitions in cases 
like interdependent networks \cite{Buldyrev:2010,baxter:2012}, networks with cooperative {\it nodes} 
\cite{Bizhani:2012}, core percolation \cite{liu:2012}, or explosive percolation 
\cite{Achlioptas:2009,radicchi:2009}. In all these cases 
other random networks either show first order transition or extremely sharp
transitions which at least superficially look like first order. This strongly suggests 
that the percolation transition for cooperative coinfections is also continuous on the 
BA model.

Indeed, simulations showed no sign of any first order transition. This is also easily understood
heuristically. For a first order transition one should have weak cooperativity at early times
(e.g. because the network is tree-like), but high cooperativity due to many long loops at 
later times. In the BA model most loops are short, and due to the strong hubs there is no 
bottleneck in the growth that could lead to nucleation.

\subsection{Small world networks}

Real-world networks not only have hubs, but they also show the ``small-world effect" (nodes are connected 
by very short paths \cite{Milgram}, so that the diameter of the network increases $\sim \log N$ or even slower), 
and they are strongly clustered. Such networks are called ``small world networks". The most popular model 
for small world networks is the Watts-Strogatz model \cite{Watts:1998}. 

In this model one starts with a regular lattice (typically with $d=1$ or $d=2$) and replaces a small
fraction $p_r$ of nearest neighbor links by random links. This creates short-cuts which, even for 
arbitrarily small $p_r$, lead to small world networks in the limit $N\to\infty$. Accordingly, critical 
phenomena like the Ising model or OP show mean field behavior, provided $p_r >0$ is kept fixed in the 
thermodynamic limit. The same is true for the Newman-Watts model where random links are just 
added to the nearest neighbor links \cite{newman1999scaling}, instead of replacing them. In order 
to find critical behavior different from mean field, one has to take $p_r \to 0$ in the limit $N\to\infty$.

This is easily understood heuristically. Assume that for some finite $L$ the effective cross-over value
between finite-$d$ and random-network behavior is at $p_r = p_r^{(L)}$. Let us now double the size,
$L\to 2L$. In models with local interactions, not much would change. Except for boundary effects, 
$p_r^{(2L)} \approx p_r^{(L)}$. But the situation is completely different in the Watts-Strogatz or 
Newman-Watts models, because links that were random on $L^d$ are no longer random on $(2L)^d$. Thus,
all these links have to be re-assigned again, and will become in average twice as long and thus also 
much more efficient in transmitting infections. If we assume that a link of length $r$ effectively helps 
to connect nodes in a region $\sim r^d$, the renormalization by a factor 2 increases the effectiveness 
by a factor $2^d$ and thus $p_r^{(2L)} \approx p_r^{(L)/2^d}$ or \cite{newman1999renormalization}
\be
    p_r^{(L)} \sim L^{-d}.     \label{ws}
\ee

Notice that this argument is rather generic and thus should apply not only to OP, but to our 
coinfection model as well.
We made some preliminary tests which indicated that this is indeed true. For fixed $N$
we found that there exists a cross-over value $p_r^\times$ such that the percolation transition 
seems to be continuous for $p_r < p_r^\times$ and discontinuous for $p_r > p_r^\times$. In 
\cite{cai2015avalanche} it was thus claimed erroneously that there is a tricritical ($L$-independent) 
value  $p_{r,c}$ at which the percolation transition changes from discrete to continuous. This 
is wrong, and the correct behavior is indeed given by Eq.~(\ref{ws}).

\section{Asymmetric Diseases and Coinfection via Multiplex Networks}
\label{multiplex}

So far we have only discussed two diseases with identical properties, which moreover spread on the same network. 
Thus they not only can infect the same nodes, but they actually use the same links. 
This is not the most relevant situation. Much more often different infections use different mechanisms for spreading
and thus also different links. For instance, HIV spreads mainly through sex contact while TB is transmitted
by coughing, speaking or sneezing via small aerosol droplets. Technically, these two diseases therefore spreads 
on two overlaid (or `multiplex') networks. Multiplex networks can of course also be used by two symmetric diseases,
but this is of minor practical interest and will not be discussed further.

This time one has many more possibilities than for symmetric diseases spreading on a single network, and 
therefore a similarly complete analysis as in the previous section is impossible. Instead, we shall discuss here 
only one simple case which demonstrates that the basic features are unchanged, and first order (or, rather, 
hybrid) transitions should be expected in many cases.

The model studied here lives on a set of $N = L\times L$ nodes, which are both connected by nearest neighbor links 
on a square lattice with helical boundary conditions and by an ER network of $\langle k\rangle N/2$ random links with 
$\langle k\rangle =4$. Disease $A$ spreads on the former, $B$ on the latter. Although both networks have the same 
(average) degree, their thresholds for single epidemics are different: $p_{c,A} = 1/2$ and $p_{c,B} = 1/4$. For 
simplicity we still assume that the primary and secondary infection rates $p$ and $q$ are the same for both diseases.

The behavior is qualitatively different for the three cases $p<p_{c,B}, p_{c,B} < p < p_{c,A}$, and 
$p> p_{c,A}$. In the last case both single diseases are supercritical, and the same is of course also true when 
they act cooperatively. In spite of this, any one (or both) of them can die, so we have four different outcomes
analogous to those shown in Fig.~\ref{ER-joint_masses} for ER networks. For $p<p_{c,B}$, at the other extreme,
both epidemics can survive -- if at all -- only due to cooperation. Actually, as we shall see, they always die,
even if cooperation is as strong as possible. Finally, for $p_{c,B} < p < p_{c,A}$ we can have large $m_a$ only 
when $m_{ab}$ and $m_b$ are also large, but $m_b$ can be large without a large $A$ outbreak.

The only case of interest for us is thus $1/4 < p< 1/2$, and we restrict ourselves also to instances where the 
$B$ epidemic is large. Furthermore, for simplicity we assume initial configurations with $N_0$ randomly located 
doubly infected sites, with $N_0 \ll N$ and $N\to\infty$ (although much of the analysis given below holds also for more 
general initial conditions). 

With these assumptions, $N_B(t)$ and $N_b(t)$ initially increase exponentially,
while $N_A(t)$ decreases exponentially. Since $N_A(0)/N$ is already small, disease $A$ cannot have therefore 
an influence on the growth of $B$. The latter stops when the density of $b$ sites, $\rho_b(t) = N_b(t)/N$,
reaches a finite value given by the positive solution of the equation \cite{Newman:2001}
\be
   \rho_b = 1 - e^{-p\langle k\rangle\rho_b}. \label{ER-rho_b}
\ee
Notice that this final $b$ density does not fluctuate in the limit $N\to\infty$ and is independent of $N_0$,
as long as $N_0/N < \rho_b$.

\begin{figure}[]
\includegraphics[width=8cm]{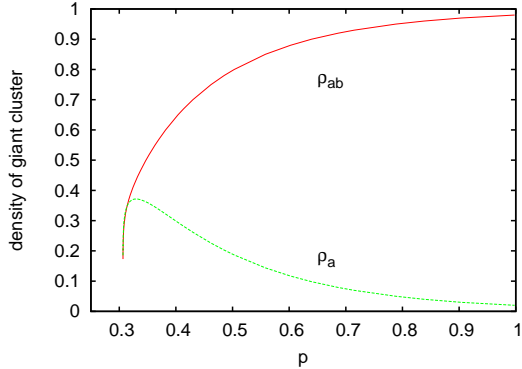}
\vglue -4mm
\caption{(color online) Plots of the $a$ and $ab$ densities of giant clusters, for multiplex networks with $q=1$.
   Within the accuracy of the plot, identical results were obtained by direct simulations and by simulating 
   disease $A$ on randomly located $b$'s with density given by Eq.~(\ref{ER-rho_b}). System sizes were between
   $2^{20}$ and $2^{25}$ sites, with no visible dependence on the size.}
\label{multiple}
\end{figure}

The fate of epidemic $A$ depends on $p,q,$ and $N_0$. If $p$ is too small (for given $q$ and $ N_0$), $A$ simply 
dies out. If $p$ is sufficiently large, however, the initial decrease of $N_A(t)$ can be turned around by the 
increased cooperativity induced by the growth of $\rho_b(t)$. A necessary condition for the latter -- and 
thus a lower bound $p^*(c)$ on the threshold $p_c(q)$ -- is found 
easily, using the fact that the $b$ sites are randomly distributed on the lattice. Thus $A$ evolves at very 
late times like mixed bond-site percolation in a frozen random background, where a fraction $\rho_b$ of sites 
can be infected with probability $q$ by any infected neighbor, while the remaining fraction $1-\rho_b$ can be 
infected with probability $p$. For given $(p,q)$ it is easy to evaluate $p^*(q)$ numerically.
We just have to solve first Eq.~(\ref{ER-rho_b}), and then we must see whether the modified 2-d percolation
problem is sub- or supercritical. For $q=1$, in particular, we obtain $p^*(1) = 0.30654(1)$. For all $p 
\in [p^*(q), 1/2]$ we can also calculate the density $\rho_a$ of the mixed percolation problem.  Values of $\rho_a$ and 
 $\rho_{ab}$ for events with giant clusters are shown in Fig.~\ref{multiple} for $q=1$. They were calculated 
both by the strategy outlines above and by direct simulations, with identical results within the line widths.
For $p$ very close to $p^*(1)$ they both satisfy the standard power law for OP
$\rho_a,\rho_{ab} \sim (p-p^*(1))^\beta$, although this is not evident from Fig.~\ref{multiple} due to 
the smallness of the scaling region.

On the other hand, it is easy to see that $A$ always dies out in the limit $N\to\infty$ -- without creating a 
giant epidemic -- if $N_0/N\to 0$ in this limit. This follows simply from the fact that it takes then infinitely 
long until $\rho_b(t)$ reaches any non-zero value. If $p<1/2$, disease $A$ has already died out by this time.

One might thus conclude that the bound $p^*(q)$ is irrelevant, but this is not true. Indeed, for any $p>p^*(q)$
there exists a critical value $\rho^*$ of the initial $B$ density, such that a giant $A$ epidemic is possible for 
$\rho_B(0) \equiv N_B(0)/N > \rho^*$. When $\rho_B(0)$ is exactly equal to $\rho^*$, then the density of 
the giant cluster is equal to $\rho_a$, whence the percolation transition at $\rho_B(0)=\rho^*$ is discontinuous 
for $p>p^*$.

For $p\to p^*(q)$ the $A$ cluster in the mixed percolation problem becomes critical, i.e. $\rho_a\to 0$. Thus 
in this limit the transition is continuous. Qualitatively, all this is similar to the situation in the 
mean field model \cite{Chen:2013}.

\section{Summary and Discussion}

Basically, we verified the mean field prediction that cooperativity in coinfections (or syndemics) can easily lead to 
very instable situations and thus also to first order transitions at the thresholds where these epidemics just barely 
can spread. This is of course not surprising, since cooperativity is akin to positive feedback which is well known 
to lead to increased instability.

\subsection{Surprises and open problems}

What is surprising, however, is the very rich phenomenology that we found, even in the simplest case of symmetric 
diseases using the same networks. For asymmetric diseases using overlay networks we expect an even richer scenario, 
which we only have scratched so far. 

One of the main surprises is that the behavior on trees and on ER networks are very different. Usually, they show
identical critical behavior. One might argue that the difference is not so surprising, because we are dealing here
with first order transitions, and there is no reason why they should be the same on trees and on ER graphs -- even 
though no such case was known before. But all first order transitions which we studied in this paper in detail 
are actually hybrid \cite{Dorogovtsev:2008,Goltsev:2006,Bizhani:2012}, and thus are both first order {\it and
critical}. This is clearly demonstrated e.g. in Fig.~\ref{ER_survive}, where we find a more or less standard 
finite time scaling which has no analogy in the case of Cayley trees. 

Another (and less well understood) surprise is the behavior in three dimensions, where two different microscopic 
implementations of the model give completely different results. In an implementation with strictly zero latency 
we found no first order transition (and critical behavior in the OP universality class), while for an implementation 
with non-zero but finite latency we found strong dependence on the lattice type and even on the initial condition. 
More precisely, on the simple cubic lattice -- which is bipartite, i.e. sites can be classified as even and odd -- 
we found a continuous transition, if the seed of the epidemic includes both even and odd sites, while a first order 
transition was found, if the seed consisted only of sites of one parity. For one non-bipartite lattice (a modified bcc lattice) 
the transition was continuous (and in the OP universality class), while for another one it was found to be discontinuous. 
More work is needed to clarify when the transition is first or second order.  Notice that bipartivity was recently found to 
be crucial to understand non-universal behavior in another percolation problem (`agglomerative percolation') 
\cite{christensen:2012,Lau:2012}, but this does not seem to be closely related to the present model.

When starting with an entire infected (hyper-)plane, any model with a first order transition (as seen from 
the bulk properties) provides a model for critically pinned driven interfaces in isotropic random media.
Such interfaces have been studied intensively during the last decades. The standard field theoretical model for them
\cite{le_Doussal:2002,tang:2009} assumes that overhangs can be neglected. This gives an upper critical 
interface dimension $d_u=4$ (i.e., the upper bulk dimension is $d=5$), and critical exponents calculated 
by renormalization group methods that agree 
well with numerical simulations \cite{Nattermann:1992,Leschhorn:1997,Rosso:2003}. But it is not clear 
whether the no overhangs assumption is relevant or not, i.e. whether this model describes also realistic 
cases where overhangs occur. Both in three and in four dimensions, it seems that our model (which does of course
allow overhangs) gives critical exponents that do {\it not} agree with 
\cite{le_Doussal:2002,tang:2009,Nattermann:1992,Leschhorn:1997,Rosso:2003}. In $d=3$ the surfaces 
seem to be more rough: On finite base surfaces of size $L\times L$ with periodic boundary conditions, the 
height fluctuations are proportional to $L$ at the critical point. But more studies are needed to clarify
the situation.

A related observation is that the bulk below the surface does contain holes, which make up a substantial 
fraction of all sites. At criticality, the fraction of doubly infected sites in the bulk is less than 60
per cent in $d=3$, and is even $\approx 40 \%$ in $d=4$. These densities show a number of non-trivial 
scaling laws, none of which had been considered previously in the literature. It is an open question
whether the distribution of hole sizes shows any non-trivial scaling.

\subsection{The Role of Loops}

While the above show that there are still a large number of open questions -- indeed, like any other major work
it poses more open questions than does it solve -- there are also some features that stand out very clearly.
One of these is the role of loops. We find that loops are necessary for first order transitions, as demonstrated
clearly for Cayley trees. Indeed, it is immediately clear that two diseases starting at a single site can never 
lead to a first order transition, since they have to use the same paths of infection. Thus they can only spread
together, if each one of them could already spread by itself. Cayley trees and (sparse) ER graphs share the 
absence of {\it short} loops, which is in most cases sufficient for both to show the same critical behavior. 
But in the present case this is not true, because of the {\it long} loops present in ER graphs. 

Indeed, we claim that a necessary condition for first order transitions in the present model is the 
predominance of long loops over short ones. This explains immediately why there are first order transitions on 
$4-d$ lattices, while there are none on the square lattice with short range infection -- even if we include
infection between next and next-next nearest neighbors. Only when we allow long range infection with a 
sufficiently slow fall-off of the infection probability, we do find first order transitions in low dimensions. 
We have not shown any data, but we had also studied small-world networks \cite{Watts:1998}. There, any non-zero 
probability for long range rewiring leads to an abundance of long loops, which outnumber by far the short ones. 
As a consequence we found also there first order transitions.

The paucity of short loops together with the abundance of long loops leads essentially to a bottleneck. For 
short times both diseases cannot spread easily, and moreover they cannot effectively cooperate. Thus they start
off to propagate into different directions. But if both 
succeed to survive until long loops become important, then they will meet at some time and then any region 
infected by $A$ will be easy prey for $B$ and vice versa. For $d=4$ this bottleneck leads to phenomena
reminiscent of nucleation in metastable systems. Strangely, this seems to be completely different for 
the bottlenecks in $d=3$ which are characterized by power laws instead of the (stretched) exponentials
typical for nucleation.

\subsection{Other models: SIS, SIC, interdependent networks, and cooperative binary vapor deposition}

In the present paper we have only treated the case where both diseases by themselves are of SIR type. 
Thus, after a short infectious period the agents become healthy again and immune to the disease they already 
had -- but with increased susceptibility for the other disease. 

Alternatively, we could have assumed SIS dynamics either for one or for both diseases. In that case there
is no immunization, and the epidemic can live {\it in situ} forever. Somewhat more subtle is the case where 
$A$, say, is SIR and $B$ is SIS. In that case $B$ can thrive in situ, while $A$ has to spread for survival.
If $A$ does not survive, then $B$ is only locally affected by it and will show only the OP transition. 
On the other hand, if $A$ does survive, then $B$ will see an increasingly large favorable environment,
and the situation will resemble the model with long range memory of \cite{grassberger:1997,dammer:2004}.
It is however not a priori clear whether critical exponents will be the same.

As a next step consider a model, called `SIC' in the following (for `susceptible-infected-coinfectious') where 
hosts never heal completely. After a short illness they may not show any symptoms and they may live without 
any problems. But they still carry within themselves the pathogen in a dormant and non-aggressive form.
If another individual comes in contact with this one having a dormant $A$, the outcome of the encounter depends 
on whether the second individual has (or has had) already disease $B$ or not. If she has not, the first individual is 
unable to infect her/him. But if the second individual had already (had) $B$ (and is thus sufficiently weakened), 
then the pathogen in the first individual is 
sufficiently virulent to infect her/him. So the first host is not {\it infective}, but {\it coinfective}. 
We expect that also in this model there should be a rich zoo of possible phase transitions.

We should point out that all these models are closely related to the  model of 
interdependent networks of 
Ref.~\cite{Buldyrev:2010}. There, one considers a multiplex network where, in the simplest case, each node
consists of two subnodes which are connected by different links. If one of the subnodes dies, the other 
dies also. This is illustrated in \cite{Buldyrev:2010} by a country where each electrical power station
has associated with it in the same city also a computer station needed to control it. Power stations and 
computer stations are connected among themselves by different links. If one power station breaks down, in 
principle other stations should take over. But since also the local computer is dead, it cannot transmit 
the information. This leads then to other power stations to break down, to more computers to fail, and 
finally to a catastrophic cascade resulting in an all-encompassing black-out.

The analogy with coinfection is based on the observation that a power station failure is akin to a
disease $A$ of a city, while a computer failure is another disease denoted as $B$. In the network dependency
model, one disease immediately leads also to the other, while in our model(s) one disease only leads 
to an enhanced susceptibility for the other. In that sense, the formulation in terms of cooperative 
coinfections allows much more flexible interactions between different types of failures than the strict 
dependency assumed in \cite{Buldyrev:2010}. Working out the detailed relationships between these two
classes of models might lead to valuable insights, both into coinfection and interdependent network models.

Finally, when starting with entire infected hyperplanes, our model can be seen as describing surface growth 
near a (de-)pinning transition. In this interpretation of ordinary percolation spreading from an extended
source, it is more natural to consider the elementary growth step not as an infection of a susceptible 
site, but as occupation of an empty site by an adsorbing particle. In our case, this would lead to an 
adsorption process with two different adsorbing species, where adsorption of both together becomes more 
probable than absorption of only one of the species. While such surface growth processes are of course 
common and even technologically of interest, a special feature of our model is that a particle of species
$A$ can only adsorb on a surface that contains already $A$. This is a severe restriction and might limit
the interest in this interpretation. On the other hand, the fact that we can easily generate in this way 
interfaces with extremely ``spongy" phases below them is intriguing and might warrant further study.

\bibliography{paper}

\end{document}